\documentclass{aa} 
\usepackage{times,graphicx,color}
\usepackage[colorlinks=true,allcolors=blue]{hyperref}
\usepackage{natbib}
\bibpunct{(}{)}{;}{a}{}{,} 
\usepackage{parskip}
\usepackage{multirow}
\usepackage{enumitem}
\usepackage{array}
\usepackage{stfloats}
\usepackage{hhline}
\usepackage[tableposition=top]{caption}
\usepackage{arydshln}
\usepackage[mathscr]{euscript}
\usepackage{threeparttable}
\begin{document}

\title{Towards a fully consistent Milky Way disk model $-$ IV. The impact of Gaia DR2 and APOGEE}

\author{K. Sysoliatina\inst{1} and A. Just\inst{1}
}

\mail{Sysoliatina@uni-heidelberg.de}

\institute{$^{1}$Astronomisches Rechen-Institut, Zentrum f\"{u}r Astronomie der Universit\"{a}t Heidelberg, 
M\"{o}nchhofstr. 12--14, 69120 Heidelberg, \mbox{Germany}}

\date{Printed: \today}

\abstract 
{}
{We present an updated version of the semi-analytic Just-Jahreiß (JJ) model of the Galactic disk 
and constrain its parameters in the Solar neighbourhood.}
{The new features of the JJ model include a simple two-component gaseous disk, 
a  star-formation rate  (SFR) function of the thick disk that has been extended in time, 
and a correlation between the kinematics of molecular gas and thin-disk populations. 
Here, we study the vertical number density profiles and $W$-velocity distributions 
determined from $\sim2 \cdot 10^6$ local stars of the second \textit{Gaia} data release (DR2). 
We also investigate an apparent Hess diagram of the \textit{Gaia} DR2 stars selected in a conic volume towards the Galactic poles.
Using a stellar evolution library, we synthesise stellar populations with 
a four-slope broken power-law initial mass function (IMF), the SFR,  and an age-metallicity relation (AMR). 
The latter is consistently derived with the observed 
metallicity distribution of the local Red Clump (RC) giants from the Apache Point Observatory Galactic Evolution Experiment (APOGEE). 
Working within a Bayesian approach, we sample the posterior probability distribution 
in a multidimensional parameter space using the Markov chain Monte Carlo (MCMC) method. 
}
{We find that the spatial distribution and motion of the \textit{Gaia} DR2 stars
imply two recent SF bursts centered at ages of \mbox{$\sim0.5$ Gyr} and \mbox{$\sim3$ Gyr} and characterised by a 
\mbox{$\sim30$\%} and \mbox{$\sim55$\%} SF enhancement,  respectively, relative to a 
monotonously declining SFR continuum. The stellar populations associated with this SF excess 
are found to be dynamically hot for their age: they have $W$-velocity dispersions of 
\mbox{$\sim12.5$ km s$^{-1}$} and \mbox{$\sim26$ km s$^{-1}$}.  
The new JJ model is able to reproduce the local star counts with an accuracy of \mbox{$\sim5$-\%}. 
}
{Using \textit{Gaia} DR2 data, we self-consistently constrained 22 parameters of the updated JJ model.
Our optimised model predicts two SF bursts within the last $\sim4$ Gyr, which may point to recent episodes of gas infall. 
}


\authorrunning{K. Sysoliatina and A. Just}
\titlerunning{The local model based on Gaia DR2 and APOGEE}

\maketitle 

\defcitealias{just10}{Paper~I}
\defcitealias{just11}{Paper~II}
\defcitealias{rybizki15}{Paper~III}

\section{Introduction}\label{sect:intro}

Our current understanding of the morphology and kinematics of the Milky Way, 
and of the physical processes that shape its structure and govern its evolution, 
is deepening  to a greater extent and faster than ever. 
Since the second data release of the astrometric mission \textit{Gaia} in April 2018 (DR2, \citealp{gaia16,gaia18}),
many new details on Galactic morphology have been revealed by studies based on the high-precision \textit{Gaia} astrometry,
in addition to studies of the data from the available high-resolution and large spectroscopic surveys, such as 
the Sloan Digital Sky Survey (SDSS), Apache Point Observatory Galactic Evolution Experiment (APOGEE, \citealp{eisenstein11,majewski17}), 
the Galactic Archaeology with HERMES (GALAH, \citealp{martell17}), 
and The Large Sky Area Multi-Object Fibre Spectroscopic Telescope Survey (LAMOST, \citealp{cui12}). 
At the same time, many of the previously known properties of the Galaxy have been re-confirmed 
and investigated to a much greater extent than ever before. 

A number of studies of local kinematics with \textit{Gaia} DR2 have reported the presence of numerous 
arc-like substructures in the velocity space ($V_\mathrm{r}$,$V_\mathrm{\phi}$)\footnote{Here and further 
($R,\phi,z$) and ($V_\mathrm{r},V_\mathrm{\phi},V_\mathrm{z}$) are coordinates and velocities in the cylindrical Galactocentric coordinate system.} 
\citep{katz18,hunt19,khanna19}.  
These kinematic signatures of the local stellar streams and moving groups are qualitatively reproduced in N-body simulations 
with a transient spiral structure in the presence of a bar \citep{hunt19,khanna19}.
Another example is a newly discovered `phase spiral' in the ($V_\mathrm{z}$,$z$) space, 
whose origin may be related to external perturbations \citep{bhawthorn19} or the secular disk evolution \citep{khoperskov19}. 
These kinematic features give us strong evidence of ongoing phase mixing processes in the present-day Galaxy, 
implying that the latter is currently out of an equilibrium state. Another non-equilibrium feature was found by \citet{bennett19},
who examined the south-north asymmetries in the vertical star count and kinematic trends in the Solar neighbourhood. 
The authors found a wave-like plane-asymmetric pattern both in star counts and stellar kinematics; 
as they are similar for all stellar populations, these waves may be a disk response to the recent interaction 
with a satellite since such disk-satellite interactions have been predicted to excite bending and breathing modes in the disk \citep{widrow14}.    

Additionally, many new stellar aggregates are identified in the \textit{Gaia} DR2 data. \citet{meingast19} 
detected a local stream located within \mbox{400 pc} from the Sun 
that consists of dynamically cold stars with a three-dimensional (3D) velocity dispersion of only \mbox{1.3 km s$^{-1}$}. 
New catalogues of OB stars and open clusters \citep{cantat-gaudin20,chen19,liu19}
allow for a better tracing of the Milky Way spiral pattern. The extensive samples of \textit{Gaia} DR2 
A- and F-type stars, as well as lower main-sequence (MS) stars, 
were used to test the plausibility of the different formation scenarios of spiral arms in the Galaxy \citep{griv20}. 
Also, the five-dimensional (5D) dynamic information from \textit{Gaia} DR2 revealed mostly young filamentary stellar groups in the Solar neighbourhood 
that reflect the last \mbox{$\sim1$ Gyr} of star formation (SF) in the disk \citep{kounkel19}.

These examples clearly show that the \textit{Gaia} data 
reveal the most realistic picture of the Milky Way that has ever been available to astronomers. 
This new, complex picture enables us to expand our knowledge of the Galaxy formation and evolution. 
However, despite of the abundance of this new observational information, 
recent years yielded relatively few studies attempting to reconstruct the global star-formation rate (SFR) and the dynamic evolution of the Milky Way. 
With our present-day understanding of the complexity of the physical processes guiding Galactic evolution, 
the task of building a concordance model of the Milky Way seems more challenging than ever.

Many modern semi-analytic Milky Way models derive the Galactic gravitational potential 
self-consistently with the assumed matter distribution, based on the combined Poisson-Boltzmann equation. 
The first model of this class described the Solar neighbourhood in terms of a multicomponent disk 
consisting of a set of isothermal stellar populations in the presence of a spheroidal halo \citep{bahcall84}. 
By introducing a parametrised SFR, a disk heating function in the form 
of an age-velocity dispersion relation (AVR), as well as an initial mass function (IMF) and an age-metallicity relation (AMR), 
we can then fully constrain the local disk evolution and predict the observed star counts. 
When applied to the different radial disk zones independently, this approach allows us to build a global model of the Galaxy. 
In such a way, for example, a multicomponent disk is described in the Besan\c con Galaxy model 
(BGM, \citealp{robin3,robin12,robin17,czekaj14,bienayme18}). The BGM is widely used in a variety of Galactic studies. 
\citet{ban16} used the model to constrain the distribution of free-floating planets towards the Galactic centre (GC).  
\citet{cabral19} relied on the BGM stellar population synthesis to derive the chemical composition of the planetary building blocks. 
In comparison with the large-scale surveys, BGM was used to constrain the IMF, local SFR, and matter density \citep{mor18} 
and proved helpful in studies of the disk formation \citep{robin16,nasello18}. It was also used to create a mock stellar catalogue
at the preparatory step to \textit{Gaia} DR2 \citep{rybizki18a}. 
Thus, such Galactic models built of independent radial zones are handy tools that allow us to investigate 
key aspects of the evolution of the Milky Way.

However, there is a growing store of evidence indicating that different radial zones of the disk may 
actively interact with each other during its evolution history. 
For example, works based on the exploration of the disk metallicity gradient \citep{wielen96}, 
elemental abundances of the local interstellar medium (ISM), and young stars \citep{nieva12}, 
as well as local kinematics \citep{sellwood02,schoenrich09}, suggest that the Sun might have migrated from 
the inner disk up to several kpc during its 4.5 Gyr lifetime. 
There are two types of processes that can cause radial redistribution of the stars of the disk. 
Firstly, gravitational scattering leads to the diffusion of orbits in phase space, which is often referred as blurring. 
Secondly, stars may resonantly interact with non-axisymmetric disk structures, such as the bar and spiral arms.
Resonant interactions mainly result in quick changes of orbital radii with essentially no change in eccentricity  
-- this is referred to as the churning mechanism. Unfortunately, 
the relative weights of the two processes as well as their overall importance 
in disk evolution still remain highly uncertain. 
The spatial resolution of modern hydrodynamic simulations of Milky Way-like galaxies does not yet allow for 
the tracing of these processes in the disk and, therefore, radial migration cannot be unambiguously quantified from the theoretical point of view. 
Several attempts have been made to constrain radial migration based on reasonable assumptions about the 
disk evolution and the observed present-day metallicity distributions across the disk \citep{minchev18,frankel18,frankel20}. 
They report a strong radial migration across the Galactic disk: for example, according to \citet{frankel20}, 32\% of 6 Gyr old stars 
have migrated more than 2.6 kpc from their birthplace. However, these studies rely on uncertain stellar ages, 
which may suffer from systematic errors, and they use a number of assumptions that are known to be simplifications 
or that may not necessarily be valid (e.g. ISM enrichment due to stellar evolution takes place at the stellar birthplace; 
SFR shape is exponentially declining).

At the moment, there is no global Milky Way model that is able to predict the detailed radial and vertical disk structure, 
incorporate stellar evolution to produce star counts, and simultaneously take into account the radial migration. 
Indeed, adding the radial migration mechanisms in their parametrised form to a BGM-like model 
adds a significant level of complexity to the whole machinery of a stellar population synthesis: 
the AMR can no longer be used as an input, but instead, a full chemical evolution model must be added to predict the ISM 
composition at each radius and each moment of time. 
We would also have a number of additional free parameters to be constrained, 
which cannot be done based solely on the local data, requiring instead an extended data sample.
In practice, the radial migration is often ignored in Galactic studies, 
such that the results derived from the data covering some range of Galactocentric distances can be interpreted as `averaged' 
due to migration over a larger distance range.

Some of the above-mentioned observational evidence from \textit{Gaia} DR2 suggests that the Milky Way experienced disk-satellite interaction episodes
that left imprints on the spatial distribution and kinematics of its stars. At the same time, disk-satellite interactions are 
expected to have an impact on the SFR, both through the direct accretion of gas and stars from the impactor and  by triggering 
SF processes in the disk by tidal forces and shocks.     
Therefore, it is plausible that the Milky Way's SFR would exhibit measurable deviations from a simple exponential 
or power-law decline during the last $\sim8$ Gyr.  
With these considerations taken into account, \citet{mor19} used $\sim2.9 \cdot 10^6$ \textit{Gaia} DR2 stars over the Hess diagram
to constrain the Galactic SFR. The SFR was given in a non-parametric form in nine age bins in the framework of BGM. 
As the SFR cannot be disentangled from the IMF, 
they also varied the slopes of the three-slope broken power-law IMF. As a result, \citet{mor19} found that the data demonstrate an excess of 
young A and F stars with respect to models having an exponentially declining SFR, such that their best model includes a SF burst \mbox{$2-3$ Gyr} ago. 
However, the obtained constraints on the burst shape and position are not very strong and, in principle, a model with almost constant SFR with a quick 
drop during the last $\sim1$ Gyr cannot be ruled out either. Also, the authors did not check the consistency between the observed kinematics and 
the dynamic heating prescribed by the BGM. In addition, the assumed dynamical heating of the disk has its own impact 
on the predicted star counts and  should, therefore, also be adapted to obtain robust SFR parameters. 

These findings motivated us to put a reliable constraint on the shape of the thin-disk SFR using our 
local semi-analytic disk model. In this study, we aim to optimise the SFR self-consistently not only with the IMF, 
but also with the AVR and AMR functions.
In previous papers of this series, we presented a semi-analytic Just-Jahrei{\ss} disk model (\mbox{JJ model} hereafter). 
The model describes the thin disk as a set of isothermal mono-age populations 
whose evolution is represented by four input functions given in analytic form: SFR, IMF, AVR, and AMR. 
The thick disk, gas, dark matter (DM), and stellar halo are also added to the total mass budget. 
The model is based on an iterative solving of the Poisson-Boltzmann equation in a thin-disk approximation.
The JJ model does not aim to reproduce the south-north and azimuth asymmetries in the local disk structure and kinematics, 
but traces the overall trends in stellar spatial distribution, vertical kinematics, and metallicity gradients averaged over large enough volumes. 
This approach allows to keep the number of model parameters relatively small and, at the same time, to describe the main properties of the 
chemical, stellar, and dynamic evolution of the disk. 

In \citeauthor{just10} (\citeyear{just10}, hereafter \mbox{\citetalias{just10}}), the kinematic parameters of the JJ model were 
calibrated against the vertical kinematics of the \mbox{\textit{Hipparcos}} MS stars \mbox{\citep{vanleeuwen07}},
which were complemented with the sample taken from the Fourth Catalogue of Nearby Stars (CNS4, \mbox{\citealp{jahreiss97}}). 
Additionally, F and G stars from the Geneva-Copenhagen Survey (GCS1, \citealp{nordstrom04}) were used to constrain the AMR shape.  
In \citeauthor{just11} (\citeyear{just11}, \mbox{\citetalias{just11}}), the SFR and thick-disk parameters were constrained via calibration 
against the SDSS star counts in the cone towards the northern Galactic pole. Their best fit corresponded to the thin-disk SFR monotonously declining 
after a peak \mbox{$\sim10$ Gyr} ago and to the thick-disk vertical density profile close to the $sech^\alpha$ law of an isothermal stellar population. 
The IMF parameters were optimised with the local \mbox{\textit{Hipparcos}} data combined with an updated version of the CNS4 
(\mbox{\citealp{rybizki15}}, \mbox{\citetalias{rybizki15}}). The latest version of the model IMF 
is a four-slope broken power-law function \citep{rybizki18}. 
Recently, the JJ model was tested in the Solar cylinder against the sample of stars 
selected from the fifth release of the Radial Velocity Experiment (RAVE DR5, \citealp{kunder17}) 
and the \textit{Tycho-Gaia} Astrometric Solution catalogue \citep{lindegren16} of the first \textit{Gaia} data release (DR1, \citealp{gaia16}). 
This helped to identify a number of non-critical, albeit non-negligible, model-to-data discrepancies \citep{sysoliatina18}. 

As we showed in \citet{sysoliatina18}, a forward modelling approach with the reddening and distance errors included in the model 
makes the model-to-data comparison challenging and computationally expensive. Therefore, to perform an efficient exploration of the 
multidimensional parameter space for this study,
we chose a Bayesian approach and sampled the posterior probability distribution with the Markov chain Monte Carlo (MCMC) method. 
To optimise model parameters, we use essentially complete \textit{Gaia} DR2 data samples selected in the local volume with simple colour-magnitude cuts
and corrected for the reddening and extinction with the 3D extinction maps from \citet{lallement18} and \citet{green19}.
We adapt the local surface density normalisations for all Galactic components, IMF and AVR parameters, 
as well as a set of parameters governing the shape of SFR. 
The latter is allowed to have two additional peaks at ages $\lesssim 4$ Gyr. Each of them is characterised by a vertical velocity dispersion that is 
independent from the value prescribed by the AVR function for the thin-disk populations of the same age. 
The model-to-data comparison is performed in terms of the vertical number density profiles of different stellar populations, 
their velocity distribution functions, and an apparent Hess diagram. 
We additionally use the APOGEE RC sample to constrain the local AMR and then keep it 
constant during the MCMC posterior probability distribution sampling. 
This two-step update of parameters is possible because the chemical enrichment history has only a small 
impact on the predicted star counts and disk kinematics and, therefore, the AMR parameters can be updated post factum. 
A closure iteration over the AMR reconstruction procedure is performed in the end to achieve a full consistency between the dynamic and chemical 
parts of the model. 

We note here that as we are currently modelling only the Solar neighbourhood and working solely with the local data, and 
also due to the difficulty in quantifying the radial migration processes,
we do not add complexity to the JJ model by adding the radial migration. Instead, we keep in mind 
that the local data used for our analysis (and therefore our derived model parameters) partly represent other disk radial zones than just the local.
In other words, the lack of the radial migration in the model affects the interpretation of its functions that describe the disk evolution. 
Our model SFR can be viewed as the present-day age distribution corrected for stellar evolution and averaged over R$_\odot\pm\Delta$R, 
where $\Delta$R depends on the strength of the radial migration and may be as large as several kpc. 
The same interpretation is applicable to the heating function AVR and a simple chemical enrichment model in the form of the AMR.  

This paper has the following structure.  
\mbox{Section \ref{sect:data}} describes the construction of the local \textit{Gaia} DR2 samples 
of different stellar populations, as well as our selection of the local APOGEE RC sample.  
\mbox{Section \ref{sect:model}} presents a summary of the improvements introduced to the model. 
\mbox{Section \ref{sect:method}} describes the stellar population synthesis procedure and explains our parameter optimisation method.
\mbox{Section \ref{sect:results}} contains the results of the MCMC posterior sampling. 
Then we discuss the obtained results in \mbox{Section \ref{sect:discus}} and present our conclusions in \mbox{Section \ref{sect:final}}. 
Additional material on \textit{Gaia} DR2 TAP queries, the full posterior probability distribution, 
and an estimate of the impact of the 3D extinction maps on the data selection  
are presented in \mbox{Appendices \ref{sect:append_tap}-\ref{sect:append_mcmc}}.

\section{Data}\label{sect:data}

In order to improve the JJ model, which\ provides a detailed insight into the spatial structure and vertical kinematics 
of the Galactic disk in the Solar neighbourhood (Section \ref{sect:model}), 
we require a local stellar sample with the full 6D dynamical information available. Moreover, these data  
have to be representative for the overall underlying distribution of stellar populations; alternatively, 
the data selection function has to be well understood. 
There is no doubt that the best data of this kind that is presently available comes from \textit{Gaia} DR2 \citep{gaia18}, with its  
high-quality astrometric, photometric, and spectroscopic measurements. Therefore, we chose \textit{Gaia} DR2 as a base 
for our analysis of the spatial distribution and motions of the local stellar populations. The procedure is as follows.

First, we define our samples with 
simple colour-magnitude and distance cuts to build essentially complete samples of the local stars. 
With these criteria, we select \textit{Gaia} DR2 stars with three known components of phase space (i.e. spatial coordinates) and use them to construct 
the vertical number density profiles. To study the vertical velocity distributions, 
we further select subsets with known proper motions and radial velocities. 
Our data selection procedure also includes astrometric and photometric quality cuts, de-reddening, 
and correction for the incompleteness of the selected samples. 

Additionally, we use spectroscopic information from the APOGEE RC catalogue \citep{bovy14}, which 
provides a well-defined and mostly unbiased sample of bright stars across the Galactic disk. 
We use APOGEE RC robust distances, metallicities, and $\alpha$-element abundances to constrain the chemical evolution of the disk in terms of the AMR 
(Section \ref{sect:method_amr}); in the next step, this simple chemical enrichment model is applied
to perform a stellar population synthesis for the local volume.

\subsection{Gaia samples}\label{sect:data_gaia}

\subsubsection{Absolute CMD}\label{sect:data_gaia_cmd}

A detailed understanding of data completeness is critical for their adequate modelling and, thus, for improving model parameters. 
Therefore, before starting the data selection procedure, we thoughtfully address this question 
and then use our understanding of data completeness as our first guide to formulating reasonable sample selection criteria.

As presented in \citet{gaia18}, \textit{Gaia} DR2 is expected to be essentially complete in the 
apparent magnitude range of $12 \ \text{mag} < G < 17 \ \text{mag}$. At the very bright end, $G \lesssim 4 \ \text{mag}$, 
star counts are known to be unreliable because of image saturation effects, 
but for stars with an apparent brightness of $7 \ \text{mag} \lesssim G < 12 \ \text{mag,}$ 
the catalogue completeness is significantly improved in \textit{Gaia} DR2 with respect to DR1 
(see \mbox{Figure 1} in \citealp{gaia18}). An estimate of the \textit{Gaia} DR2 
completeness was recently derived as a result of its comparison to the Two Micron All Sky Survey (2MASS, \citealp{skrutskie06})   
and implemented in the Python module \texttt{gdr2\_completeness}\footnote{\url{https://github.com/jan-rybizki/gdr2_completeness}}. 
\textit{Gaia} DR2 and 2MASS star counts were compared in three $G$-magnitude bins; this comparison shows 
that only when stars as faint as $G \approx 17-18 \ \text{mag}$  
enter the sample, \textit{Gaia} DR2 completeness starts to suffer in the near-plane regions, especially in the direction towards the GC,
where completeness deteriorates due to strong extinction and stellar crowding.  Even though in this case \textit{Gaia} DR2 completeness can be as low as \mbox{$\sim70$\%} for certain lines-of-sight, its typical value in the Galactic plane still remains as high as \mbox{$\sim95$\%}. 

Taking this information into account, together with the fact that the local stars predominantly appear to be intrinsically bright because of their proximity to the Sun, we decided to aim at local stars with apparent magnitudes in the range of $7 \ \text{mag} < G < 17 \ \text{mag}$. 
This selection allows us to construct abundant local samples that include both bright and faint stars 
and, at the same time, to simplify the sample incompleteness modelling by assuming that the raw data 
retrieved from the \textit{Gaia} DR2 catalogue are essentially complete.

\begin{figure}
\includegraphics[scale=0.55]{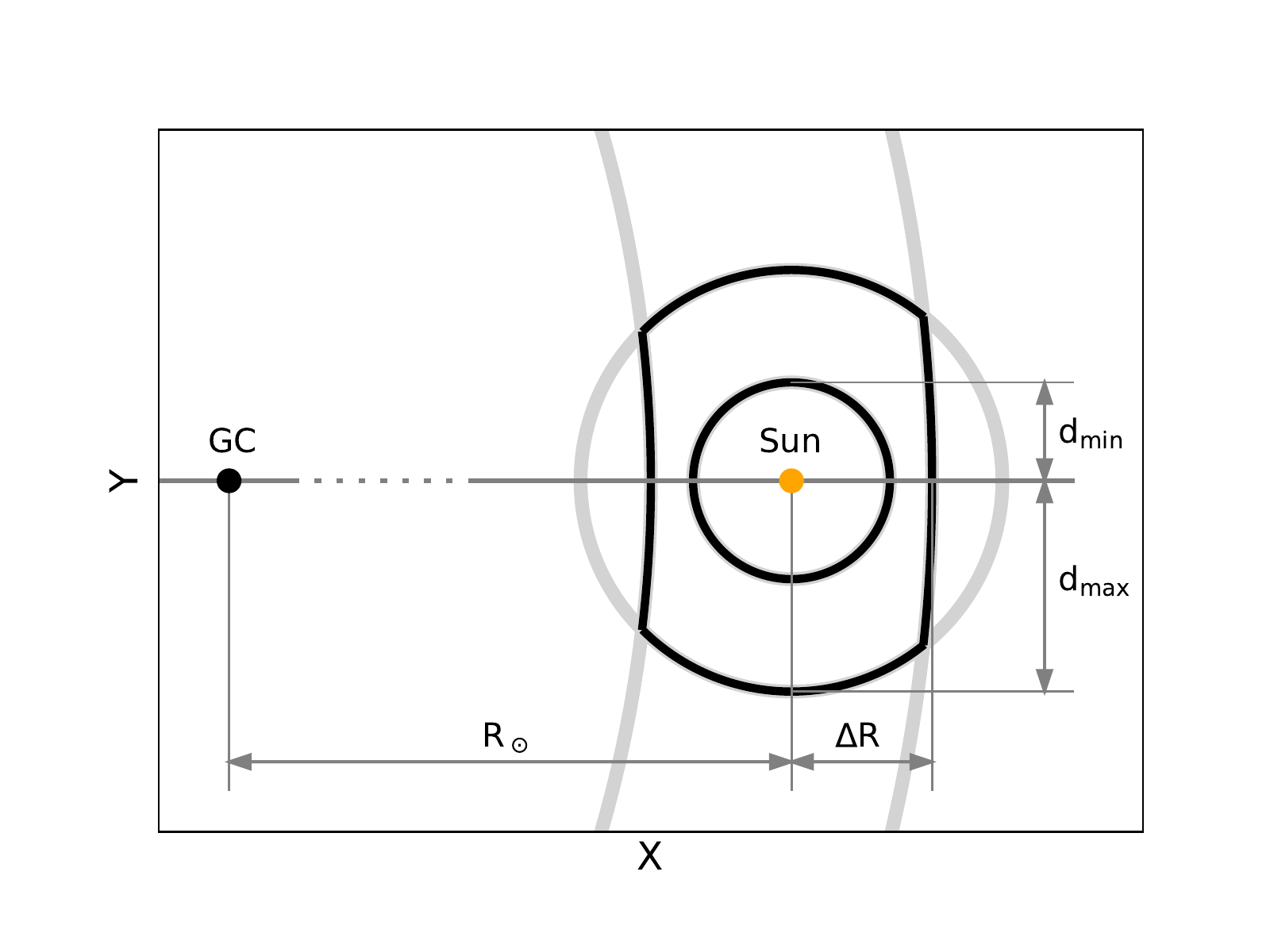}
\caption{Spatial geometry of the \textit{Gaia} samples in XY projection (apart from the conic sample; see Section \ref{sect:data_gaia_cone}). 
The Galactic centre (GC) is marked with a black point. 
Each \textit{Gaia} sample is selected in a Solar-centred spherical shell with the inner and outer radii of $d_\mathrm{{min}}$ and $d_\mathrm{{max}}$ 
from Table \ref{tab:data} (see columns $d_3$ and $d_6$ for the full samples and kinematic subsamples, respectively). 
Additionally, only stars belonging to the local annulus $R_\odot \pm \Delta R$ with \mbox{$R=8.2$ kpc} and 
\mbox{$\Delta R=150$ pc} are taken into consideration (thick grey arcs).
}
\label{fig:data_geom}
\end{figure}

To constrain properties of the stellar populations in a robust way,
both young and old stars must be represented in the data. 
Therefore, our data selection criteria are additionally motivated by our pre-knowledge about age distributions of different stellar classes. 
For this reason, we targeted several areas of the colour-magnitude diagram (CMD) constructed with the \textit{Gaia} 
\mbox{$G_\mathrm{BP}-G_\mathrm{RP}$} colours and absolute magnitudes $M_\mathrm{G}$. The latter are calculated from \textit{Gaia} 
apparent magnitudes $G$ and parallax distances and are de-reddened with the local 3D extinction map 
(\mbox{Sections \ref{sect:data_gaia_geom} and \ref{sect:data_gaia_compl}}, and also \mbox{Appendix \ref{sect:append_dust}}). 
The six samples that we target through the simple colour-magnitude cuts (columns $M_\mathrm{G}$ and \mbox{$G_\mathrm{BP}-G_\mathrm{RP}$} 
in \mbox{Table \ref{tab:data}}, see also \mbox{Figure \ref{fig:hess_abs_compl}}) 
can be classified into the following three groups according to their age coverage:
\begin{itemize}
 \setlength\itemsep{0.25em}
 \item[] \textit{Young stars.} They are represented by A and F samples. \mbox{A-type} stars 
 trace the last \mbox{$\sim1$ Gyr} of the SF processes in the Galactic disk, and F stars have ages younger than \mbox{$\sim6$ Gyr} 
 with a peak \mbox{at $\sim3.5$ Gyr} ago (see modelled age distributions of the samples in \mbox{Figure \ref{fig:loc_age}}).   
 \item[] \textit{Old stars.} We selected three MS samples that contain G and K dwarfs, as well as a mixture of both. These samples cover 
 a full range of stellar ages and are dominated by old long-lived stars (50\% of G and K dwarfs are older than \mbox{$\sim8$ Gyr} and 
 \mbox{$\sim9$ Gyr}, respectively). 
 \item[] \textit{Mixed ages.} Additionally, we used a sample of giants selected in a colour-magnitude range dominated by the RC stars. 
 This sample of bright giants consists of a mixture of stars of different ages spanning over almost the whole age range, \mbox{$\tau \gtrsim 0.5$ Gyr}. 
 The RC stars are known to include a significant fraction of young stars, with their age distribution showing a strong peak at \mbox{$\sim1.5-2$ Gyr}.
\end{itemize}

\subsubsection{Spatial geometry}\label{sect:data_gaia_geom}

\setlength{\extrarowheight}{0.25em}
\setlength{\tabcolsep}{4pt}
\begin{table*}[t!]
\centering
\tiny
\caption{Data selection criteria and the \textit{Gaia} DR2 sample statistics.}
\begin{tabular}{l|cc|cc|cccc|cc} 
\hhline{===========}
\multicolumn{3}{m{2.0cm}|}{{\begin{minipage}[t][0.35cm][t]{2.0cm} {} \end{minipage}}} 
& \multicolumn{2}{m{2.5cm}|}{\begin{minipage}[t][0.35cm][t]{2.7cm} \center{$G=[7, 12]$} \end{minipage}}
&\multicolumn{4}{m{4.0cm}|}{\begin{minipage}[t][0.35cm][t]{6.5cm} \center{$G=[7, 17]$} \end{minipage}}
&\multicolumn{2}{m{2.0cm}}{\begin{minipage}[t][0.35cm][t]{2.7cm} \center{$G=[12, 17]$} \end{minipage}}\\ \hline
Sample & $M_\mathrm{G}$, mag & $G_\mathrm{BP}-G_\mathrm{RP}$, mag & $d_{6}$,  pc & $\mathscr{N}_{6}$ & $d_{3}$,  pc 
& $\mathscr{N}_{3}$ & $\mathscr{N}_{3}\big|_\mathrm{no \ qual.}$ & $\mathscr{N}_3\big|_\mathrm{G<12 \ \& \ d_{6}}$ 
& $\mathscr{N}_{2}$ & $\mathscr{N}_{2}\big|_\mathrm{no \ qual.}$ \\ [4pt] \hline
Full sample & [-2, 12] & [0, 3] & [0, 600] & 514 595 & [0, 600] & 4 093 283 & 4 510 085 & 608 759 &  & \\
A stars & [0, 2] & [0, 0.3]  &  &  & [255, 600] & 11 851 & 12 213 &  &  & \\ 
F stars & [3, 4.3] & [0.48, 0.7]  & [65, 345] & 68 156 & [65, 600] & 188 362 & 193 491 & 76 789 &  & \\ 
RC stars & [-0.2, 0.8] & [0.9, 1.5] & [275, 600] & 14 336 & [275, 600] & 14 640 & 14 698 & 14 631 &  & \\ 
G dwarfs& [4.8, 5.5] & [0.7, 1.2]  & [30, 200] & 45 199 & [30, 600] & 352 988 & 368 983 & 48 783 &  & \\ 
G/K dwarfs & [5.9, 6.7] & [0.9, 1.6] & [20, 115] & 13 093 & [20, 600] & 449 229 & 476 562 & 14 324 &  & \\ 
K dwarfs & [7, 8.5] & [1.3, 2.2]  & [10, 50] & 2 588 & [10, 500] & 785 745 & 846 056 & 2 924 &  & \\ \hdashline
Cone &  & [0, 3] &  &  &  &  &  &  & 262 616 & 266 735 \\ 
\hhline{===========}
\end{tabular} 
\label{tab:data}
\end{table*}

As a next step, for each sample defined on the CMD, as given in \mbox{Table \ref{tab:data}},
we use our adopted absolute and apparent magnitude limits to transform them to the limiting heliocentric distances: 
\begin{equation}
 d_\mathrm{i} = 10^{0.2(G_\mathrm{i} - M_\mathrm{G,i})+1} \quad \text{with} \ \mathrm{i}=\{\text{min},\text{max}\} 
\label{eq:dminmax}
.\end{equation}
This defines a spherical shell with the inner and outer radii of $d_\mathrm{min}$ and $d_\mathrm{max}$, respectively, 
where a stellar population limited by the absolute magnitudes $[M_\mathrm{G,min},M_\mathrm{G,max}]$ 
falls into the range of apparent magnitudes $[G_\mathrm{min},G_\mathrm{max}]$ (\mbox{Figure \ref{fig:data_geom}}). 
By setting $G_\mathrm{min}$ and $G_\mathrm{max}$ to the adopted boundaries of the \textit{Gaia} DR2 completeness range
of 7 mag and 17 mag, respectively, we built a set of essentially complete samples in the Solar neighbourhood. 
As we can see from \mbox{Eq. (\ref{eq:dminmax})}, for many of our samples from \mbox{Table \ref{tab:data},} the outer 
radius of the shell $d_\mathrm{max}$ can be as large as \mbox{$\sim2$ kpc}. 
However, the median relative parallax error of the \textit{Gaia} DR2 stars with $7 \ \text{mag} < G < 17 \ \text{mag}$ 
is as high as $\sim10\%$ at \mbox{2 kpc} from the Sun.   
In this case, the inverse parallax is not a robust estimate of the true distance. 
In order to stay on the safe side, we set an absolute limit to $d_\mathrm{max}$ of \mbox{600 pc},  
as at this distance scale \textit{Gaia} DR2 stars of $7 \ \text{mag} < G < 17 \ \text{mag}$ 
have median relative parallax errors of $\lesssim 2.5\%$ only 
and, thus, our use of parallax distances is justified (see also \mbox{Section \ref{sect:discus_lim_dist}}).  

Furthermore, we keep our samples very local in terms of Galactocentric distances 
in order to avoid  interference on the part of the radial variations of the disk density and dynamical heating with parameters of the local JJ model. 
Thus, we selected only stars with Galactocentric distances $R_\odot-\Delta R < R < R_\odot+\Delta R$, where \mbox{$\Delta R=150$ pc} 
(\mbox{Figure \ref{fig:data_geom}}). 
Galactocentric distances are calculated from parallax distances with the adopted Solar position at \mbox{$R_\odot=8.2 \pm0.05$ kpc} (consistent 
with the most recent value of \mbox{$8.178 \pm 0.013_\mathrm{stat} \pm 0.022_\mathrm{sys}$ kpc} from \citealp{gillessen19}) and 
\mbox{$z_\odot = 20 \pm 3$ pc}
(motivated by the value of \mbox{20.8 $\pm$ 0.3 pc} from \citealp{bennett19}). 

\subsubsection{Reddening, quality cuts, and completeness}\label{sect:data_gaia_compl}

We calculate de-reddened magnitudes and colours using the local 3D dust map from \citet{green19}. 
As it covers only about 3/4 of the sky, we complement it with the lower-resolution map from \citet{lallement18} to cover the remaining sky area
(\mbox{Figure \ref{fig:ext_maps}} in \mbox{Appendix \ref{sect:append_dust}}). 
To calculate the colour excess $E_\mathrm{BP-RP}$ and extinction $A_\mathrm{G}$, we use the colour transformations from \citet{evans18}. The impact of the extinction and reddening on our sample statistics is discussed in 
\mbox{Section \ref{sect:discus_lim_dust}} and \mbox{Appendix \ref{sect:append_dust}}. 

\begin{figure*}
\begin{minipage}[c]{0.67\textwidth}
 \includegraphics[scale=0.5]{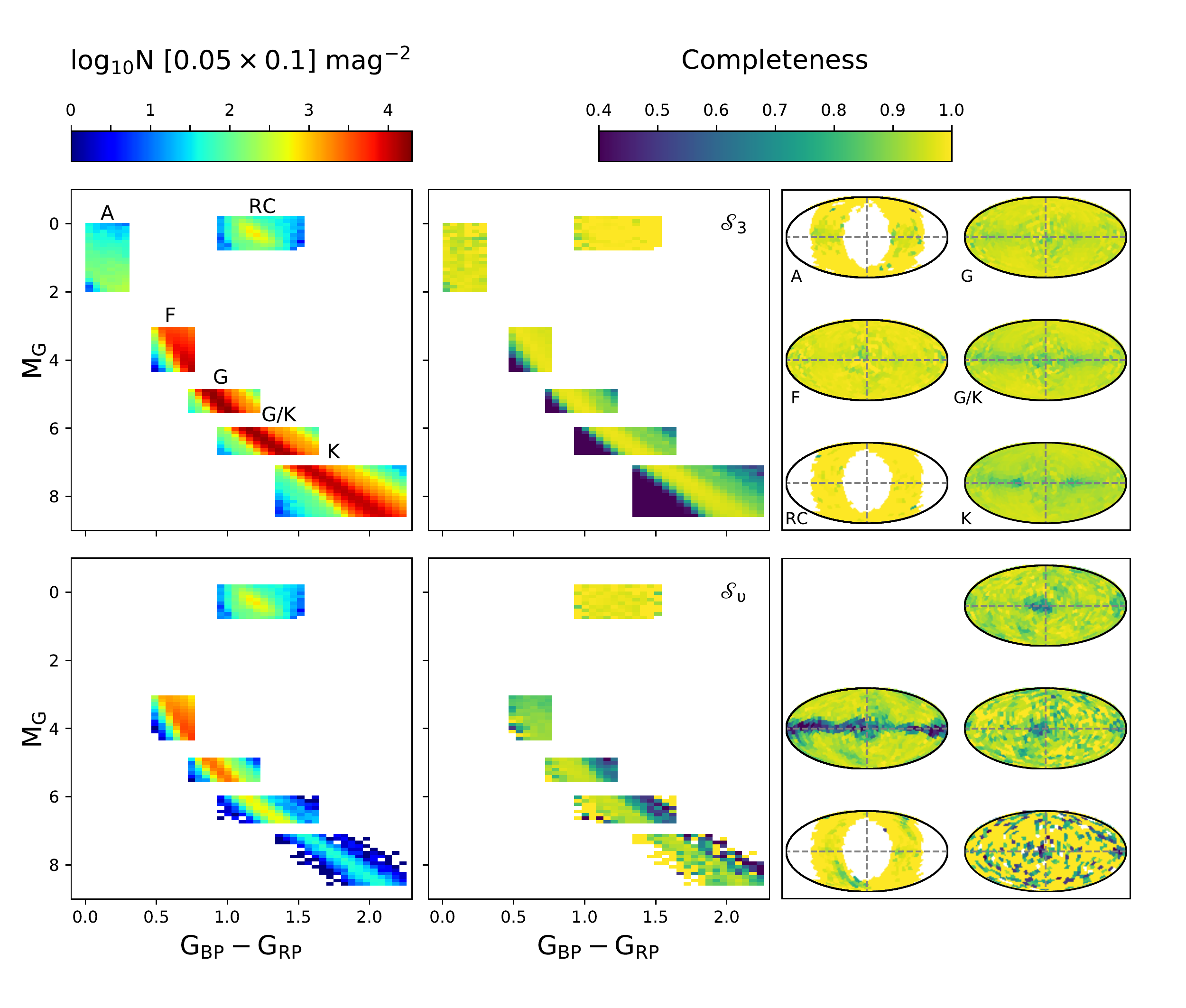}
\end{minipage}\hfill
\begin{minipage}[c]{0.3\textwidth}
    \caption{\textit{Top row:} Selected samples of the local A, F, and RC stars and G and K dwarfs on the absolute CMD 
    constructed with the de-reddened \textit{Gaia} DR2 colours and magnitudes (left). 
    Completeness of these samples as given by \mbox{Eq. (\ref{eq:samples_sf1})} over the CMD (middle) and across the sky (right). 
    \textit{Bottom row:} Kinematic subsamples selected from the samples from the top row, except for A stars (left). 
    Illustration of an additional incompleteness 
    introduced by selecting only stars with available radial velocities (see \mbox{Eq. (\ref{eq:samples_sf2})}) and variation of this incompleteness  
    over the CMD (middle) and across the sky (right).}\label{fig:hess_abs_compl}
\end{minipage}
\end{figure*}

\ifx
\begin{figure*}
\floatbox[{\capbeside\thisfloatsetup{capbesideposition={right,center},capbesidewidth=4.5cm}}]{figure}[\FBwidth]
{\caption{\textit{Top row:} Selected samples of the local A, F, and RC stars and G and K dwarfs on the absolute CMD 
constructed with the de-reddened \textit{Gaia} DR2 colours and magnitudes (left). 
Completeness of these samples as given by \mbox{Eq. (\ref{eq:samples_sf1})} over the CMD (middle) and across the sky (right). 
\textit{Bottom row:} Kinematic subsamples selected from the samples from the top row, except for A stars (left). 
Illustration of an additional incompleteness 
introduced by selecting only stars with available radial velocities (see \mbox{Eq. (\ref{eq:samples_sf2})}) and variation of this incompleteness  
over the CMD (middle) and across the sky (right).}
\label{fig:hess_abs_compl}}
{\includegraphics[scale=0.5]{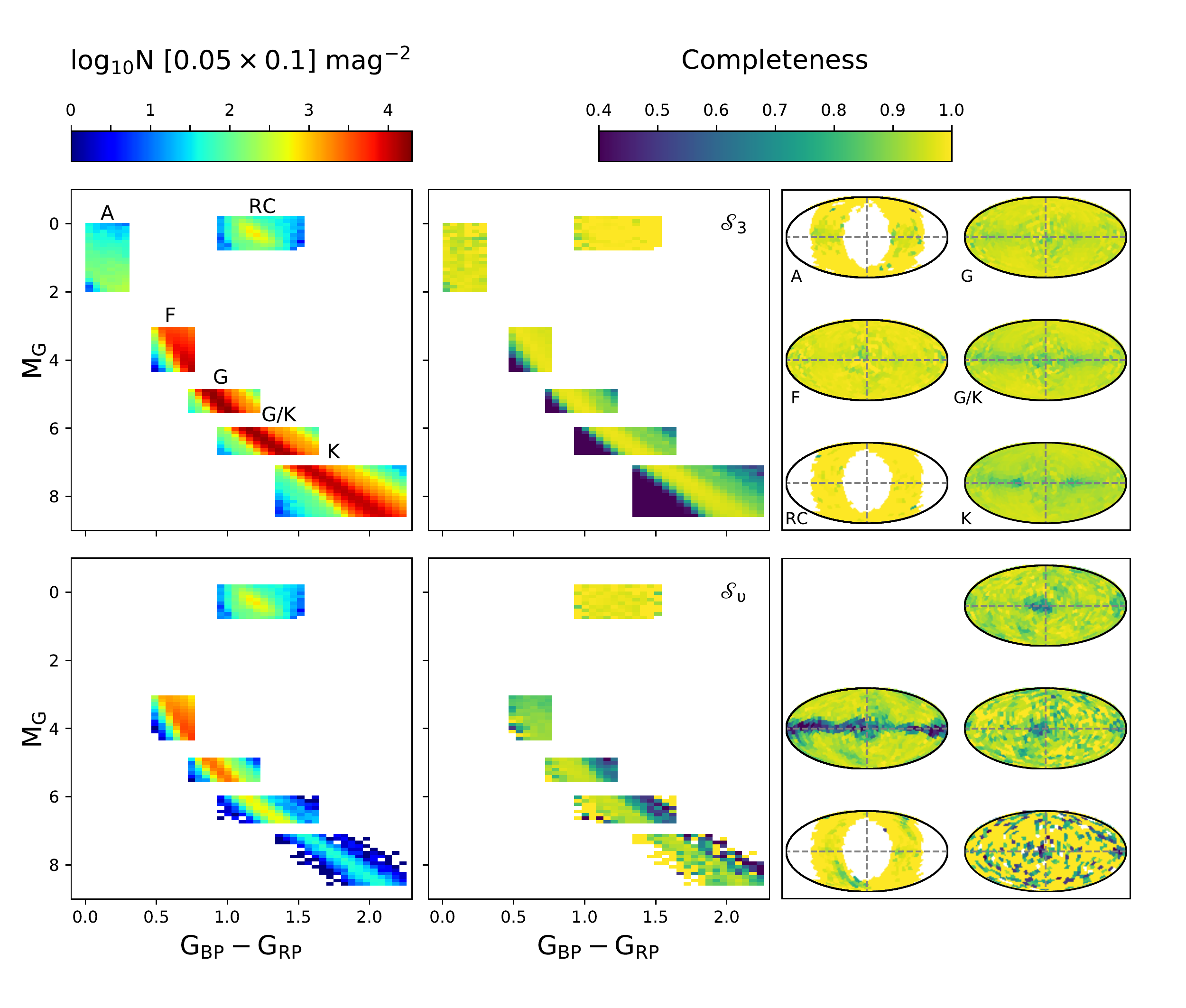}}
\end{figure*}
\fi

Additionally, we apply two cleaning cuts to remove stars with potentially unreliable astrometric and photometric parameters 
(based on discussions in \mbox{\citealp{lindegren18}} and \mbox{\citealp{evans18}}):
\begin{align}
\label{eq:data_cleaning1}
\mathtt{astrometric\_excess\_noise} &< 1 ,\\
\label{eq:data_cleaning2}
\mathtt{phot\_bp\_rp\_excess\_factor} &< 1.3+0.06 \, (G_\mathrm{BP}-G_\mathrm{RP})^2. 
\end{align}
The first cut cleans our data of astrometric binaries and the second one removes stars whose photometry was affected by saturation effects 
in the crowded fields. As these stars are real astrophysical objects, 
we add an additional correction to include the correct local mass and gravitational potential in the model.
Together, these cuts remove up to \mbox{$\sim5$\%} from our samples. 
The final number of stars in each sample is given in column $\mathscr{N}_3$ of 
\mbox{Table \ref{tab:data}}\footnote{Indices 2, 3, and 6 in the column names of \mbox{Table \ref{tab:data}} 
refer to the number of the phase space components known for a given sample.}. 
Column $\mathscr{N}_3\big|_\mathrm{no \ qual.}$ contains the number of stars in the samples before the cleaning cuts were applied. 
We find that for all samples the astrometric quality cut, \mbox{Eq. (\ref{eq:data_cleaning1})}, is the main contributor to the fraction of removed stars. 
The incompleteness introduced by applying \mbox{Eqs. (\ref{eq:data_cleaning1}) and (\ref{eq:data_cleaning2})} can be calculated simply as a ratio:
\begin{equation}
\label{eq:samples_sf1}
 \mathscr{S}_3 = \frac{\mathscr{N}_3}{\mathscr{N}_3\big|_\mathrm{no \ qual.}}. \\
\end{equation}
The left panel of the top row in \mbox{Figure \ref{fig:hess_abs_compl}} shows the absolute CMD of the final clean samples 
with three known components of phase space (i.e. the coordinate vector). 
The middle and right plots of the top row illustrate how the completeness factor $\mathscr{S}_3$ varies across the CMD and over the sky 
for each sample. In line with our expectations, the lowest completeness on the CMD have areas below the MS where binaries with a 
white dwarf companion can be located but a very low number of single MS stars. 
We also see a weak correlation between the completeness value and Galactic latitude, with the lower completeness area aligned with the Galactic plane.  
We compensate for the introduced incompleteness in a natural way, which also saves us computational time as is has to be applied only once: 
each star in a given sample is assigned with a weight $1/\mathscr{S}_3(l,b)$, where $\mathscr{S}_3(l,b)$ is completeness 
over the sky as shown at the top right panel of \mbox{Figure \ref{fig:hess_abs_compl}}. These weights are later used during calculation of 
the observed number density profiles of the \textit{Gaia} samples. 

\begin{figure}
\centering
\centerline{\resizebox{\hsize}{!}{\includegraphics{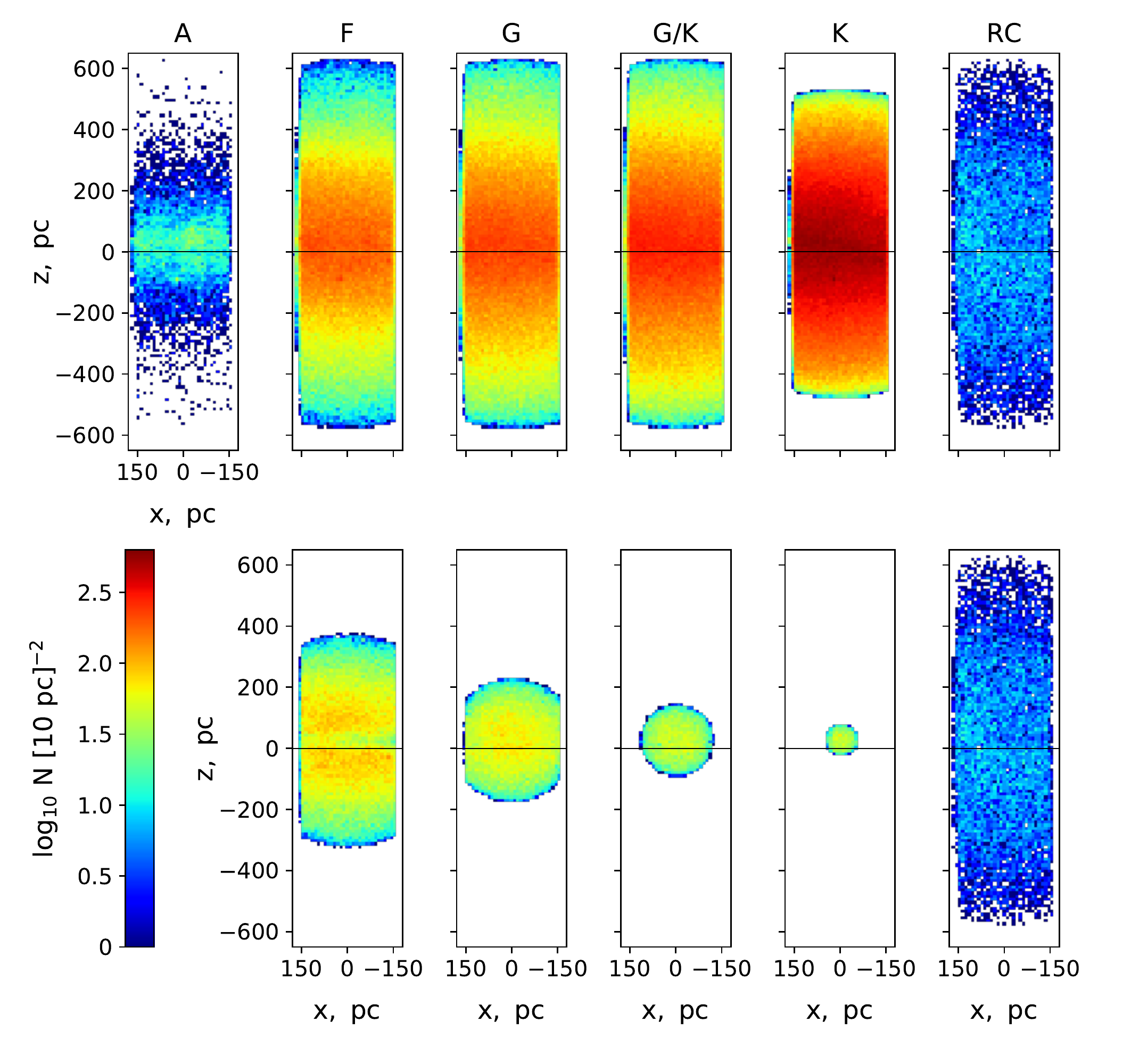}}}
\caption{Spatial distribution of the six \textit{Gaia} samples (top row) and 
their five kinematic subsets (bottom row) plotted in XZ Cartesian coordinates. The X axis points to the GC.}
\label{fig:data_sample}
\end{figure}

To study the disk kinematics we also need velocities, therefore we additionally use the
\textit{Gaia} DR2 7.2-million radial velocity catalogue (RVC, \citealp{katz19}). 
In practice, this means that we select subsets with known radial velocities from our previously defined colour-magnitude samples.
So far, radial velocities have been measured by \textit{Gaia} for relatively bright stars only, $G \lesssim 12-13 \ \text{mag}$.  
Thus, our kinematic subsets are naturally restricted to have a faint limit around this value, we set $G_\mathrm{max} = 12 \ \text{mag}$.  
As for the bright apparent magnitude limit, we use the same value as before, namely, $G_\mathrm{min} = 7 \ \text{mag}$.  
In consequence, our kinematic samples occupy smaller volumes than the full colour-magnitude samples 
(compare their $d_\mathrm{max}$ values in \mbox{Table \ref{tab:data}}, columns $d_3$ and $d_6$; 
also see \mbox{Figure \ref{fig:data_sample}}). 
There is an additional limitation related to the stellar effective temperatures: RVC includes stars with temperatures up to $6 \, 900$ K only. 
As a result, stars of spectral class A are completely missed in the RVC being too hot. 
In the case of the K-dwarf sample, we additionally clean the data from the core members of the Hyades cluster \citep{roeser19}. 
For the stars in our remaining five kinematic subsamples, we calculate the 
vertical component of the spatial velocity, $W,$ using the corresponding component of the Solar peculiar velocity 
\mbox{$W_\odot = 7.25$ km s$^{-1}$} \citep{schoenrich10}. We also calculate $W$-velocity errors taking into account the full error covariance 
matrix provided in \textit{Gaia} DR2.

In order to ensure that these RVC subsamples (column $\mathscr{N}_6$ in \mbox{Table \ref{tab:data}}) are kinematically unbiased, 
we estimate their completeness relative to the complementary samples built from stars with only three phase-space components known. 
For each kinematic subsample $\mathscr{N}_6$ we select stars from the corresponding final colour-magnitude sample $\mathscr{N}_3$
with apparent magnitudes $\mathrm{G}<12$ and stars that are located withing the volume of the kinematic subsample 
(column $\mathscr{N}_3\big|_\mathrm{G<12 \& d_6}$ in Table \ref{tab:data}). 
The overall completeness of the radial velocity samples ranges from \mbox{$\sim70$\%} for F stars to \mbox{$\sim99$\%}
for the RC sample and is expressed as:
\begin{equation}
\label{eq:samples_sf2}
 \mathscr{S}_6 = \mathscr{S}_3 \times \mathscr{S}_v, \quad \text{where} \quad 
                        \mathscr{S}_v = \frac{\mathscr{N}_6}{\mathscr{N}_3\big|_\mathrm{G<12 \ \& \ d_{6}}}
.\end{equation}
The left panel of the bottom row in Figure \ref{fig:hess_abs_compl} shows the logarithmic 
number density of the final kinematic subsamples at the absolute CMD. 
The corresponding middle and right panels illustrate the variation of the factor 
$\mathscr{S}_v$ over the CMD and across the sky. The lowest completeness 
is seen near the Galactic plane and in the direction towards the GC. The completeness factor is essentially homogeneous 
over given areas of the absolute magnitudes and colours, except for the regions above the MS where binary stars are located. 
The correction of this incompleteness is performed exactly in the same way as in the case of their parent samples: 
weights $1/\mathscr{S}_6(l,b)$ are used during the calculation of velocity distribution functions from the kinematic subsamples 
(\mbox{Section \ref{sect:method}}). \mbox{Figure \ref{fig:data_sample}} shows the spatial distribution in XZ projection 
of both full  A-, F-, RC-star and G-, G/K-, K-dwarf samples (top row), and their 
kinematic subsamples (except of A stars, bottom row).

\subsubsection{Full local sample}\label{sect:data_gaia_loc}

For subsequent testing of the updated JJ model (\mbox{Section \ref{sect:results_loc}}),
we also use the full local sample consisting of all stars with apparent magnitudes \mbox{$7 \ \text{mag} < G < 17 \ \text{mag}$}, 
colours \mbox{$0 \ \text{mag} < G_\mathrm{BP}-G_\mathrm{RP} < 3 \ \text{mag}$}, 
and absolute magnitudes \mbox{$ -2 \ \text{mag} < M_\mathrm{G} < 12 \ \text{mag}$} (\mbox{Table \ref{tab:data}}). 
The full sample is selected in the local 600-pc sphere truncated by the cut \mbox{$R_\odot-\Delta R < R < R_\odot+\Delta R$} 
with \mbox{$\Delta R=150$ pc}, as before.   
The quality cuts from \mbox{Eqs. (\ref{eq:data_cleaning1}) and (\ref{eq:data_cleaning2})}, as well as de-reddening, are 
applied in the full analogy to the colour-magnitude samples described above. 

\subsubsection{Conic sample}\label{sect:data_gaia_cone}

Additionally, we select the \textit{Gaia} DR2 sample with a spatial geometry different from that shown in \mbox{Figure \ref{fig:data_geom}}. 
As all samples constructed previously are very local, with \mbox{$d_\mathrm{max} < 600$ pc}, they do not contain enough information 
to constrain the thick-disk and halo model parameters. According to the JJ model from \citetalias{just11},  
the thick disk starts to dominate over the thin disk in terms of the mass density 
at \mbox{$|z| \approx 0.8-1$ kpc}, and the halo takes over the thick-disk density at \mbox{$|z| \approx 3$ kpc}. 
Therefore, we need an additional sample that also target faint distant stars. 

To build such a sample, we select all \textit{Gaia} DR2 stars in two cones with $20^\circ$ 
opening angle towards both northern and southern Galactic poles. 
The range of apparent magnitudes and colours is set to \mbox{$12 \ \text{mag} < G < 17 \ \text{mag}$} 
and \mbox{$0 \ \text{mag} < G_\mathrm{BP}-G_\mathrm{RP} < 3 \ \text{mag}$}. As before, we also 
clean the sample from stars with bad photometry by applying \mbox{Eq. (\ref{eq:data_cleaning2})}. 
The astrometric cut, \mbox{Eq. (\ref{eq:data_cleaning1})},
is not used here as we are not interested in proper motions and parallaxes.
The extinction map is applied differently for different stars in this sample.
When the parallax is known, we use the inverse parallax as a distance estimate and take extinction and colour excess values 
corresponding to this distance at a given line-of-sight. If the distance to a star exceeds 
the maximum distance available in the extinction map for this line-of-sight, we apply the maximum available reddening and extinction,
as lower limits of these quantities. Similarly, we take the maximum reddening and extinction for a given line-of-sight 
when the parallax is unknown (\mbox{$\sim1$\%} of the sample only), implicitly assuming that the star is located far away. 
Finally, we clean our sample from extragalactic sources using a catalogue from \citet{bailer-jones19}. 
The final cone sample contains $262 \, 616$ stars. The 1.5-\% incompleteness introduced by the photometric quality cut 
is compensated for by weights $1/\mathscr{S}_2(l,b)$ calculated fully analogously to \mbox{Eq. (\ref{eq:data_cleaning1})}, 
but with values $\mathscr{N}_2$ and $\mathscr{N}_2\big|_\mathrm{no \ qual.}$. 
\mbox{Figure \ref{fig:cone_hess}} shows the apparent Hess diagram of the conic sample, constructed with a colour-magnitude resolution of 
\mbox{$\Delta (G_\mathrm{BP}-G_\mathrm{RP}) = 0.05$ mag} and \mbox{$\Delta G = 0.1$ mag} 
and smoothed with a window of \mbox{$3 \times \{\Delta (G_\mathrm{BP}-G_\mathrm{RP}), \Delta G \} =  \{0.3, 0.15\}$ mag}.

\subsubsection{Summary of the selection}\label{sect:data_gaia_sum}

We summarise the whole procedure of the \textit{Gaia} samples selection as follows: 

\begin{figure}[t]
\centering
\includegraphics[scale=0.4]{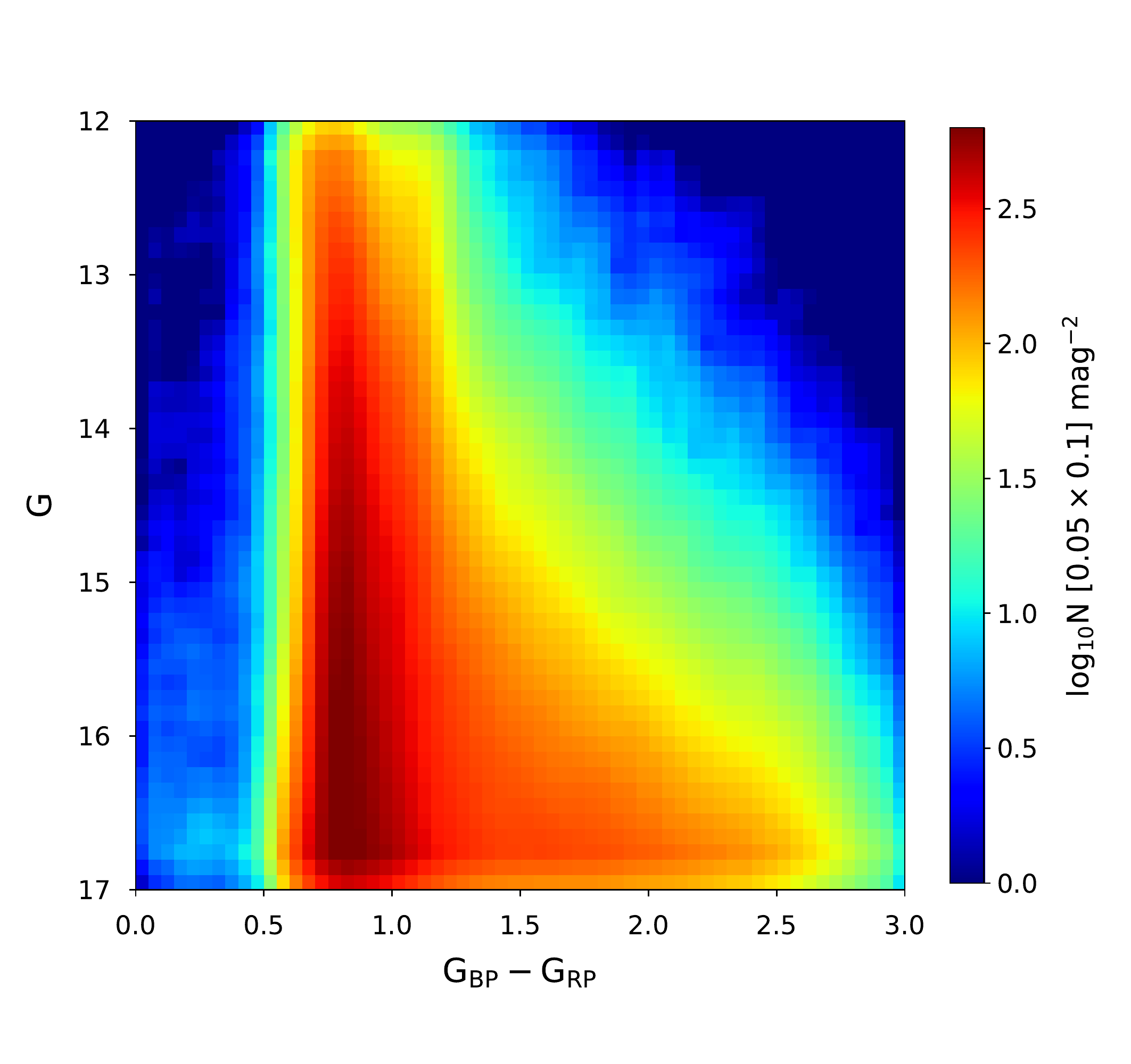}
\caption{Smoothed apparent Hess diagram of the conic sample constructed with 
the de-reddened \textit{Gaia} DR2 colours and apparent magnitudes.}
\label{fig:cone_hess}
\end{figure}

\begin{itemize} 
\renewcommand{\labelitemi}{-}
 \item Firstly, we pre-select\footnote{We use \textit{Gaia} ARI TAP services at \url{http://gaia.ari.uni-heidelberg.de/tap.html}}  
 all stars in the truncated local 600-pc sphere
 with apparent magnitudes of \mbox{$7 \ \text{mag} < G < 17 \ \text{mag}$} 
 and within the colour range \mbox{$0 \ \text{mag} < G_\mathrm{BP}-G_\mathrm{RP} < (3 + \epsilon) \ \text{mag}$} 
 (\mbox{Section \ref{sect:data_gaia_loc}}). 
 The TAP queries used for this step are given in \mbox{Appendix \ref{sect:append_tap}}. 
 Then we apply more specific distance and colour-magnitude cuts: \mbox{$d_\mathrm{min} < d < d_\mathrm{max}$}, 
 $M_\mathrm{G,min} < M_\mathrm{G} < M_\mathrm{G,max}$, and $(G_\mathrm{BP}-G_\mathrm{RP})_\mathrm{min} < G_\mathrm{BP}-G_\mathrm{RP} < 
 (G_\mathrm{BP}-G_\mathrm{RP})_\mathrm{max} + \epsilon$.
 This pre-selects different local populations as defined in \mbox{Table \ref{tab:data}}. 
 The colour range is extended towards the red part by \mbox{$\epsilon = 0.5$ mag} in order to include stars that are reddened being 
 located behind the nearby molecular clouds.
 The conic sample is pre-selected similarly, but with constraint on latitudes rather than distances, $|b|>80^\circ$.
\item Secondly, for all samples, we calculate de-reddened colours and magnitudes, applying the combined 3D extinction map 
 from \citet{green19} and \citet{lallement18}.  
\item Afterwards we apply the adopted colour-magnitude cuts (\mbox{Table \ref{tab:data}}) to the derived de-reddened colours. 
\item The data are further cleaned from stars with unreliable astrometric parameters and photometry by applying cuts from 
\mbox{Eqs. (\ref{eq:data_cleaning1}) and (\ref{eq:data_cleaning2});} 
in the case of the conic sample, only \mbox{Eq. (\ref{eq:data_cleaning2})} is used. 
The numbers of stars in the final samples are given in columns $\mathscr{N}_3$ and $\mathscr{N}_2$ in \mbox{Table \ref{tab:data}}. 
\item For all samples defined on the absolute CMD we further select kinematic subsamples with known 6D phase space information 
and calculate the $W$-velocity component and its error for each star. 
The kinematic subsamples constitute \mbox{$\sim10$\%} of the full colour-magnitude samples and 
the corresponding numbers of stars are listed in column $\mathscr{N}_6$ of \mbox{Table \ref{tab:data}}.
\item Finally, we estimate the incompleteness introduced to each sample by our quality cuts and, in the case of kinematic subsamples, 
by the removal of stars without radial velocities. By comparing the number of stars in the final samples and
in the samples before the radial velocity and/or quality cut were applied,   
we define a selection function $\mathscr{S}_\mathrm{i}$ given by \mbox{Eqs. (\ref{eq:samples_sf1}) and (\ref{eq:samples_sf2})}. 
The correction factor $1/\mathscr{S}_\mathrm{i}(l,b)$ is used as a weight for each star in the sample during the calculation 
of the quantities of interest. 
\end{itemize}

\subsection{APOGEE Red Clump sample}\label{sect:data_apogee}

\begin{figure}[t]
\centerline{\resizebox{\hsize}{!}{\includegraphics{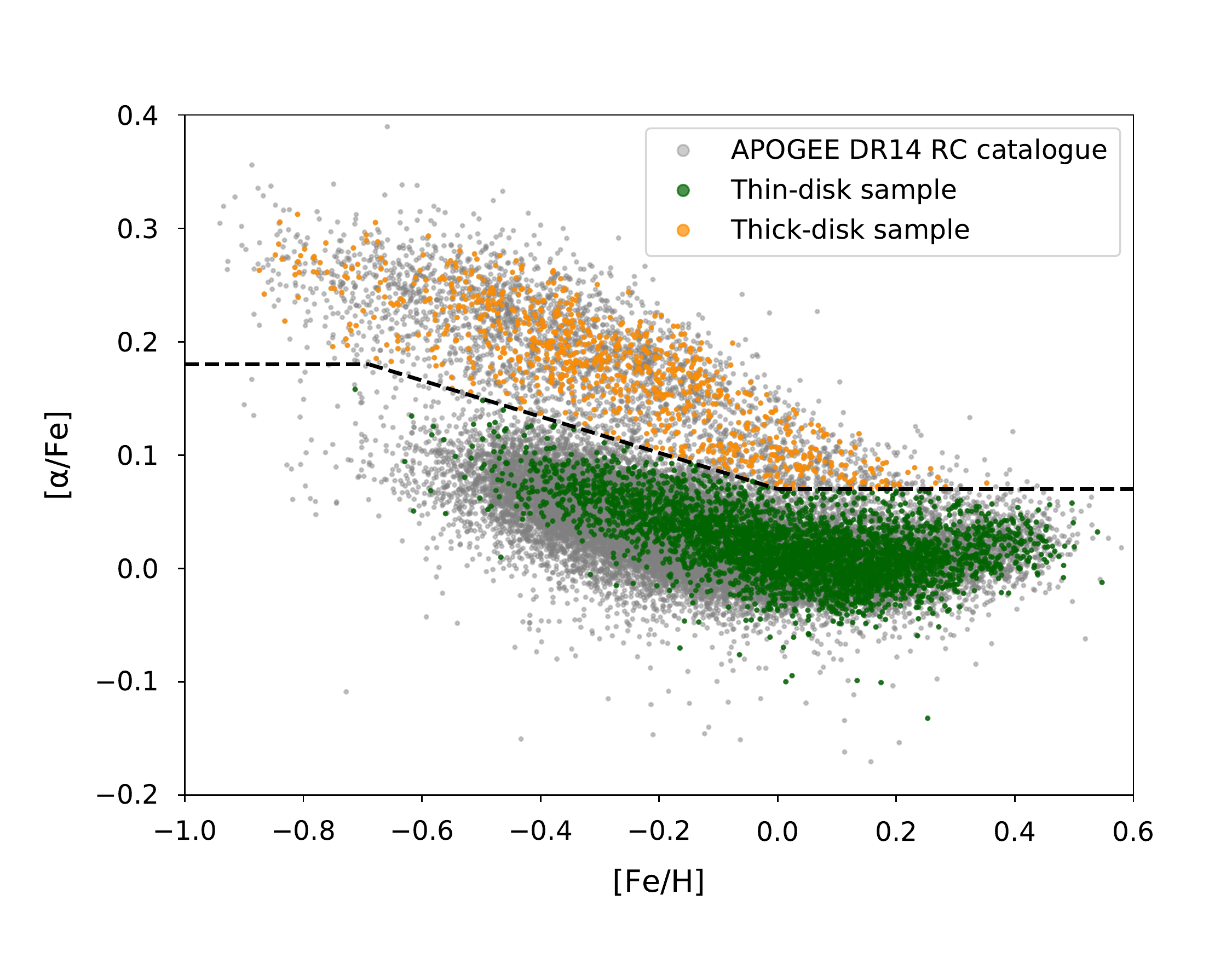}}}
\caption{Local low-$\alpha$ thin-disk (green) and high-$\alpha$ thick-disk (orange) samples selected for the analysis.
The full APOGEE RC sample is shown in grey. 
The dashed black line marks the adopted boundary between the two populations, as in \mbox{Eq. (\ref{eq:chem_disks})}.}
\label{fig:rc_data}
\end{figure}

In addition to the \textit{Gaia} data, we use the fourteenth data release (DR14) of the kinematically unbiased APOGEE RC catalogue, 
its earlier version was presented and discussed in \citet{bovy14}. 
As the quality of the spectroscopic data is intrinsically high, we chose not to 
use any additional cuts with respect to the signal-to-noise ratio ($S/N$): its lowest value in the catalogue is 20, but only 2\% of the stars 
have $S/N < 50$.  

We split \mbox{29\,502 APOGEE RC stars} into the low- and high-$\alpha$ populations, 
which we later associate with the modelled thin and thick disk, respectively. 
To separate the populations, we use [Fe/H] and [$\alpha$/Fe] provided in the APOGEE RC catalogue and define the boundary between the
two populations in the following way (\mbox{Figure \ref{fig:rc_data}}):
\begin{equation}
\mathrm{[\alpha/Fe]} = 
  \begin{cases}
      \ 0.07, & \mathrm{[Fe/H]} > 0 \\ 
      \ -0.16  \mathrm{[Fe/H]} + 0.07, & -0.69 < \mathrm{[Fe/H]} < 0 \\
      \ 0.18, & \mathrm{[Fe/H]} < -0.69. 
  \end{cases}
\label{eq:chem_disks}
\end{equation}
As RC stars are bright giants with only a small scatter in luminosity, their spectrophotometric distances
are very precise, the relative distance errors in the catalogue are \mbox{$\sim5$\%} only. 
We convert these distances to Galactocentric cylindrical coordinates with 
the same values of (R$_\odot$, z$_\odot$) as in \mbox{Section \ref{sect:data_gaia_geom}}. 
Then we select stars in the range of Galactocentric distances $R_\odot-\Delta R < R < R_\odot+\Delta R$ with \mbox{$\Delta R=0.5$ kpc}. 
In the case of APOGEE, we could not keep the sample extremely local as we did with \textit{Gaia} by setting \mbox{$\Delta R = 150$ pc}: 
there are only $\sim2 \, 000$ APOGEE RC stars located in such a narrow Solar annulus and this low number of stars is not 
enough for a reliable analysis. 
We further select low-$\alpha$ stars at distances from the Galactic plane \mbox{$|z|<1$ kpc}, and high-$\alpha$ stars at 
\mbox{$|z|<2$ kpc}. 

To define the RC sample in the model, we use cuts on the synthetic effective temperatures $T_{\mathrm{eff}}$, $\log{g}$, 
and luminosities $L$, \mbox{Eqs. (\ref{eq:rc_select1}) and (\ref{eq:rc_select2})}. 
Therefore, for the sake of consistency, we also apply the same cut 
on the effective temperature to the APOGEE data, selecting stars with $4 \, 250 \ \text{K} < T_\mathrm{eff} < 5 \, 250  \ \text{K}$ 
(the adopted cuts on $\log{g}$ and $L$ mimic the selection criteria applied during the construction of the RC catalogue, 
so they do not need to be applied to the data).
The final local low-$\alpha$ and high-$\alpha$ samples contain 3 910 and 847 stars, respectively (\mbox{Figure \ref{fig:rc_data}}). 

In order to derive the AMR, we aim to reproduce the APOGEE RC metallicity distribution functions (MDFs)
within our model framework (\mbox{Section \ref{sect:method_amr}}). As we do not try to predict the actual RC star counts 
but work only with the normalised MDFs, we are not interested in completeness of the selected APOGEE RC samples 
and do not include it into our analysis. 

We also note, that though the individual precision of metallicity obtained by the APOGEE Stellar Parameter and Chemical Abundances Pipeline, ASPCAP, 
is very high, \mbox{$0.01-0.02$ dex}, the overall uncertainty, including systematic errors may be as high as \mbox{0.1 dex} \citep{jofre19}. 
This will inevitably add to the final uncertainty budget of the derived AMR.

\section{JJ model update}\label{sect:model}

The most recent version of the JJ model describes the Galactic thin disk as a set of isothermal mono-age stellar populations, 
whose evolution is governed by the four input functions: a declining SFR with a peak at old ages, a four-slope broken power-law IMF, 
and an AVR and AMR monotonously increasing with age \citep{sysoliatina18}. A detailed description of the model was presented in \citetalias{just10}, 
and the model machinery generalised for the whole Galactic disk will be summarised in an upcoming work \mbox{(Sysoliatina and Just, in prep.)}. 
In this study we only concentrate on the new or updated model components. 

\subsection{Thin and thick disk}\label{sect:model_disks}

In this section, we explain our motivation for introducing a new treatment of the thick disk component: 
instead of a single mono-age subpopulation, we opt to use a set of subpopulations 
that represent a thick-disk SFR extended in time. As a result, we also have to update an analytic form of the thin-disk SFR function, 
such that the updated JJ model is able to reproduce our old results.

It is generally recognised that the high-$\alpha$ metal-poor population observed in our Galaxy
mostly consists of old stars formed on a short time scale and constitutes a thick-disk component. 
In comparison to the low-$\alpha$ thin disk, it 
is more extended perpendicularly to the Galactic plane, less extended radially, and more dynamically heated. 
In \citetalias{just10}, the thick disk was modelled in a simple way as a single-age isothermal component, 
that is equivalent to an assumption that the thick disk was formed during a single SF burst event. 
Yet many Galactic models treat the thick disk formation as a process extended in time, 
with a corresponding time scale ranging from as short as \mbox{0.1 Gyr} (two-infall chemical models from \citealp{grisoni18})
to \mbox{$4-5$ Gyr} as assumed in \citet{haywood13}.
It is not easy to probe the thick-disk age distribution with observational data, 
as stellar age determination is known to be very challenging and stellar ages obtained with different techniques 
have large uncertainties and may suffer from significant systematic errors. 
Nevertheless, available age maps provide a useful insight into the real age distribution in the Galaxy. 
Age maps constructed with the supervised classification method known as \mbox{\textit{The Cannon}} \citep{ness15}, 
indicate that the age dispersion of the high-$\alpha$ APOGEE RC population may reach \mbox{$\sim1-2$ Gyr}, 
although individual age uncertainties were as high as \mbox{$\sim40$\%} in this analysis. 
A more recent study on isochrone ages implies \mbox{$\sim1$ Gyr} age dispersion for the same stars, 
with individual age uncertainties reduced to \mbox{$\sim20$\%} \citep{sanders18}. 
Therefore, we now allow a non-zero age spread for the thick disk in our model. 
To do so, we introduce the following thick-disk SFR as a function of Galactic time:  
\begin{align}
\label{eq:sfrt}
 SFR_\mathrm{t}(t) &= \left<SFR_\mathrm{t}\right> sfr_\mathrm{t} (t), \quad \text{with} \\
 &sfr_\mathrm{t}(t) = (t+t_\mathrm{t1})^\gamma (e^{-\beta t} - e^{-\beta t_\mathrm{t2}}) \ \ \text{and} \\ \nonumber
 &\left<SFR_\mathrm{t}\right> = \frac{\Sigma_\mathrm{t}}{\int_0^{t_\mathrm{p}} sfr_\mathrm{t}(t) g_\mathrm{t}(t) dt} \nonumber. 
\end{align}
Here, $sfr_\mathrm{t}(t)$ prescribes the shape of the SFR function: 
parameters $\gamma$ and $\beta$ control the SFR growth and steepness of its decline.
The parameter $t_\mathrm{t1}$\footnote{To avoid a confusion between similar thin-disk and thick-disk parameters and component-related quantities, 
we use subscripts `d' and `t' to specify that a given parameter or quantity refers to the thin or thick disk, respectively.} 
determines the starting value of the SFR as $sfr_\mathrm{t}(0 \ \text{Gyr}) = t_\mathrm{t1}^\gamma (1-e^{-\beta t_\mathrm{t2}})$, 
and $t_\mathrm{p}-t_\mathrm{t2}$ is the maximum age of the thick-disk population with $t_p$ being the disk's present-day age. 
The value of $t_\mathrm{p}$ is set to 13 Gyr
to be consistent with the typical age estimates derived for the oldest globular clusters in the Galactic bulge and halo \citep{fuente15,ortonali19}.
With a time resolution of \mbox{$\Delta t = 25$ Myr}, this results in 160 isothermal subpopulations for the thick disk. 
The $SFR_\mathrm{t}(t)$ function is normalised to the present-day thick-disk surface density $\Sigma_\mathrm{t}$ 
to describe the thick-disk formation history in physical units of \mbox{M$_\odot$ pc$^2$ Gyr$^{-1}$}. 
An additional time-dependent function, $g_\mathrm{t}(t),$ is necessary to account for 
the mass loss due to stellar evolution as not all of the mass converted into the thick-disk subpopulations 
at a given time, $t_\mathrm{i}$, survived to the present day in the form of stars or remnants. The mass loss function can be expressed as:
\begin{equation}
 \label{eq:mass_loss}
 g(t) = 1 - \int_0^{t_\mathrm{p}} \int_{m(t)}^\infty \Delta m(\mathrm{[Fe/H]}(t),m) \frac{dn}{dm} dm \ dt
.\end{equation}
In this definition $\Delta m(\mathrm{[Fe/H]}(t),m)$ is a yield or total fraction of mass returned to the interstellar medium (ISM) in the form of 
processed material, which naturally depends on stellar mass and metallicity. The latter is prescribed by the model AMR function, [Fe/H]$(t)$. 
The differential ratio, $dn/dm,$ corresponds to the model IMF and $m(t)$ is a limiting stellar mass, 
such that only stars with masses larger than this limiting value 
contribute to the stellar feedback at a given time. In practice, the evaluation of the mass loss function, 
given the IMF and the thick-disk AMR, is performed with the 
chemical evolution code \textit{Chempy}\footnote{\url{https://github.com/jan-rybizki/Chempy}}\citep{rybizki17}, 
where yields of asymptotic giant branch (AGB) and sypernovae (SNe) Type II stars are taken from \citet{karakas10} and \citet{nomoto13}, 
the ISM enrichment due to SNe Type Ia explosions are described by yields and delay time distribution (DTD) from \citet{maoz10}, and 
stellar lifetimes are adopted from \citet{argast00}. 
The thick-disk SFR parameters fixed in this work are given in \mbox{Table \ref{tab:fixed_param}}.  

In order to be consistent with the earlier versions of the JJ model, we also re-define the thin-disk SFR: 
\begin{align}
\label{eq:sfrd0}
  SFR_\mathrm{d}(t) &= \left< SFR_\mathrm{d} \right> sfr_\mathrm{d}(t), \quad \text{where} \\ 
  sfr_\mathrm{d}(t) &= \frac{(t^2-t_\mathrm{d1}^2)^\zeta}{(t + t_\mathrm{d2})^\eta} \ \  \text{and} \ \  
  \left< SFR_\mathrm{d} \right> = \frac{\Sigma_\mathrm{d}}{\int_0^{t_\mathrm{p}} sfr_\mathrm{d}(t) g_\mathrm{d}(t) dt}. \nonumber
\end{align}
Parameters $t_\mathrm{d2}$, $\zeta$, and $\eta$ are chosen such that the cumulative SFR of the total disk closely follows 
the same function predicted by the \mbox{JJ model} used in \citet{sysoliatina18}. 
With $SFR_\mathrm{t}$ parameters given in \mbox{Table \ref{tab:fixed_param}}, 
we find the most suitable $(t_\mathrm{d2},\zeta, \eta)=(7.8 \ \text{Gyr}, 0.8, 5.6)$. 
The maximum of this thin-disk SFR function can be calculated according to the formula:
\begin{align}
\Sigma_\mathrm{max} &= SFR_\mathrm{d}(t_\mathrm{max}), \\ \nonumber
t_\mathrm{max} &= \frac{\eta t_\mathrm{d2}}{\zeta - 2 \eta} \left( 1 + \sqrt{1 + \frac{\zeta^2 - 2 \eta \zeta}{\eta^2} 
                \frac{t_\mathrm{d1}^2}{t_\mathrm{d2}^2}} \right)
                \bigg\rvert_{t_\mathrm{d1} = 0 \ \text{Gyr}} = \\ \nonumber 
&= \frac{2 \eta t_\mathrm{d2}}{\zeta - 2\eta}.
\end{align}
Motivated by the model-to-data inconsistencies already observed in \citet{sysoliatina18} (see \mbox{also Section \ref{sect:results}}), 
we add two additional Gaussian peaks to the thin-disk SFR:
\begin{align}
\label{eq:sfrd}
SFR_\mathrm{d}^\prime(t) &= \left < SFR_\mathrm{d}^\prime \right > sfr_\mathrm{d}^\prime(t), \quad \text{where} \\
 sfr_\mathrm{d}^\prime(t) &= sfr_\mathrm{d}(t) + \nonumber \\
   &sfr_\mathrm{d}(t_\mathrm{p} - \tau_\mathrm{p1}) \frac{\Sigma_\mathrm{p1}}{\Sigma_\mathrm{max}}
                    \exp{\left(-\frac{(t_\mathrm{p} - \tau_\mathrm{p1}-t)^2}{2 (d\tau_\mathrm{p1})^2} \right)} + \nonumber \\
   &sfr_\mathrm{d}(t_\mathrm{p} - \tau_\mathrm{p2})\frac{\Sigma_\mathrm{p2}}{\Sigma_\mathrm{max}} 
                    \exp{\left(-\frac{(t_\mathrm{p} - \tau_\mathrm{p2}-t)^2}{2 (d\tau_\mathrm{p2})^2} \right)},  \nonumber \\
   \left< SFR_\mathrm{d}^\prime \right> &= \frac{\Sigma_\mathrm{d}}{\int_0^{t_\mathrm{p}} sfr_\mathrm{d}^\prime(t) g_\mathrm{d}(t) dt}. \nonumber 
\end{align}
Here $(\tau_\mathrm{p1},d\tau_\mathrm{p1})$ and $(\tau_\mathrm{p2},d\tau_\mathrm{p2})$ are the mean ages and dispersions of the additional SFR peaks 
and peak parameters $\Sigma_\mathrm{p1}$ and $\Sigma_\mathrm{p2}$ are related to their real amplitudes through 
$\Sigma_\mathrm{p,i}^{\prime} = sfr_\mathrm{d}(t_\mathrm{p}-\tau_\mathrm{p,i}) \cdot 
            \left< SFR_\mathrm{d}^\prime \right>/\Sigma_\mathrm{max} \cdot \Sigma_\mathrm{p,i}$.

\subsection{Gas model}\label{sect:model_gas}

In our previous works on the local disk model, starting from \citetalias{just10} to the most recent testing of the model against 
the TGAS-RAVE data in \citet{sysoliatina18}, the gas component was defined in full analogy to the thin disk: 
we used a set of 180 isothermal gas subpopulations whose vertical kinematics was given by a scaled thin-disk AVR. 
In this paper, we use an easier and more physically motivated gas model:  
we introduce a two-component ISM consisting of molecular and atomic gas, H$_2$ and HI, with 
the prescribed surface densities from \citet{mckee15} and scale heights from \citet{nakanishi16}. 

At the Solar circle, the ratio of gas-to-stellar disk surface densities is $\sim0.3$, therefore a robust model of the gas component
is crucial for the modelling of the local gravitational potential. 
At the same time, it is not easy to determine the local gas density from observations.  
In the case of atomic gas observed with the \mbox{21-cm} line, it is challenging for several reasons: (1) overall number of molecular clouds 
is low which makes it difficult to determine their spatial distribution; 
(2) special corrections for optical thickness and presence of He and heavier elements must be applied. 
For molecular gas, two additional problems arise: (3) H$_2$ is not observed directly, but is traced via CO, 
and CO-to-H$_2$ conversion factor introduces a considerable 
uncertainly; (4) some fraction of molecular hydrogen can be `dark' due to lower local concentration of CO as a tracer element.
Additionally, the vertical fall-off of different gas components is usually assumed to be Gaussian and this simplification adds to the total 
uncertainty of the local gas density estimates. 

For the local surface density of the molecular gas, we adopt \mbox{1.7 M$_\odot$ pc$^{-2}$} from \citet{mckee15}, 
which is the value from \citet{flynn06} 
updated with the new CO-to-H$_2$ ratio recommended by \citet{bolatto13}. This is the surface density for the Solar circle, 
and \citet{mckee15} recommends the use of a
lower density of \mbox{1.0 $\pm$ 0.3 M$_\odot$ pc$^{-2}$} for the Solar neighbourhood as the Sun is located in the Local Bubble. However, we do not 
apply this correction as we aim to apply our model to extended data samples. For the atomic gas we use \mbox{10.86 M$_\odot$ pc$^{-2}$} which is 
a sum of surface densities of the cold and warm atomic gas components from \citet{mckee15}. 
Our total gas surface density sums up to \mbox{12.86 $\pm$ 1.42 M$_\odot$ pc$^{-2}$}. 
The vertical velocity dispersions of H$_2$ and HI 
are then adjusted in our model to reproduce the observed thickness of the gas from \citet{nakanishi16} (\mbox{Table \ref{tab:fixed_param}}). 

Finally, we link the vertical kinematics of molecular gas to the vertical kinematics of the zero-age thin-disk subpopulation, 
as the latter is expected to inherit the vertical velocity dispersion of the material it was formed from. 
As before, the AVR is given as a power law, and is most conveniently defined in terms of age (Eq. (30) in \citetalias{just10}):
\begin{equation}
\label{eq:avr}
 \sigma_\mathrm{W}(\tau) = \sigma_\mathrm{e} \left( \frac{\tau + \tau_0}{t_\mathrm{p} + \tau_0} \right)^\alpha, 
            \  \text{where} \quad \tau = t_\mathrm{p} - t
.\end{equation}
This link between the gas and thin-disk vertical kinematics reduces the number of the model's free parameters as it 
sets a constraint on one of the AVR parameters:
\begin{equation}
\label{eq:avr_h2}
 \sigma_\mathrm{W}(0 \, \mathrm{Gyr}) \equiv \sigma_\mathrm{W,H_2}: \quad \tau_0 = \frac{t_\mathrm{p}}
                    {\left( ^{\sigma_\mathrm{W,H_2}}/_{\sigma_\mathrm{e}} \right)^{-1/\alpha}-1}
.\end{equation}

\begin{table}
 \centering
 \tiny
 \caption{JJ model parameters fixed in this work.}
 \begin{tabular}{l|l} \hhline{==}
 {\begin{minipage}[c][0.5cm][c]{2.0cm} \center{Solar position} \end{minipage}} &  
 {\begin{minipage}[c][0.5cm][c]{6.0cm} \center{$R_\odot=8.2$ kpc, $z_\odot=20$ pc} \end{minipage}} \\ \hdashline
                                                                  
 {\begin{minipage}[c][1.0cm][c]{2.0cm} \center{Thick-disk SFR} \end{minipage}} &  
 {\begin{minipage}[c][1.0cm][c]{6.0cm} \center{$t_\mathrm{t1} = 0.1 \ \text{Gyr}$, $t_\mathrm{t2} = 4 \ \text{Gyr}$, \\[5pt]
                                $\gamma = 3.5$, $\beta = 2 \ \text{Gyr}^{-1}$} \end{minipage}} \\ \hdashline

 {\begin{minipage}[c][1.0cm][c]{2.0cm} \center{Thin-disk SFR} \end{minipage}} &  
 {\begin{minipage}[c][1.0cm][c]{6.0cm} \center{$t_\mathrm{d1} = 0 \ \text{Gyr}$, $t_\mathrm{d2} = 7.8 \ \text{Gyr}$} \end{minipage}} \\ \hdashline

 {\begin{minipage}[c][1.0cm][c]{2.0cm} \center{Gas model} \end{minipage}} &  
 {\begin{minipage}[c][1.0cm][c]{6.0cm} \center{$\Sigma_\mathrm{H_2} = 1.7 \ \text{M}_\odot \ \text{pc}^{-2}$, 
                                        $\Sigma_\mathrm{HI} = 10.86 \ \text{M}_\odot \ \text{pc}^{-2}$, \\[5pt]
                                        $h_\mathrm{H_2} = 61$ pc, $h_\mathrm{HI} = 195$ pc} \end{minipage}} \\ \hdashline

 {\begin{minipage}[c][1.5cm][c]{2.0cm} \center{Stellar halo profile \\ and kinematics} \end{minipage}} &  
 {\begin{minipage}[c][1.5cm][c]{6.0cm} \center{$\alpha_\mathrm{in} = 0.7$, $q_\mathrm{in}=-2.5$, \\[5pt] 
                                $\alpha_\mathrm{out}=0.8$, $q_\mathrm{out}=-4$ \\[5pt]
                                $\sigma_\mathrm{sh} = 100$ km s$^{-1}$, $\sigma_\mathrm{dh} = 140$ km s$^{-1}$} \end{minipage}} \\ \hdashline

 {\begin{minipage}[c][0.5cm][c]{2.0cm} \center{Time resolution} \end{minipage}} &  
 {\begin{minipage}[c][0.5cm][c]{6.0cm} \center{$t_\mathrm{p}=13$ Gyr, $\Delta t = 25$ Myr} \end{minipage}} \\ \hdashline

 {\begin{minipage}[c][0.5cm][c]{2.0cm} \center{Spatial resolution} \end{minipage}} & 
 {\begin{minipage}[c][0.5cm][c]{6.0cm} \center{$z_\mathrm{max}=2$ kpc, $\Delta z$ = 2 pc} \end{minipage}} \\  \hhline{==}
 \end{tabular}
 \label{tab:fixed_param}
\end{table}

\subsection{Stellar halo}\label{sect:model_halo}

In \citetalias{just11}, a flattened spherical halo was used to fit the SDSS star counts towards the northern Galactic pole. 
In this study we treat the stellar halo as a single isothermal component of the age of 13 Gyr characterised 
by the vertical velocity dispersion of \mbox{100 km s$^{-1}$} and
the local surface density that is allowed to be optimised. 
At \mbox{$|z| \leq 2$ kpc}, where the Poisson-Boltzmann equation is solved, 
the vertical density profile of the halo is derived self-consistently with the 
total gravitational potential $\Phi(|z|)$. However, to simulate our conic \textit{Gaia} sample, we need to assume the vertical density profile 
of the stellar halo at higher distances from the Galactic plane, \mbox{$|z|> 2$ kpc}. 
For this purpose, we adopt a broken power law. 
We define a radial Galactocentric coordinate, $r(q) = \sqrt{R^2 + z^2/q^2}$, 
where $q$ is a flattening parameter. As in practice our modelling is performed in the Cartesian grid, 
we write the halo density profile as a function of distance from the Galactic plane $|z|$. 
A breaking height that sets a boundary between the inner and outer halo is related 
to a breaking radius as $r_\mathrm{br}=\sqrt{R^2+z_\mathrm{br}^2/q_\mathrm{in}^2}$, 
the value of \mbox{$r_\mathrm{br} = 25$ kpc} is adopted from \citet{bhawthorn16}. 
Then the general form of the halo vertical profile can be written as   
\begin{equation}
\rho_\mathrm{sh}(|z|) = 
  \begin{cases}
      \ \left\{\rho_\mathrm{sh}(|z|),\Phi(|z|) \right\}, & |z| \leq 2 \ \text{kpc} \\ 
      \ \rho_\mathrm{sh,0}^\mathrm{in} \left( {r(q_\mathrm{in})}/{R_\odot} \right)^{-\alpha_\mathrm{in}}, & 2 \ \text{kpc} < |z| \leq z_\mathrm{br} \\
      \ \rho_\mathrm{sh,0}^\mathrm{out} \left( {r(q_\mathrm{out})}/{R_\odot} \right)^{-\alpha_\mathrm{out}}, & |z| > z_\mathrm{br} 
  \end{cases}
\label{eq:shalo}
.\end{equation}
Here $(q_\mathrm{in},\alpha_\mathrm{in})$ and $(q_\mathrm{out},\alpha_\mathrm{out})$ are 
flattening and slope of the inner and outer halo, respectively. 
Their values taken from \citet{bhawthorn16} are listed in \mbox{Table \ref{tab:fixed_param}}. 
The scaling parameters, $\rho_\mathrm{sh,0}^\mathrm{in}$ and $\rho_\mathrm{sh,0}^\mathrm{out},$ are chosen  
such that they ensure continuity of the profile at \mbox{$|z| = 2$ kpc} and $|z| = z_\mathrm{br}$. Division by $R_\odot$ is added as 
the scaling densities are given for the Solar position.

\section{Method}\label{sect:method}

With the specified input functions $\{\text{SFR, IMF, AVR, AMR}\}$, as well as several additional model parameters (\mbox{Table \ref{tab:fixed_param}}),
the Poisson-Boltzmann equation is solved iteratively. 
This returns a self-consistent pair of the gravitational potential and the vertical density law, \{$\Phi(|z|),\rho(|z|)$\}. 
Then the derived density profiles of the thin- and thick-disk and halo mono-age subpopulations are used to model the local 
samples with different photometric properties. Using a stellar evolution library, we locate our mono-age subpopulations in 
the 3D age-metallicity-mass space (\mbox{Section \ref{sect:method_stpop}}). Finally, with photometric information added from the isochrones, we 
are able to reproduce the spatial distribution of samples selected on the CMD, as well as model different selection effects 
(\mbox{Sections \ref{sect:method_nzfw}, \ref{sect:method_cone}, and \ref{sect:method_amr}}).

\subsection{Stellar population synthesis}\label{sect:method_stpop}

We follow the scheme described in our previous study \citep{sysoliatina18}. 
For the sake of completeness, we summarise the main points of this procedure here. 

We populate the 3D age-metallicity-mass parameter space using age-metallicity pairs prescribed by the adopted AMR 
and stellar masses given in isochrones. Following \citet{sysoliatina18}, we use the term `stellar assembly' to refer 
to a set of stars characterised by the same age, metallicity, and mass. 
We take a set of 62 isochrone tables generated with the PAdova and TRieste Stellar Evolution Code (PARSEC v.1.2S; \citealp{bressan12})
and the COLIBRI code (v. S{\_}35; \citealp{marigo17}) for the thermally pulsing asymptotic giant branch (\mbox{TP-AGB}) 
phase\footnote{\url{http://stev.oapd.inaf.it/cgi-bin/cmd}}. 
The selected isochrones cover a wide metallicity range, 
\mbox{[Fe/H]$\, =-2.6...+0.47,$} and are generated with a linear age step of \mbox{$\Delta \tau = 50$ Myr}.  
The total number of stellar assemblies of the thin disk, thick disk, and halo sum up to $\sim6.6 \cdot 10^5$. 
Each of our mono-age isothermal subpopulations of the disk is characterised by metallicity drawn from the AMR function 
(\mbox{Section \ref{sect:method_amr}}). 
In the case of the halo we use a set of metallicities drawn from the Gaussian metallicity distribution (\mbox{Section \ref{sect:method_cone}}).
Then, for a given age-metallicity pair we chose an isochrone with the closest age and metallicity values. 
Finally, each of the age-metallicity-mass stellar assemblies from the chosen isochrone 
is further assigned with a surface number density prescribed by the SFR and IMF. 

In order to check the impact of the stellar library on our results, we also use an alternative set of isochrones from  
MESA (Modules and Experiments in Stellar Astrophysics, \mbox{\citealp{paxton11,paxton13,paxton15}}) 
Isochrones and Stellar Tracks (MIST\footnote{\url{http://waps.cfa.harvard.edu/MIST/}}, v.1.2; \citealp{dotter16,choi16})
(see \mbox{Sections \ref{sect:results_mcmc2} and \ref{sect:discus_lim_stel}}). 
The MIST isochrones are used with the same metallicity and age grid as the isochrones generated by PARSEC.

\subsection{Vertical profiles and velocity distributions}\label{sect:method_nzfw}

To simulate the samples of A, F, and RC stars and G, G/K, and K dwarfs, we 
apply colour-magnitude cuts from \mbox{Table \ref{tab:data}} to the set of stellar assemblies, fully analogously to criteria applied to the data.
Here we use synthetic Gaia passbands from \citet{maiz-apellaniz18}.  

At this point, the construction of the number density profiles and velocity distribution functions of the mock samples is quite straightforward. 
The number density profiles are calculated according to Eq. (4) from \citet{sysoliatina18} with \mbox{$|z|$-resolution} of \mbox{20 pc}. 
The velocity distribution functions $f(|W|)$ are derived as a sum of Gaussian distributions with known dispersion 
weighted by the number of stars in a given volume. Here the $W$-velocity dispersions of the thin-disk populations is prescribed by the AVR, 
\mbox{Eq. (\ref{eq:avr})},
and parameters $\sigma_\mathrm{p1}$ and $\sigma_\mathrm{p2}$ that are associated with the recent periods of the SF enhancement 
(more in \mbox{Section \ref{sect:method_mcmc_prior}}).  
The thick-disk and halo velocity dispersions, $\sigma_\mathrm{t}$ and $\sigma_\mathrm{sh},$ are listed in \mbox{Table \ref{tab:fixed_param}}. 
We trace the $f(|W|)$ shape up to \mbox{$|W|_\mathrm{max}=60$ km s$^{-1}$} and set 
the velocity resolution to \mbox{$\Delta |W|=2$ km s$^{-1}$} as this bin size is larger than a typical velocity error in the bin 
which is \mbox{$\sim0.7$ km s$^{-1}$}.

\subsection{Conic sample modelling}\label{sect:method_cone}

In order to model our third quantity of interest, the Hess diagram of the conic sample, we need to make several additional steps. 

As explained in \mbox{Section \ref{sect:method_amr}}, we use APOGEE RC stars to constrain the AMR functions of the thin and thick disk. 
In the case of the stellar halo, we assume a Gaussian metallicity distribution with mean metallicity and dispersion 
$(\left< \text{[Fe/H]} \right>,\sigma_\text{[Fe/H]})=(-1.5,0.4)$. 
From this distribution, we draw nine metallicity values and use them in combination with 
halo age of 13 Gyr to select isochrones. The total number density assigned to the halo stellar assemblies is normalised to its present-day 
surface density up to $|z|=2$ kpc. 

In order to optimise computation time needed to produce the mock Hess diagram of the conic sample, we use 
a grid with a variable $|z|$-step. To define such a grid, we use the condition 
\begin{equation}
z_\mathrm{i+1} = z_\mathrm{i} 10^{0.2 \Delta G}, 
\label{eq:cone_zgrid}
\end{equation}
where \mbox{$\Delta G = 0.1$ mag} is the apparent magnitude resolution of the Hess diagram. 
Close to the Galactic plane, we also check that the vertical resolution of our grid is not better than \mbox{2 kpc}, 
as this is the vertical resolution, $\Delta z,$ adopted for the reconstruction of the vertical gravitational potential 
(see \mbox{Table \ref{tab:fixed_param}}). Defined by \mbox{Eq. (\ref{eq:cone_zgrid})}, the vertical step of such a grid grows linearly with $|z|$, 
and reaches \mbox{$\sim6$ kpc} at height \mbox{$|z|=130$ kpc} that is a height up to which we model the halo profile. 

We also ignore the vertical offset of the Sun and assume that it is located exactly in the Galactic plane. 
In the case when \mbox{$z_\odot = 20$ pc} is taken into account, 
the summation of the stellar densities in the directions to the northern and southern Galactic poles
must be performed separately up to \mbox{$\sim0.9$ kpc} (further away from the Galactic plane south-north differences in the apparent magnitudes 
of stellar assemblies become smaller than the adopted resolution $\Delta G$). This results in smoothing of the Hess diagram along the vertical axis, 
apparent magnitude $G$, although all the features change only slightly. At the same time, Hess diagram computation time almost doubles, 
which may have a great impact on the final calculation time of our MCMC sampling (\mbox{Section \ref{sect:results_mcmc_setup}}). 
As we smooth the mock Hess diagram later, the effect of $z_\odot$ is essentially masked and, therefore, we decided to ignore the Solar offset 
in this case. We note, however, that this is done only for the conic sample and in the case 
of the number density profiles and vertical velocity distributions, $z_\odot,$ value from \mbox{Table \ref{tab:fixed_param}} 
is properly taken into account during the selection of the stellar assemblies representing the \textit{Gaia} samples.  

As the self-consistent density-potential pair $\{\rho(|z|), \Phi(|z|)\}$ is calculated within the model up to \mbox{2 kpc} only 
(see parameter $z_\mathrm{max}$ in \mbox{Table \ref{tab:fixed_param}}), 
we need to extrapolate the derived density profiles to larger heights in order to model all stars within the conic volume. 
The thin disk is a plane-concentrated component, 
therefore we model its populations up to \mbox{2 kpc} only.
To include the thick disk, we fit its vertical profile with a $sech^{\alpha_\mathrm{t}}$ law of a single isothermal population. 
As in \citetalias{just11}, 
we find that the vertical profile of the thick-disk component follows the $sech^{\alpha_\mathrm{t}}$ law very closely 
as it is kinematically homogeneous. 
So we use such an extrapolation to model the thick-disk stellar assemblies up to \mbox{12 kpc} from the Galactic plane. 
The vertical profile of the halo is chosen to be a broken power law and is given by \mbox{Eq. (\ref{eq:shalo})}. 
Although the opening angle of the cone is relatively small, $\theta = 20^\circ$, at large heights above the Galactic plane 
the conic sample is not local any more. Therefore, at each $|z|$ we take a mean halo density in this horizontal slice. 
Also, at each $|z|$ stellar assemblies of all components are exposed to the same cut on apparent $G$-magnitudes as was applied to the data, 
$12 \ \text{mag} < G < 17 \ \text{mag}$ (\mbox{Table \ref{tab:data}}).

\subsection{Model optimisation procedure}\label{sect:method_optim}

Before explaining how we constrain the AMR function, 
we outline the overall scheme of the model optimisation carried out in this work. 

Once they have been constructed for the initial input functions $\{\text{SFR, IMF, AVR, AMR}\}_0$, 
stellar assemblies can be used to simulate model predictions with the modified SFR and IMF.
This is possible because changes in shapes of these functions can be taken into account simply 
by adding corresponding weights sensitive to stellar ages and masses. 
However, a modification of the AMR function implies that for each modelled mono-age population a new isochrone has to be chosen, 
and the whole process of weighting the SFR with IMF functions has to be repeated for the new stellar assemblies. 
As this process is relatively time-consuming, it is not possible to update the AMR on-the-fly 
as only a short model-to-data comparison time is acceptable 
for an exploration of the multidimensional parameter space with the MCMC method.
Among the four model input functions, including the AVR (which implicitly prescribes how the populations are distributed perpendicular 
to the Galactic plane), the weakest impact on the predicted self-consistent density-potential pair is from the AMR function. 
Taking this into account, we developed the following model optimisation scheme: 
\setlength\itemsep{0.3em}
 
Firstly, we constrain the chemical evolution history given initial $\{\text{SFR, IMF, AVR}\}_0$. To do so, we simulate the local RC sample 
 (\mbox{Section \ref{sect:data_apogee}}), and iteratively converge to AMR$_0$ such that the observed metallicity distributions 
 of the high- and low-$\alpha$ 
 \mbox{APOGEE RC} stars are best reproduced within this model (Section \ref{sect:method_amr}). 

Secondly, we use the MCMC technique to adapt the most important model parameters self-consistently (Section \ref{sect:method_mcmc}). 
 At this step, the shape of the AMR$_0$ function is not varied. 
 
Finally, using the new functions $\{\text{SFR, IMF, AVR}\}^\prime$ obtained in the previous step, 
 we update the chemical evolution part of the model following the strategy from step (1). 
This sequence of steps is repeated until the process converges to a fully self-consistent set of the model functions 
$\{\text{SFR, IMF, AVR, AMR}\}^\prime$. 
In practice, we find that all changes to the model parameters become negligible as soon as just after the second iteration.

\subsection{Age-metallicity relation with APOGEE RC}\label{sect:method_amr}

\begin{figure*}[t]
\centerline{\resizebox{\hsize}{!}{\includegraphics{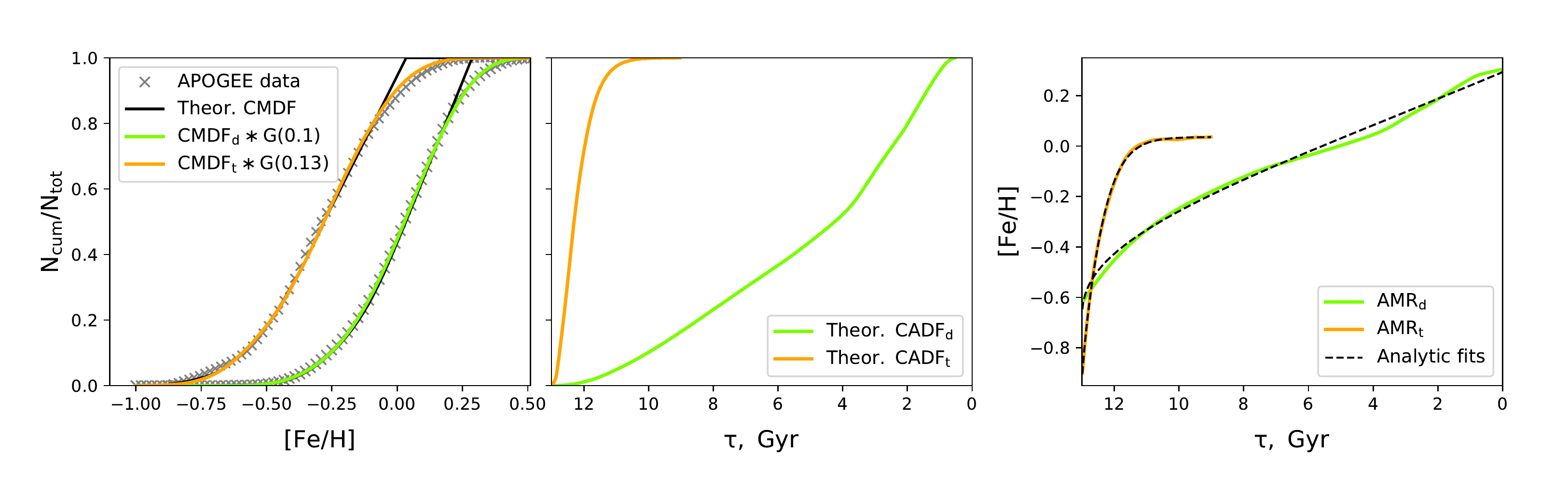}}}
\caption{AMR reconstruction. All model predictions shown here are generated with the best-fit parameters of $\mathrm{MCMC1}$ model 
(\mbox{Section \ref{sect:results}}). 
\textit{Left}: Normalised CMDFs of the high- and low-$\alpha$ APOGEE RC samples. Grey crosses show the smoothed data, 
thick black lines are the `deconvolved' CMDFs (see text), and the full coloured lines illustrate 
the result of the convolution of the `deconvolved' CMDF with Gaussian kernels. 
\textit{Middle}: Modelled CADFs for the thin and thick disk RC populations. 
\textit{Right:} Resulting AMR for the thin and thick disk. Analytic fits are over-plotted with black dashed curves.}
\label{fig:get_amr}
\end{figure*}

In order to constrain the AMR function, we perform a direct comparison between the observed metallicity distribution of the 
high- and low-$\alpha$ APOGEE RC stars (\mbox{Section \ref{sect:data_apogee}}) 
and the predicted age distributions of the thin- and thick-disk RC populations in the model. 

The left panel of \mbox{Figure \ref{fig:get_amr}} shows the cumulative metallicity distribution functions 
(CMDFs) calculated for two selected RC samples and 
smoothed with the Savitzky-Golay filter (grey crosses). 
A highly non-linear behaviour of the observed CMDFs in the low- and high-metallicity regimes 
can be associated with metallicity uncertainties or systematic errors. 
Otherwise, under the assumption that the low and high metallicity values correspond directly to the old and young stellar ages,
this picture implies a very rapid enrichment in the beginning of the Galactic evolution and during the last \mbox{$\sim1-2$ Gyr} of 
both thin- and thick-disk formation.
The first of these epochs of fast ISM enrichment is naturally associated with an early phase 
of the Galactic evolution characterised by intense SF processes. 
However, a recent rise of the ISM enrichment rate, as implied by the low-$\alpha$ CMDF, 
contradicts a variant of the mediocrity principle that assumes that the time we live in is not special.  
Alternatively, the most metal-rich stars in both samples may be stars that migrated to the Solar radius from the inner Galaxy, 
thus, they are not necessarily the youngest populations in the samples. 
Within the naive approach applied here, we do not aim to distinguish between the two above-mentioned interpretations 
of the non-linear, metal-rich parts of the observed CMDFs. 
We rather aim to reconstruct the AMR that: (1) predicts a monotonous increase of metallicity with Galactic time; and 
(2) does not demonstrate any special behaviour in the end of both thin- and thick-disk evolution phases. 
The residual deviations from such AMR functions are included in the form of an error model.

We assume that the shape of the metal-rich parts of both observed CMDFs can be expressed   
as a result of the convolution of a linear function with a Gaussian kernel. Here, the former corresponds to 
observations fully consistent with our AMR of interest and the latter represents the effect of observational 
errors or the presence of migrated stars. Analytically, this convolution $C(x)$ can be expressed as:
\begin{align}
 \label{eq:analyt_convol}
 C(x) & = \nonumber \\ 
 & \frac{1}{2}(kx+b) \cdot \left[\mathrm{Erf} \left(\frac{kx+b}{\sqrt{2}k\sigma} \right) - \mathrm{Erf} \left(\frac{kx+b-1}{\sqrt{2}k\sigma} \right) \right] + \nonumber \\
 & \frac{k\sigma}{\sqrt{2\pi}}\left[\exp{\left(-\frac{(kx+b)^2}{2k^2\sigma^2}\right)} - \exp{\left(\frac{(kx+b-1)^2}{2k^2\sigma^2}\right)}\right] + \nonumber \\
 & \frac{1}{2}\left[1 + \mathrm{Erf} \left(\frac{kx+b-1}{\sqrt{2}k\sigma}\right)\right], \quad \text{where} \quad x=\mathrm{[Fe/H].}
\end{align}
Here, $k$ and $b$ are slope and offset of the CMDF linear part, and $\sigma$ is a the Gaussian dispersion. 
We use \mbox{Eq. (\ref{eq:analyt_convol})} to fit the upper parts of the observed CMDFs of the high- and low-$\alpha$ APOGEE RC samples, 
$N_\mathrm{cum}/N_\mathrm{tot}>0.5$. 
We use their full linear parts $N_\mathrm{cum}/N_\mathrm{tot}=0.3...0.7$ to calculate parameters $k$ and $b$.  
Then values of $\sigma$ needed to reproduce the metal-rich tails of CMDFs are found to be 
\mbox{$\sim0.1$ dex} and \mbox{$\sim0.13$ dex} for the thin and thick disk, respectively. 
The `deconvolved' linear trends of the metal-rich parts of the observed CMDFs are shown with black thick lines in 
\mbox{Figure \ref{fig:get_amr}}. These `deconvolved' CMDFs are later used for the AMR reconstruction. 

\begin{table}
 \centering
 \tiny
 \caption{AMR fit parameters for Eqs. (\ref{eq:amr_fit}) and (\ref{eq:amr_fit_f}).}
 \begin{tabular}{l|c|c|c|c|c|c} \hhline{=======}
 & {\begin{minipage}[t][0.35cm][t]{1.2cm} \center{$\mathrm{[Fe/H]_0}$} \end{minipage}} 
     & {\begin{minipage}[t][0.35cm][t]{1.2cm} \center{$\mathrm{[Fe/H]_\mathrm{p}}$} \end{minipage}} 
     & {\begin{minipage}[t][0.35cm][t]{0.8cm} \center{$\mathrm{r_\mathrm{d}}$} \end{minipage}}  
     & {\begin{minipage}[t][0.35cm][t]{0.8cm} \center{$\mathrm{q}$} \end{minipage}} 
     & {\begin{minipage}[t][0.35cm][t]{0.8cm} \center{$\mathrm{r_\mathrm{t}}$} \end{minipage}} 
     & {\begin{minipage}[t][0.35cm][t]{0.8cm} \center{$\mathrm{t_0}$} \end{minipage}} \\ \hline
 Thin disk & $-0.7$ & 0.29 & 0.34 & $-0.72$ & &  \\
 Thick disk & $-0.94$ & 0.04 & & & 0.77 & 0.97 \\[2pt] \hhline{=======}
 \end{tabular}
 \label{tab:amr}
\end{table}

In order to proceed, we assume that there is a unique correspondence between metallicities in the `deconvolved' CMDFs 
and stellar ages in the Solar neighbourhood. This is known to be a simplification, 
but in \mbox{Section \ref{sect:discus_lim_amr},} we argue that this approach is nevertheless reasonable enough for our analysis.  
We synthesise the stellar assemblies that represent thin and thick disk as explained above. 
To do so, we specify the model functions $\{\text{SFR, IMF, AVR}\}_0$, and assume initial functions AMR$_{\text{ini}}$. 
From the generated stellar assemblies, we select RC stars by applying the following cuts: 
\begin{align}
 4 \, 250 \ \text{K} &< T_\mathrm{eff} < 5 \, 250 \ \text{K} \quad \& \quad 1.6 < \log{L/L_\odot} < 1.85  \label{eq:rc_select1} ,\\ 
 \log{g} &< min(2.9, 22.5 \log{T_\mathrm{eff}} - 80.1 + 0.5 \mathrm{[Fe/H]}).  \label{eq:rc_select2}
\end{align}
Here, Eq. (\ref{eq:rc_select1}) selects stars in the temperature and luminosity range corresponding to the location of RC 
at the Herzsprung-Russel diagram (HRD). 
The general form of the second cut given by \mbox{Eq. (\ref{eq:rc_select2})} is taken from 
the analysis of the APOGEE giants performed in \citet{bovy14}, 
where the authors used a similar criterium to clean the RC sample from the red giant branch (RGB) (see their \mbox{Eqs. (2) and (3)}). 
The exact coefficients of \mbox{Eq. (\ref{eq:rc_select2})} are different to those given in \citet{bovy14}, 
as the authors used an older version of Padova isochrones, 
so their selection cannot properly separate RC and RGB stars in the $\{\log{g}, \log{T_\mathrm{eff}}, \mathrm{[Fe/H]}\}$ parameter space
of the updated stellar library used in this work. We also find that \mbox{Eq. (\ref{eq:rc_select2})} is not useful 
when the MIST isochrones are used and it needs to be modified also in this case. 
For the sake of completeness, we also show our second version of \mbox{Eq. (\ref{eq:rc_select2}),} which can be used 
together with the MESA stellar evolution library: 
\begin{equation}
 \log{g} < min(2.9, 22.1 \log{T_\mathrm{eff}} - 78.4 + 0.6 \mathrm{[Fe/H]})  
 \label{eq:rc_select2b}
.\end{equation}

Using our mock RC stellar assemblies and modelled vertical gravitational potential, we predict the vertical number density profiles of the RC stars
up to the heights of \mbox{$|z|=1$ kpc} and \mbox{$|z|=2$ kpc} for the thin and thick disk, respectively 
(by analogy to the data selection criteria in Section \ref{sect:data_apogee}). 
The cumulative age distribution functions (CADFs) calculated for these mock RC samples are shown in the middle panel of 
\mbox{Figure \ref{fig:get_amr}}. 

Finally, each metallicity from the `deconvolved' CMDF is assigned with an age from the corresponding modelled CADF. 
The resulting AMR usually differs from the initial function AMR$_{\text{ini}}$, unless the initial guess is very good. 
To bring the derived AMR into the full consistency with the assumed $\{\text{SFR, IMF, AVR}\}_0$, 
the generation of the RC age distributions is repeated three to four times. During each iteration, 
we use the AMR derived in the previous cycle to update our stellar assemblies, as well as the
mass loss function. In the end, the procedure converges to our target AMR$_0$.  

For convenience, we also fit the thin- and thick-disk AMR functions with the following equation:
\begin{equation}
\label{eq:amr_fit}
\mathrm{[Fe/H]}(t) = \mathrm{[Fe/H]}_0  + (\mathrm{[Fe/H]}_\mathrm{p}-\mathrm{[Fe/H]}_0) \cdot f(t),
\end{equation}
where $\mathrm{[Fe/H]}_0$ and $\mathrm{[Fe/H]}_\mathrm{p}$ are the initial and present-day metallicity values and 
\begin{equation}
\label{eq:amr_fit_f}
f(t) = 
  \begin{cases}
      \ \log(1+q(t/t_\mathrm{p})^{r_\mathrm{d}})/\log(1+q) & \text{for \ the thin \ disk,} \\
      \ \tanh^{r_\mathrm{t}}(t/t_0)  & \text{for \ the thick \ disk.}
  \end{cases}
\end{equation}
The CADFs, as well as the thin-disk and thick-disk AMR functions shown in the right panel of Figure \ref{fig:get_amr}, are 
consistent with the updated parameters of the \mbox{JJ model} presented below in \mbox{Table \ref{tab:best_param}} ($\mathrm{MCMC1}$ model). 
The best-fit parameters of \mbox{Eqs. (\ref{eq:amr_fit}) and (\ref{eq:amr_fit_f})} are summarised in \mbox{Table \ref{tab:amr}}.

\subsection{Bayesian analysis}\label{sect:method_mcmc}

In order to improve the JJ model parameters, we use Bayes theorem.
It links posterior $\mathbb{P}(m|d)$ expressing a probability of a model based on the given data, the likelihood $\mathcal{L}(d|m)$
describing a probability to observe some data based on a given model, and prior $\mathcal{P}(m)$ 
expressing our initial knowledge about the model parameters: 
\begin{equation}
 \mathbb{P}(m|d) \propto \mathcal{L}(d|m) \mathcal{P}(m)
 \label{eq:bayes_theorem}
.\end{equation}
It is common to write \mbox{Eq. (\ref{eq:bayes_theorem})} in logarithmic scale with likelihood and posterior normalised to their maximum values: 
\begin{equation}
 \ln{\mathbb{P}} = \ln{\frac{\mathcal{L}}{\mathcal{L}_0}} + \ln{\frac{\mathcal{P}}{\mathcal{P}_0}}
 \label{eq:log_posterior}
.\end{equation}
In the subsections below, we explain how the likelihood and prior are calculated and we specify a set of model parameters chosen for optimisation.

\subsubsection{Likelihood}\label{sect:method_mcmc_likelihood}

The likelihood function is aimed at describing a model-to-data goodness of fit in terms of the three different types of data: 
vertical number density profiles, 
$W$-velocity distribution functions, and apparent Hess diagram of the conic sample. 
As there is no standard recipe for how to combine probabilities from different data types, 
we use the following method: each data type is set to correspond to a separate term in the logarithmic likelihood 
and separately normalised on its own value 
that is calculated for a set of parameters we refer to as `standard' (\mbox{Section \ref{sect:method_mcmc_prior}}). 
This approach ensures that the contributions of these three terms to the total likelihood are comparable, 
such that all data types have similar weights during the fitting procedure. 

To make the final formula for the likelihood more compact, we introduce several definitions. 
We have the indices $(z,W,H)$ corresponding to the type of quantity a given term refers to: 
vertical profiles, kinematics, and Hess diagram, respectively. 
Superscripts $(m,d)$ mark model or data origin of some value. 
We also define the number of $|z|$- and $|W|$-bins used to calculate the vertical number density profiles and 
velocity distributions from the selected \textit{Gaia} samples as $B_z$ and $B_W$, respectively.  Then
$B_\mathrm{W}$ is the same for all samples with known vertical kinematics, 
$B_\mathrm{W} = |W|_\mathrm{max}/\Delta|W|=30$, and $B_\mathrm{z}$ depends on the 
vertical extent of the sample and may vary. In the case of the apparent Hess diagram of the conic sample, 
$B_\mathrm{H}$ is the total number of colour-magnitude bins given by the adopted colour-magnitude resolution 
(\mbox{Section \ref{sect:data_gaia_cone}}): $B_\mathrm{H} = 60 \times 50 = 3 \cdot 10^3$.  
The number of samples used to fit star counts and kinematics are denoted as 
$S_\mathrm{z}=6$ and $S_\mathrm{W}=5$, respectively (\mbox{Table \ref{tab:data}}). 
Finally, notations $n_\mathrm{z}$ and $n_\mathrm{W}$ correspond to the number density of stars per 
$|z|$- or \mbox{$|W|$-bin}, 
and $N_\mathrm{H}$ is the actual number of stars in a colour-magnitude bin at the Hess diagram. 

In terms of these definitions, and with $\pmb{\theta_0}$ being the vector of standard parameter values, 
the general expression for the normalised logarithmic likelihood is written as: 
\begin{equation}
\label{eq:log_likelihood}
\ln{\frac{\mathcal{L}}{\mathcal{L}_0}} = 
 \frac{\ln\mathcal{L}_\mathrm{z}}{|\ln\mathcal{L}_\mathrm{z}(\pmb{\theta_0})|} 
    + \frac{\ln\mathcal{L}_\mathrm{W}}{|\ln\mathcal{L}_\mathrm{W}(\pmb{\theta_0})|} + 
    \frac{\ln\mathcal{L}_\mathrm{H}}{|\ln\mathcal{L}_\mathrm{H}(\pmb{\theta_0})|} 
.\end{equation}
The first term in \mbox{Eq. (\ref{eq:log_likelihood})} quantifies consistency between 
the number density profiles derived from our selected \textit{Gaia} samples 
and the number density profiles predicted by the \mbox{JJ model} for the same stellar populations: 
\begin{equation}
\label{eq:log_likelihood_z}
 \ln\mathcal{L}_\mathrm{z} = -\frac{1}{S_\mathrm{z}} \sum_i^{S_\mathrm{z}} \left( \frac{1}{B_\mathrm{z,i}} \sum_k^{B_\mathrm{z,i}} 
                \frac{|\log{n_\mathrm{z,k}^\mathrm{m}} - \log{n_\mathrm{z,k}^\mathrm{d}}|}{|\log{n_\mathrm{z,k}^\mathrm{m}}|} \right)
.\end{equation}
This formula returns a mean of the relative deviation between the predicted and observed number density per \mbox{$|z|$-bin} 
(expression in brackets), which is also averaged over all samples. 
As the actual number density values may vary by several orders of magnitude within the studied volume (see \mbox{Figure \ref{fig:nz}}), 
it is more sensible to measure the relative model-to-data deviations in logarithmic scale. 
In this case, the fitting procedure is sensitive to both the cores and tails of the density profiles. 

The second likelihood term in \mbox{Eq. (\ref{eq:log_likelihood})} gives a measure of the consistency between the observed 
and predicted $|W|$-velocity distributions.
We define it in a way similar to \mbox{Eq. (\ref{eq:log_likelihood_z})}, but using a linear scale for model-to-data 
relative deviations per velocity bin:
\begin{equation}
\label{eq:log_likelihood_w}
 \ln\mathcal{L}_\mathrm{W} = -\frac{1}{S_\mathrm{W} B_\mathrm{W}} \sum_i^{S_\mathrm{W}} \sum_k^{B_\mathrm{W}} 
                    \left( \frac{|n_\mathrm{W,k}^\mathrm{m} - n_\mathrm{W,k}^\mathrm{d}|}{n_\mathrm{W,k}^\mathrm{m}} \right)
.\end{equation}

The observed and synthetic apparent Hess diagrams are compared in terms of $\chi^2$ translated into a logarithmic space \citepalias{just11}:
\begin{equation}
\label{eq:log_likelihood_h}
 \ln\mathcal{L}_\mathrm{H} = - \frac{1}{B_\mathrm{H}} \sum_i^{B_\mathrm{H}} 
                (\log{N_\mathrm{H,i}^\mathrm{m}} - \log{N_\mathrm{H,i}^\mathrm{d}})^2 \log{N_\mathrm{H,i}^\mathrm{m}} 
.\end{equation}
By analogy to Eqs. (\ref{eq:log_likelihood_z}) and (\ref{eq:log_likelihood_w}), this formula returns a logarithmic $\chi^2$ per 
colour-magnitude bin averaged over the whole apparent Hess diagram. 

As we prefer to divide the absolute model-to-data differences  
by a smooth function, all model-to-data deviations in \mbox{Eqs. (\ref{eq:log_likelihood_z})-(\ref{eq:log_likelihood_h})} are calculated 
relative to the model predictions, that are free of noise.

\subsubsection{Prior}\label{sect:method_mcmc_prior}

We define our logarithmic prior as a sum over Gaussian probabilities:
\begin{equation}
 \ln{\frac{\mathcal{P}}{\mathcal{P}_0}} = - \sum_i^{\pmb{\theta}} \frac{1}{2} \left( \frac{\theta_\mathrm{i} - 
                    \left< \theta \right>_\mathrm{i}}{\Delta \theta_\mathrm{i}} \right)^2,
\end{equation}
where $\left< \theta \right>_i$ and $\Delta \theta_i$ are the mean and dispersion of parameter 
$\theta_\mathrm{i}$, and the sum is taken over $\pmb{\theta}$,
the full vector of parameters chosen for optimisation. 

While specifying the vector $\pmb{\theta}$, 
we tried to include all key parameters that influence predicted star counts and kinematics.  
Several new parameters were introduced in order to achieve an adequate consistency between the model and data.
At the same time, we tried to keep the number of free parameters minimal, as an increase of the parameter space dimension leads to a non-linear 
increase of the computation time needed to find the best model. 
With these considerations taken into account, we selected for the optimisation the following model parameters 
(see \mbox{Table \ref{tab:best_param}} for a summary):  
\begin{eqnarray}
\label{eq:parameters}
 \pmb{\theta} &=& \left\{ \Sigma_\mathrm{d}, \Sigma_\mathrm{t}, \Sigma_\mathrm{dh}, \Sigma_\mathrm{sh}, \right. \nonumber \\
                        && \ \sigma_\mathrm{e}, \alpha, \sigma_\mathrm{t}, \sigma_\mathrm{p1}, \sigma_\mathrm{p2}, \\
                        && \ \zeta,\eta, \Sigma_\mathrm{p1}, \tau_\mathrm{p1}, d\tau_\mathrm{p1}, 
                                         \Sigma_\mathrm{p2}, \tau_\mathrm{p2}, d\tau_\mathrm{p2}, \nonumber \\
                &&  \left. \ \alpha_\mathrm{0}, \alpha_\mathrm{1}, \alpha_\mathrm{2}, m_\mathrm{0}, m_\mathrm{1} \right\}. \nonumber 
\end{eqnarray}

The first four of these parameters refer to the surface densities of the thin and thick disk, the DM, and the stellar halo. 
These parameters play a role of the local normalisation of the model components' vertical profiles. Here, 
$\Sigma_\mathrm{d}$ and $\Sigma_\mathrm{t}$ strongly influence 
the total predicted number of stars, while the DM surface density $\Sigma_\mathrm{dh}$ impacts the shape of the vertical gravitational potential 
and, thus, affects the vertical fall-off of all populations. 
The parameter $\Sigma_\mathrm{sh}$ has very moderate impact on the vertical profiles and kinematics, 
but is important for the reconstruction of the conic sample, whose significant fraction constitutes of halo stars.  
The mean prior values of the surface densities of the thin and thick disk are 
taken close to the values from the original \mbox{JJ model} (\citetalias{just10}, \mbox{Table 2}) and allow for reasonably large dispersions $\left< \theta \right>_\mathrm{i}$. The mean value of the surface density of the stellar halo is adopted from \citet{flynn06}. 

In the case of the DM surface density, its expectation value is set to \mbox{50 M$_\odot$ pc$^{-2}$}, which is \mbox{$\sim17$\%} 
lower than the best value from \citetalias{just10}, where \mbox{$\Sigma_\mathrm{dh}=59.9$ M$_\odot$ pc$^{-2}$}. 
This change is motivated by recent results in measurements of the local DM density, $\rho_\mathrm{dh}$. 
The values of $\rho_\mathrm{dh}$ estimated from the Galactic rotation curve usually lie in the range of 
\mbox{$0.005-0.015$ M$_\odot$ pc$^{-3}$} 
\citep{read14}; similar measurements obtained from the \textit{Gaia} DR2 rotation curve suggest an even narrower range, namely, 
\mbox{$\sim0.3-0.4$ GeV cm$^{-3}$} (\mbox{$\sim0.008-0.011$ M$_\odot$ pc$^{-3}$}) \citep{salas19}. 
Assuming a roughly constant DM density up to \mbox{2.3 kpc} away from the Galactic plane at the Solar radius 
(as our old value of $\Sigma_\mathrm{dh}$ corresponds to \mbox{$|z|<2.3$ kpc}), 
this translates to \mbox{$\Sigma_\mathrm{dh} \approx 23-69$ M$_\odot$ pc$^{-2}$} or, with more conservative constraint from \citet{salas19}, 
to \mbox{$\Sigma_\mathrm{dh} \approx 37-51$ M$_\odot$ pc$^{-2}$}. Our old value lays at the high end of the range, and therefore 
is not optimal for the prior mean. With the assumed mean value of \mbox{50 M$_\odot$ pc$^{-2}$}, we still favour the heavy DM halo. 
However, the fitting procedure is allowed to extend the investigation to much lower values, as we set a large dispersion of \mbox{10 M$_\odot$ pc$^{-2}$}.  

The second row of \mbox{Eq. (\ref{eq:parameters})} lists the kinematic parameters explored in this study. 
The first two of them, $\sigma_\mathrm{e}$ and $\alpha$, are the scale parameter and the slope of the AVR function, \mbox{Eq. (\ref{eq:avr})}. 
These parameters have a direct impact on the vertical gravitational potential through the thin-disk kinematics. 
They are also known to be correlated, but we chose to constrain both 
$\sigma_\mathrm{e}$ and $\alpha$, as it is hard to determine a priori the strength of the correlation 
and whether using only one of them will result in the same model flexibility.  
We also aim to constrain the vertical velocity dispersion of the thick disk given by $\sigma_\mathrm{t}$. 
Parameters $\sigma_\mathrm{p1}$ and $\sigma_\mathrm{p2}$ prescribe the vertical velocity dispersion of the thin-disk stellar populations 
related to the additional peaks in the thin-disk SFR function, \mbox{Eq. (\ref{eq:sfrd})}. 
These are new model parameters that have been introduced to reduce the observed discrepancies between the \textit{Gaia} data and 
the \mbox{JJ model} predictions, mainly in terms of the vertical number density profiles 
of the young stellar populations (\mbox{Section \ref{sect:results_std}}). 
Their prior means $\left< \sigma_\mathrm{p1} \right>$ and $\left< \sigma_\mathrm{p2} \right>$ represent our initial guess 
motivated by a simple test (\mbox{Section \ref{sect:results_std}})  
and the corresponding dispersions are chosen such that an exploration of a reasonably wide range of velocity values is permitted. 

The third row of \mbox{Eq. (\ref{eq:parameters})} contains parameters that govern the shape of the thin-disk SFR. 
By changing the combination of slopes ($\zeta$, $\eta$), both the amplitude and position of the SFR peak at old ages can be changed, 
as well as the steepness of the SFR decline. The next six parameters describe the amplitude, position, and duration of the 
two recent peaks of the thin-disk SFR that are tested in this work. 

The first of the new peaks corresponds to the recent SF burst 
found by \citet{mor19}. From this work we adopt the mean age and mean dispersion of the peak, 
\mbox{$\tau_\mathrm{p1}=3$ Gyr} and \mbox{$d \tau_\mathrm{p1} = 0.7$ Gyr}, respectively. 
The second of the additional SFR peaks is set to the most recent past of \mbox{$\tau_\mathrm{p2} = 0.5$ Gyr} 
ago with a width of \mbox{$d \tau_\mathrm{p2} = 0.25$ Gyr}. 
This choice is motivated by the observed shape of the vertical number density profile of the young A stars that indicates 
a presence of a dynamically hot and extremely young population (\mbox{Section \ref{sect:results_std}}). 
The adopted dispersion for $\tau_\mathrm{p2}$ is set to be large, also \mbox{$0.5$ Gyr}, such that mean age of the second SF burst can 
formally turn negative (i.e., in the future). This is done in order to test the possibility of the SF enhancement happening in the present day.  
As for the amplitudes of these additional SFR peaks, we set their mean values to zero, which implies that initially 
we do not assume any recent SF bursts in the disk formation history. 
However, wide priors of these parameters allow for the fitting procedure to put them into play in case their presence improves the 
model-to-data consistency. 

Finally, the last five parameters of the vector $\pmb{\theta}$ refer to the slopes of our broken power-law IMF 
and positions of the breaks between the slopes \{$\alpha_0$ and $\alpha_1$\} and \{$\alpha_1$ and $\alpha_2$\}. 
We find it necessary to investigate the IMF shape 
self-consistently with the SFR shape normalisation as both SFR and IMF influence the present-day age distributions and star counts. 
We do not explore the fourth IMF slope $\alpha_2$ corresponding to the high-mass regime \mbox{$m>6$ M$_\odot$} as stars of these masses  
are essentially absent in our data samples as they are too bright for the applied apparent magnitude cuts. 
The mean values of these parameters are chosen close to the IMF parameters from \citetalias{rybizki15} (also \citealp{rybizki18}), 
and the adopted dispersions ensure reasonably wide priors. 

As already mentioned in \mbox{Section \ref{sect:method_mcmc_likelihood}}, 
we also define a set of standard values of model parameters $\pmb{\theta_0}$.
These standard values are either directly taken from the previous works (\citetalias{just10}-\citetalias{rybizki15}) 
or, in cases where the description of a model component was modified, 
they are chosen such that our previous results are closely mimicked. 
The set of standard parameters is also given in \mbox{Table \ref{tab:best_param}}. 

\setlength{\extrarowheight}{0.55em}
\begin{table*}[t]
\centering
\tiny
\begin{threeparttable}
\caption{List of JJ model parameters optimised in this study. All values are derived for the Solar circle $R_\odot$. 
For shortness, two recent episodes of SF enhancement \mbox{$\sim3$ Gyr} and \mbox{$\sim0.5$ Gyr} ago 
correspond to the 1$^\text{st}$ and 2$^\text{nd}$ $SFR_\mathrm{d}$ bursts, respectively.
Errors reported in the last two columns are the 16 and 84 percentiles of the parameter PDFs.}
\begin{tabular}{cc|l|cccc} 
\hhline{=======}
Parameter & Units & Description & Prior 
& {\begin{minipage}[t][0.7cm][t]{1.5cm} \center{Standard \\ value} \end{minipage}} 
& {\begin{minipage}[t][0.7cm][t]{1.5cm} \center{Best value \\ (MCMC1)} \end{minipage}}
& {\begin{minipage}[t][0.7cm][t]{1.5cm} \center{Best value \\ (MCMC2)} \end{minipage}} \\ \hline
$\Sigma_\mathrm{d}$ & M$_\odot$ pc$^{-2}$ & Thin-disk surface density & $30\pm3$ & $29.4^{(1)}$ & $29.4^{+2.7}_{-2.7}$ & $29.0^{+2.6}_{-2.6}$  \\
$\Sigma_\mathrm{t}$ & M$_\odot$ pc$^{-2}$ & Thick-disk surface density & $5\pm1.5$ & $5.3^{(1)}$ & $4.9^{+1.4}_{-1.3}$ & $5.0^{+1.3}_{-1.3}$ \\
$\Sigma_\mathrm{sh}$& M$_\odot$ pc$^{-2}$ & Stellar halo surface density for $|z|<2$ kpc & $0.6\pm0.2$ & $0.6^{(5)}$ & $0.49^{+0.14}_{-1.14}$ & $0.36^{+0.11}_{-0.11}$ \\ 
$\Sigma_\mathrm{dh}$ & M$_\odot$ pc$^{-2}$ & DM surface density for $|z|<2$ kpc & $50\pm10$ & $59.9^{(1)}$ & $51.6^{+9.3}_{-9.3}$ & $50.9^{+9.1}_{-9.2}$ \\[5pt] \hdashline
$\sigma_\mathrm{e}$ & km s$^{-1}$ & AVR scaling factor & 26 $\pm$ 3 & $25.75^{(4)}$ & $25.1^{+1.8}_{-1.7}$ & $25.4^{+1.8}_{-1.7}$ \\
$\alpha$ & & AVR power index & 0.4 $\pm$ 0.05 & $0.375^{(1)}$ & $0.409^{+0.046}_{-0.045}$ & $0.407^{+0.044}_{-0.044}$ \\
$\sigma_\mathrm{t}$ & km s$^{-1}$ & Thick-disk $W$-velocity dispersion & $45\pm5$ & $45.3^{(2)}$ & $43.3^{+3.7}_{-3.8}$ & $44.4^{+3.5}_{-3.5}$ \\
$\sigma_\mathrm{p1}$ & km s$^{-1}$ & $W$-velocity dispersion of the 1$^\text{st}$ $\mathrm{SFR_d}$ burst & $25\pm5$ & - & $26.3^{+4.4}_{-4.0}$ & $26.0^{+4.5}_{-4.1}$ \\
$\sigma_\mathrm{p2}$ & km s$^{-1}$ & $W$-velocity dispersion of 2$^\text{nd}$ $\mathrm{SFR_d}$ burst & $15\pm3$ & - & $12.6^{+3.0}_{-2.9}$ & $12.4^{+3.0}_{-2.8}$ \\[5pt] \hdashline
$\zeta$ & & $\mathrm{SFR_d}$ parameter & $0.8\pm0.1$ & $0.8^{(4)}$ & $0.83^{+0.09}_{-0.09}$ & $0.86^{+0.09}_{-0.09}$ \\
$\eta$ & & $\mathrm{SFR_d}$ parameter & $5.6\pm0.1$ & $5.6^{(4)}$ & $5.6^{+0.1}_{-0.1}$ & $5.6^{+0.1}_{-0.1}$ \\
$\Sigma_\mathrm{p1}$ & M$_\odot$ pc$^{-2}$ & Amplitude-related parameter of the 1$^\text{st}$ $SFR_\mathrm{d}$ burst & $|0\pm 4|$ & - & $3.5^{+2.4}_{-1.8}$ & $3.6^{+2.4}_{-1.9}$ \\
$\tau_\mathrm{p1}$ & Gyr & Mean age of the 1$^\text{st}$ $SFR_\mathrm{d}$ burst & $3\pm1$ & - & $3.0^{+0.8}_{-0.9}$ & $2.9^{+0.9}_{-0.9}$ \\
$d\tau_\mathrm{p1}$ & Gyr & Dispersion of the 1$^\text{st}$ $SFR_\mathrm{d}$ burst & $0.7\pm0.15$ & - & $0.7^{+0.15}_{-0.14}$ & $0.71^{+0.15}_{-0.15}$ \\
$\Sigma_\mathrm{p2}$ & M$_\odot$ pc$^{-2}$ & Amplitude-related parameter of the 2$^\text{nd}$ $\mathrm{SFR_d}$ burst & $|0\pm 2|$ & - & $1.3^{+1.3}_{-0.8}$ & $1.4^{+1.3}_{-0.8}$ \\
$\tau_\mathrm{p2}$ & Gyr & Mean age of the 2$^\text{nd}$ $SFR_\mathrm{d}$ burst & $0.5\pm0.5$ & - & $0.5^{+0.5}_{-0.5}$ & $0.5^{+0.4}_{-0.4}$ \\
$d\tau_\mathrm{p2}$ & Gyr & Dispersion of the 2$^\text{nd}$ $SFR_\mathrm{d}$ burst & $0.25\pm0.05$ & - & $0.25^{+0.05}_{-0.05}$ & $0.25^{+0.05}_{-0.05}$ \\[5pt] \hdashline
$\alpha_{0}$ & & IMF slope for $0.08 \ \text{M}_\odot < m < m_0$ & $1.25\pm0.3$ & $1.26^{(3)}$ & $1.31^{+0.28}_{-0.28}$ & $1.29^{+0.24}_{-0.26}$ \\
$\alpha_{1}$ & & IMF slope for $m_0 < m < m_1$ & $1.5\pm0.3$ & $1.49^{(3)}$ & $1.5^{+0.23}_{-0.24}$ & $1.41^{+0.22}_{-0.22}$ \\
$\alpha_{2}$ & & IMF slope for $m_1 < m < 6 \ \text{M}_\odot$ & $3.0\pm0.3$ & $3.02^{(3)}$ & $2.88^{+0.26}_{-0.23}$ & $2.85^{+0.24}_{-0.23}$ \\
$m_{0}$ & M$_\odot$ & Break point between IMF regimes $m^{-\alpha_0}$ and $m^{-\alpha_1}$ & $0.5\pm0.15$  & $0.5^{(3)}$ & $0.49^{+0.15}_{-0.15}$ & $0.49^{+0.15}_{-0.15}$ \\
$m_{1}$ & M$_\odot$ & Break point between IMF regimes $m^{-\alpha_1}$ and $m^{-\alpha_2}$ & $1.4\pm0.15$  & $1.39^{(3)}$ & $1.43^{+0.14}_{-0.14}$ & $1.43^{+0.14}_{-0.14}$ \\[5pt]
\hhline{=======}
\end{tabular} 
\begin{tablenotes}
 \item References. (1) \citetalias{just10}; (2) \citetalias{just11}; (3) \citetalias{rybizki15};
 (4) Updated model feature, the adopted parameter value ensures consistency with (1)-(3); (5) \citet{flynn06}. 
\end{tablenotes}
\label{tab:best_param}
\end{threeparttable}
\end{table*}

\section{Results}\label{sect:results}

\subsection{Setup of MCMC runs}\label{sect:results_mcmc_setup}

\begin{figure*}
  \centering
  \centerline{\resizebox{\hsize}{!}{\includegraphics{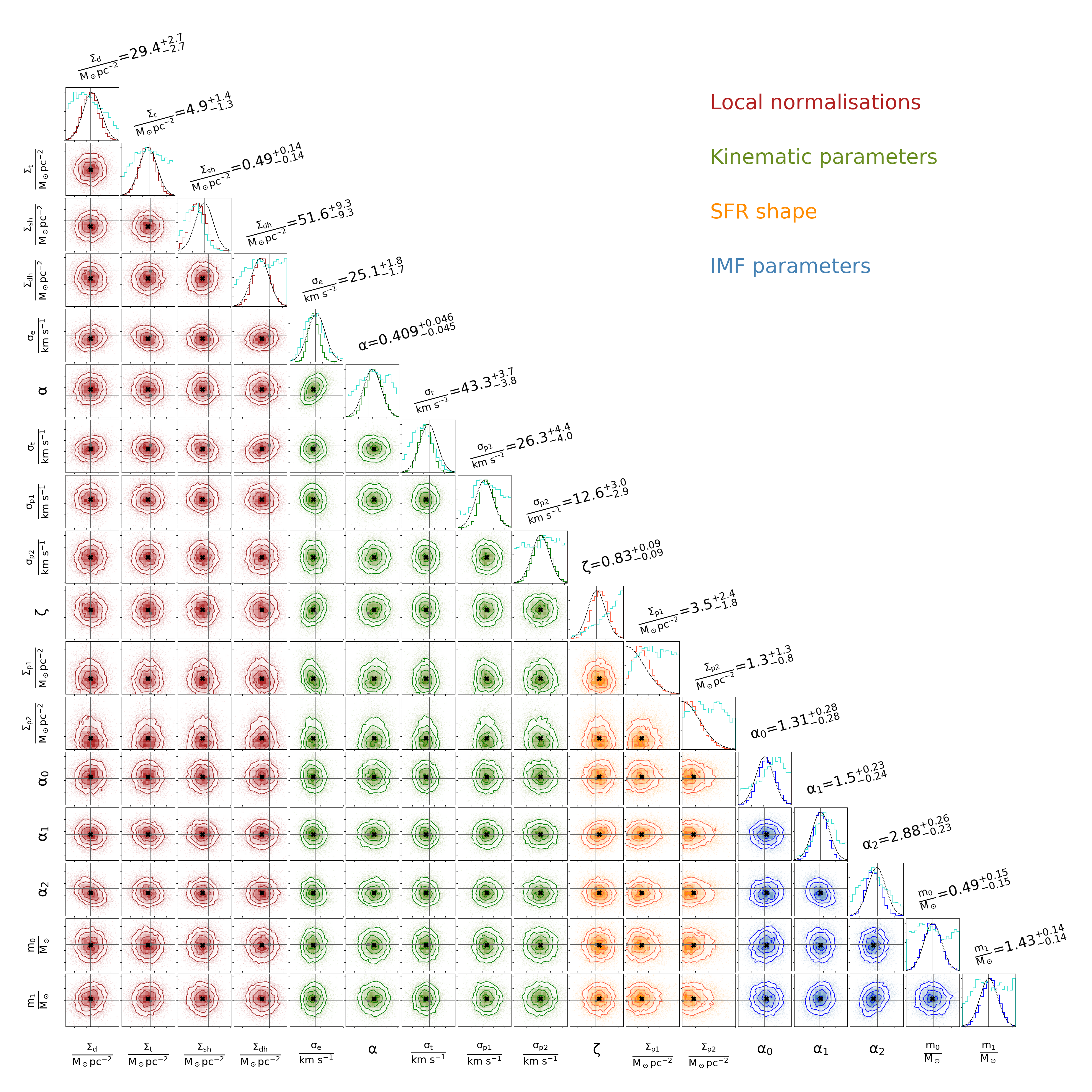}}}
\caption{Posterior PDF marginalised over the least important parameters from the set $\pmb{\theta}$.
The probability distribution is sampled during the $\mathrm{MCMC1}$ simulation.
Thin grey lines mark the standard values of parameters $\pmb{\theta_0}$ and 
black crosses show their new best values. Histograms on the plot diagonal show 
each parameter PDF (coloured steps), Gaussian priors (dashed black curves), and the likelihood PDFs (light-blue steps).}
\label{fig:sub_corner}
\end{figure*}

We use the Python realisation of the affine-invariant ensemble sampler for MCMC: the \texttt{emcee} module \citep{foreman13}. 
This code is efficiently parallelised that allows us to perform calculations on multiple CPU cores. 
The number of cores that can be used by \texttt{emcee} is related to the number of walkers as 
$n_\mathrm{core} = n_\mathrm{walker}/2$, 
and the number of walkers must be at least twice the number of parameters, $n_\mathrm{walker} \geq 2 n_\mathrm{param}$. 
In our case, $n_\mathrm{param}=22$, so we set $n_\mathrm{walker}=44$. There are several ways to initialise walkers, one of them is 
to set parameter values very close to their means, that is, to distribute them in a small cloud around $\left< \pmb{\theta} \right>$. 
Here, we use a different approach: each walker is initialised with a set of parameter values drawn randomly from the allowed intervals 
\mbox{$[\left< \pmb{\theta} \right> - 3 \Delta \pmb{\theta}, \left< \pmb{\theta} \right> + 3 \Delta \pmb{\theta}]$}. 
As recommended in \citet{foreman13}, we check the procedure performance and convergence
using the integrated autocorrelation time $\hat{\tau}$. 
This value gives an idea of the number of steps required to achieve a representative posterior sampling. 
We set the maximum number of steps in the MCMC chains to \mbox{$n_\mathrm{max}= 5 \cdot 10^4$}, 
and the convergence is usually achieved after \mbox{$(4.4-4.5) \cdot 10^4$ steps}.
We also check the mean value of the MCMC acceptance fraction, $a_\mathrm{f}$, which is a fraction of the proposed steps accepted by the MCMC routine. 
For well-behaved chains, it should remain in the range of $a_\mathrm{f}=0.2...0.5$, 
which is fulfilled during the MCMC runs performed for this study. 
In order to select independent samples from our chains, we remove the first $2 \hat{\tau}$ steps corresponding to a burn-in phase,
and select every \mbox{$\hat{\tau}/2$-th step} from each chain. With the mean autocorrelation time being approximately $\hat{\tau} \approx 450$
for our simulations, we obtain $\sim10^3$ independent samples drawn from the posterior probability distribution function (PDF) that we are interested in.

\subsection{Three models}\label{sect:results_3mod}

In this section, we describe and discuss three realisations of the \mbox{JJ model} which are presented in \mbox{Table \ref{tab:best_param}}. The first model relies on the standard parameter values $\pmb{\theta_0}$ and is referred to as the standard model. In order to investigate the model parameter space, we perform two MCMC simulations, 
as explained in \mbox{Section \ref{sect:results_mcmc_setup}}. 

Our first run explores the posterior PDF using stellar assemblies synthesised with the PARSEC stellar library. 
The sampled posterior PDF projected onto different parameter axes 
and marginalised over the least interesting model parameters is shown in \mbox{Figure \ref{fig:sub_corner}}
(the full posterior PDF is presented in \mbox{Figure \ref{fig:full_corner}}). The four colours highlight different parameter groups 
discussed in \mbox{Section \ref{sect:method_mcmc_prior}}. The colored histograms on the diagonal of \mbox{Figure \ref{fig:sub_corner}}
show the normalised PDFs for each parameter, Gaussian distributions plotted in black dashed lines correspond to the adopted prior
and the light-blue histograms present the normalised likelihood PDFs. Thin grey lines mark $\pmb{\theta_0}$, and 
the most probable parameter values are shown with large black crosses. We note that the cross positions 
can be not perfectly aligned with the visible probability maxima at the 2D projections as the real multidimensional PDF can have a complex shape. 
The best parameters, along with their 16 and 84 percentiles determined within this MCMC simulation, 
constitute a new optimised model which we henceforth refer to as $\mathrm{MCMC1}$ model (\mbox{Table \ref{tab:best_param}}). 

Additionally, in order to test how the choice of stellar evolution library impacts our results, 
we perform a complementary run of the MCMC simulation. In this case, we sample posterior distribution using the mock stellar assemblies 
calculated with the MIST isochrones. The best parameter values of this test are also listed in \mbox{Table \ref{tab:best_param}} 
($\mathrm{MCMC2}$ model, hereafter). 

We recall here that both of the described MCMC posterior samplings require an iteration (\mbox{Section \ref{sect:method_optim}}). 
At first, they are performed using stellar assemblies synthesised 
with some initial APOGEE-based AMR$_0$. Afterwards, the AMR function is updated using the new optimised set of model parameters 
(\mbox{Section \ref{sect:method_amr}}) and the stellar assemblies are re-calculated. In the end, we run a closure MCMC simulation 
based on the new stellar assemblies set. The posterior PDF shown in \mbox{Figures \ref{fig:sub_corner} and \ref{fig:full_corner}}
and the derived $\mathrm{MCMC1}$ and $\mathrm{MCMC2}$ model parameters given in \mbox{Table \ref{tab:best_param}}
are the result of such iteration, that is, they are self-consistent with the APOGEE-based AMR.
The AMR functions of the $\mathrm{MCMC1}$ and $\mathrm{MCMC2}$ models are found to be essentially identical, 
so both of the models have an AMR shown in the right panel of \mbox{Figure \ref{fig:get_amr}}.

\subsubsection{New parameters and correlations}\label{sect:results_param}

We analyse the derived multidimensional PDF shown in \mbox{Figures \ref{fig:sub_corner} and \ref{fig:full_corner}} and 
the parameter values obtained from it of the $\mathrm{MCMC1}$ model. 

As we can immediately see from \mbox{Table \ref{tab:best_param}}, almost half the investigated parameters have optimised values   
that are very close to the corresponding prior means. For a given parameter, this implies two possibilities: 
(1) our initial guess of prior was already very good, or (2) model predictions are not sensitive to this parameter, 
so it cannot be constrained within the current framework. 
We note, however, that this reasoning does not apply for correlated parameters; in this case, 
even the flat likelihood distribution does not necessarily mean that the quantity is entirely unconstrained (see below).   
Putting aside for a moment the question about correlations, we can distinguish between the two mentioned cases 
if we plot the normalised likelihood PDFs for each parameter (light-blue histograms in \mbox{Figures \ref{fig:sub_corner} and \ref{fig:full_corner}}). 
When the likelihood is sampled, our prior knowledge of the model parameters is ignored and the two above-mentioned possibilities
can be distinguished. If the mean prior value is estimated well, then also the likelihood PDF 
demonstrates a peak around the prior maximum. On the other hand, if the model predictions are only weakly sensitive to some parameter or not sensitive at all, 
then the likelihood PDF projected onto its axis is expected to be approximately flat. 

We find that we are not able to put reliable constraints on several parameters, including the width of both recent SF bursts controlled by $d\tau_\mathrm{p1}$ and $d\tau_\mathrm{p2}$, 
for which the MCMC routine returned the adopted prior means. 
The vertical velocity dispersion associated with the younger of the bursts, $\sigma_\mathrm{p2}$, 
is constrained only weakly and from the likelihood PDF, we only see that the values \mbox{$\sigma_\mathrm{p2} > 12$ km s$^{-1}$} 
are preferred (\mbox{Figure \ref{fig:full_corner}}). 
In addition, the SFR parameter $\eta$ does not appear to be important and varying the SFR parameter $\xi$ alone is enough to improve the model performance.

We also see that in most cases, pairs of the selected parameters are uncorrelated, implying 
that they are essentially independent and, therefore, well chosen. 
However, if \mbox{Figures \ref{fig:sub_corner} and \ref{fig:full_corner}} are thoughtfully inspected,  
the expected correlation between SFR and IMF can be noticed, 
as well as the (less strong) correlation between the SFR and thin- and thick-disk kinematics. 
This justifies our initial reasoning (\mbox{Section \ref{sect:intro}}) that 
in order to constrain the SFR in a robust way, its shape has to be adapted self-consistently not only with the IMF, 
but also with kinematic parameters. 
Also, a positive correlation is observed between the thin-disk kinematic parameters, $\sigma_\mathrm{e}$ and $\alpha$, which we mention in \mbox{Section \ref{sect:method_mcmc_prior}}. 
The surface density of the DM halo $\Sigma_\mathrm{dh}$ is correlated negatively with 
$\sigma_\mathrm{e}$ and the thick-disk velocity dispersion, $\sigma_\mathrm{t},$ 
and positively with the IMF parameters, $\alpha_1$ and $\alpha_2$, forming an ambiguity that we cannot resolve here. 
Thus, although $\Sigma_\mathrm{dh}$ seems to be poorly constrained at first glance (flat likelihood PDF; 
most probable value very close to the assumed prior mean), it is not entirely unknown, 
as it is implicitly constrained via correlations with other parameters. 
Finally, the IMF parameters are found to be not entirely independent from each other, 
there are moderate correlations between pairs $(\alpha_0, m_0)$, $(\alpha_1, m_1)$, and $(\alpha_0, \alpha_1)$.

\subsubsection{Age distributions}\label{sect:results_ad}

\begin{figure*}[t]
\centerline{\resizebox{\hsize}{!}{\includegraphics[scale=0.4]{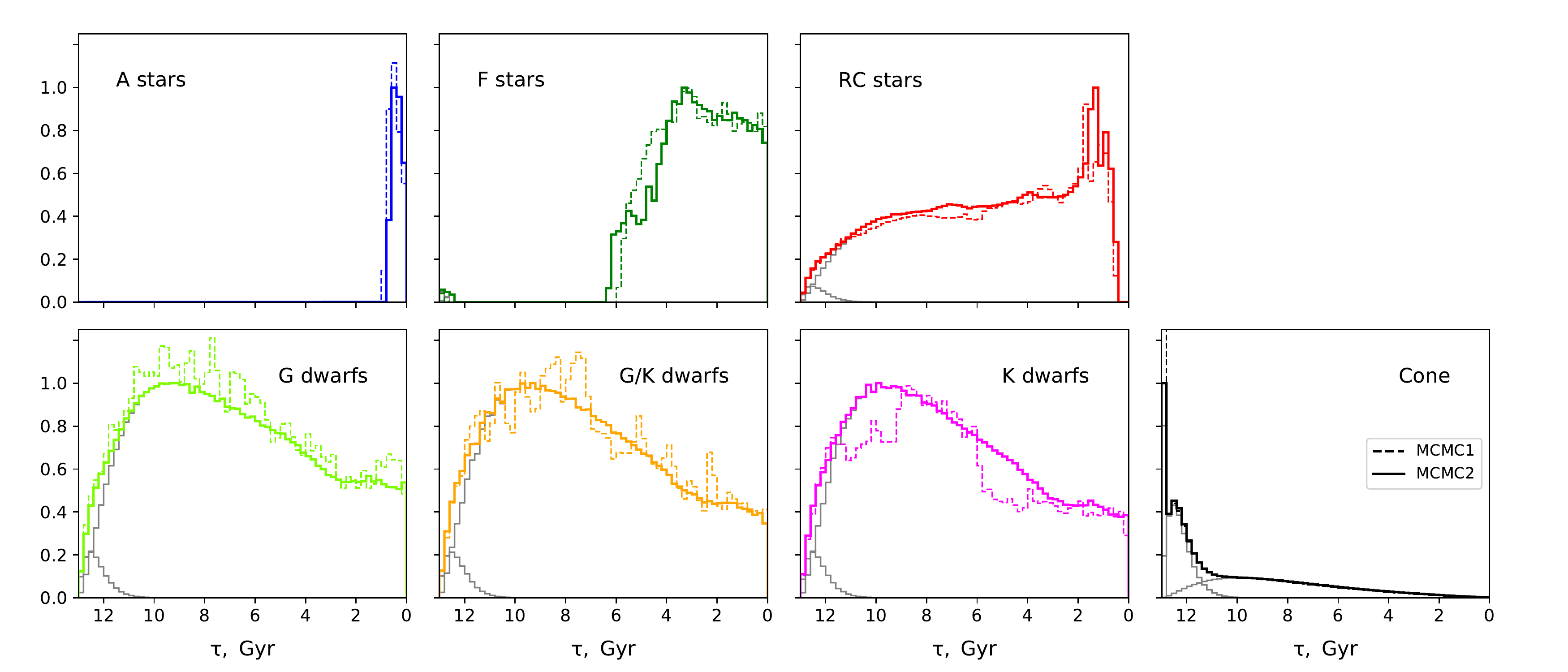}}}
\caption{Age distributions of the \textit{Gaia} samples used for the optimisation of the model parameters $\pmb{\theta}$. 
Dashed and solid coloured lines correspond to the predictions obtained with the set of $\mathrm{MCMC1}$ and $\mathrm{MCMC2}$ model 
parameters, respectively. Grey histograms show the relative contributions of the halo, thick-, and thin-disk components 
(\mbox{$\tau=13$ Gyr}, $13 \ \text{Gyr} > \tau > 9 \ \text{Gyr}$, and all ages, respectively). 
The samples are typically dominated by the thin-disk stars, e.g. 
the overall age distributions of A and F stars essentially coincide with the thin-disk contributions, 
as the thick disk and halo entirely or almost entirely miss in this case. The only sample 
with a considerable fraction of the thick-disk and halo ({$\sim 50$\%}) consists of the stars selected 
towards the Galactic poles (lower-right panel, also \mbox{see Figure \ref{fig:hess}}).  
}
\label{fig:loc_age}
\end{figure*}

Before presenting predictions of the optimised model $\mathrm{MCMC1}$ and discussing them in detail, together with the 
reference standard model and test $\mathrm{MCMC2}$ model,
we show mock age distributions for the six \textit{Gaia} samples selected on the CMD, as well as for the conic sample 
(\mbox{Table \ref{tab:data}}). 
\mbox{Figure \ref{fig:loc_age}} shows the normalised age distributions of the samples 
that correspond to the predictions of the $\mathrm{MCMC1}$ and $\mathrm{MCMC2}$ models (dashed and solid coloured lines, respectively). 
Grey histograms are the relative contributions 
from the halo (\mbox{$\tau=13$ Gyr}), thick disk (\mbox{$\tau>9$ Gyr}), and thin disk (all ages) to the overall age distributions of each sample. 
\mbox{Figure \ref{fig:loc_age}} illustrates the differences in age coverage between the selected samples  
which we referred to as part of the sample selection in \mbox{Section \ref{sect:data_gaia_cmd}}. 
We also see that the MIST isochrones are generated with a better interpolation scheme than the PARSEC isochrones, such that  
the age distributions corresponding to the $\mathrm{MCMC2}$ model look much smoother.

\subsection{Model predictions}\label{sect:results_zwh}

\subsubsection{The standard model and its weaknesses}\label{sect:results_std}

For the purposes of illustration, it is handy to present a detailed description of the standard model performance, as some of the observed model-to-data discrepancies  
served as a motivation for us to introduce the new model features described in \mbox{Section \ref{sect:model}}. 

\begin{figure*}
\centerline{\resizebox{\hsize}{!}{\includegraphics{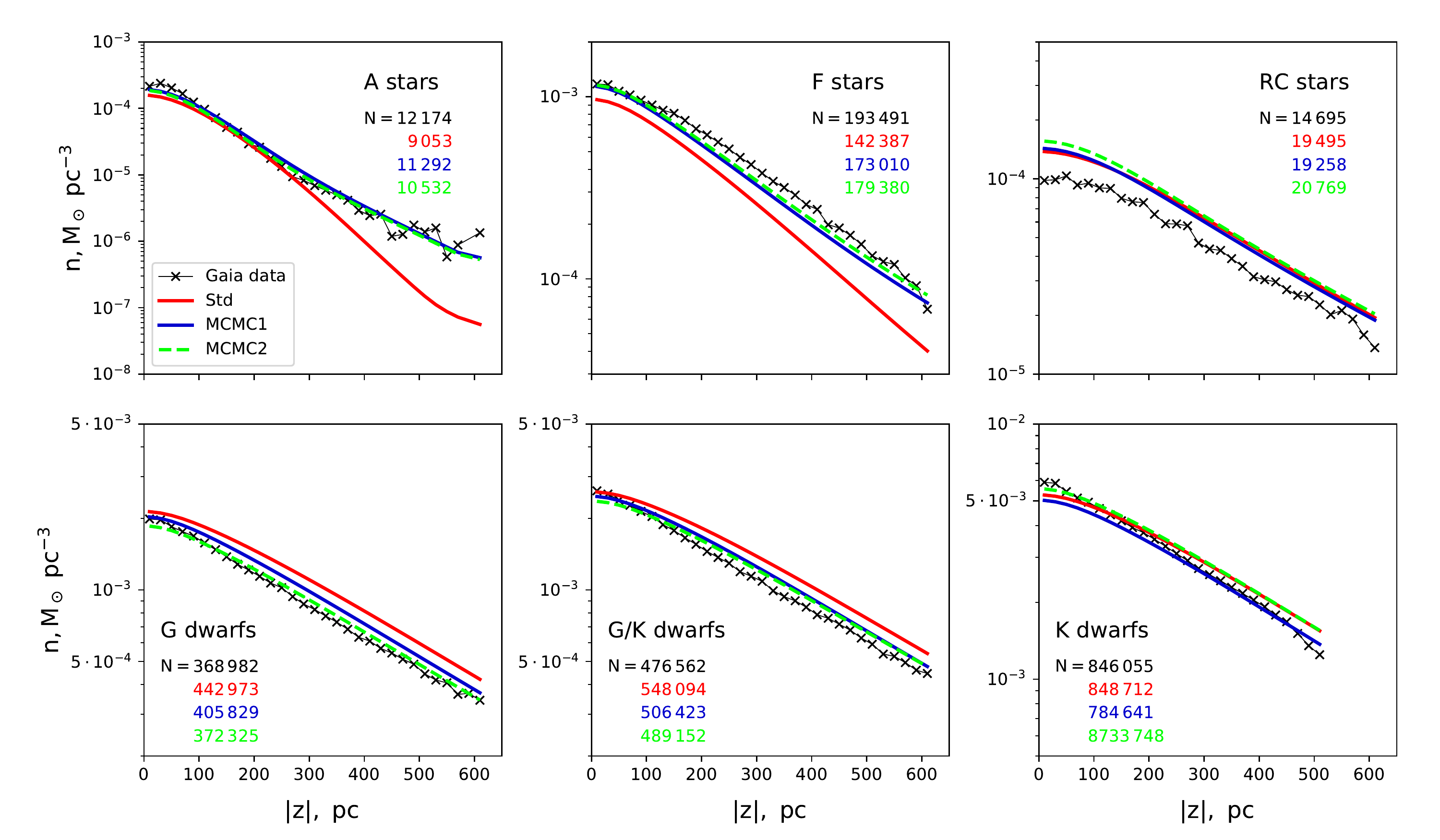}}}
\caption{Vertical number density profiles of the six local \textit{Gaia} samples. 
The data are plotted with black crosses. The solid red and blue curves show predictions 
of the standard and $\mathrm{MCMC1}$ JJ models.  
The number density profiles of the $\mathrm{MCMC2}$ model are shown with dashed green lines. 
The observed and modelled number of stars in each sample are given with the same colour coding. 
The scale and range of y-axis are chosen different for all panels in order to show clearly the shapes of all profiles. 
}
\label{fig:nz}
\end{figure*}

\begin{figure*}
\centerline{\resizebox{\hsize}{!}{\includegraphics{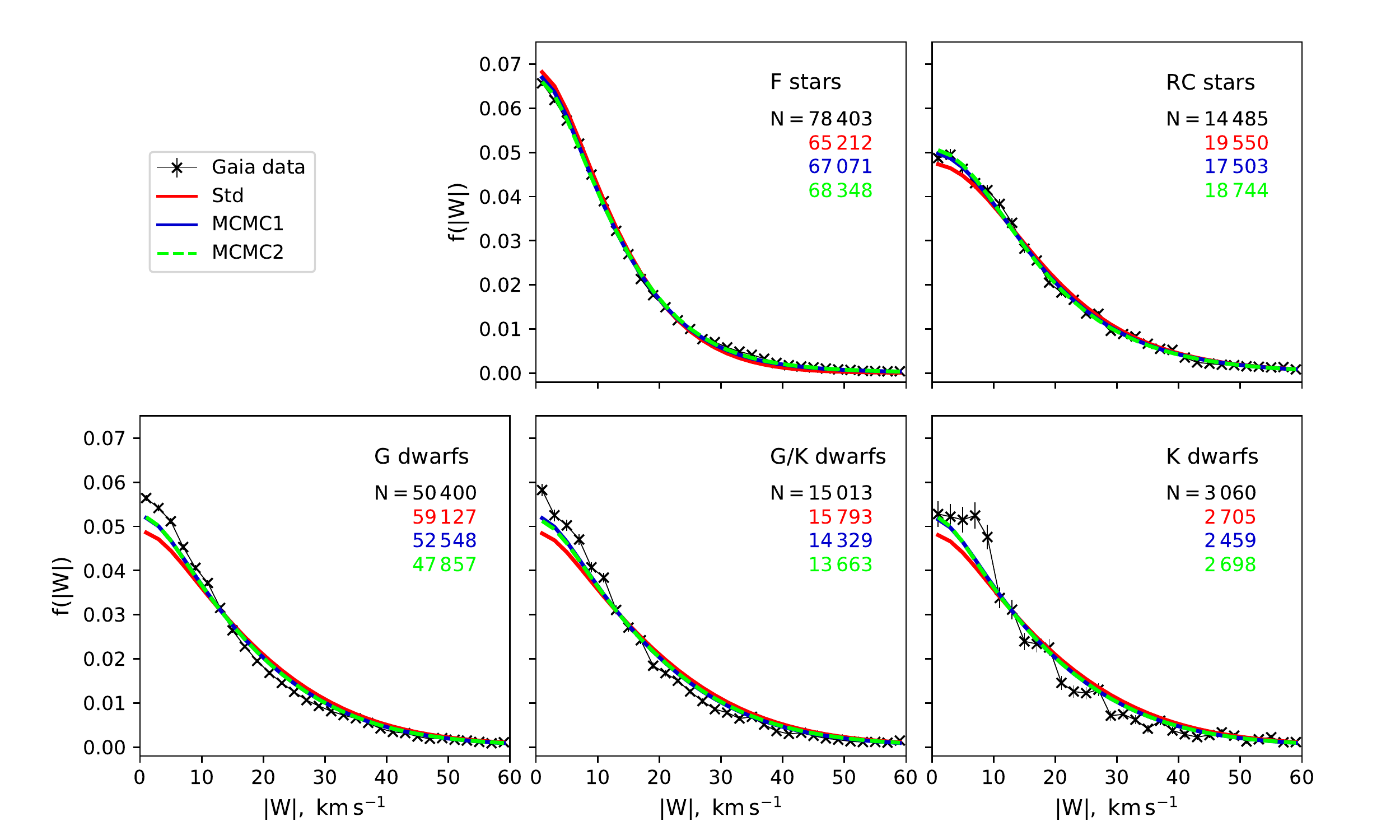}}}
\caption{Observed and predicted 
functions for the five \textit{Gaia} kinematic samples from \mbox{Table \ref{tab:data}}.
As in \mbox{Figure \ref{fig:nz}}, the data are plotted with black crosses, and the three coloured curves 
correspond to the model predictions. The observed and modelled number of stars in each sample are also given with the same colour coding.}
\label{fig:fw}
\end{figure*}

\mbox{Figure \ref{fig:nz}} shows the vertical number density laws. 
The observed profiles calculated for the six \textit{Gaia} samples (\mbox{Table \ref{tab:data}}) are shown with black crosses.  
The coloured curves correspond to the three \mbox{JJ model} realisations, the standard model predictions are plotted in red. 
There is a number of discrepancies observed in this case. First, we see significant negative or positive offsets between 
the observed and predicted density profiles for all samples except A stars and K dwarfs. 
The predicted shape of the RC number density profile is fully realistic, 
but the actual number of RC stars in the volume is overestimated by \mbox{$\sim33$\%}. 
We note that the number of stars in the data samples given in \mbox{Figures \ref{fig:nz}-\ref{fig:hess}} do not exactly match the 
sample statistics summarised in the corresponding columns of \mbox{Table \ref{tab:data}}. This happens because 
the quantities of interest are calculated from the data with the help of weights 
$1/\mathscr{S}_\mathrm{i}(l,b)$ that compensate 
for the small incompleteness effects introduced by our data selection and cleanings (Section \ref{sect:data_gaia_compl}).

For K dwarfs, the predicted number of stars in the local volume matches the observed one within \mbox{1\%}, but 
this consistency arises from a lucky combination: the lack of K dwarfs close to the Galactic plane, at \mbox{$|z| \lesssim 100$ pc}, 
is compensated by the excess of modelled K dwarfs at larger distances, \mbox{$|z| \gtrsim 200-250$ pc}.  
For the rest of the samples the observed number density profiles are not followed by the modelled ones as well.
In general, profiles of the MS stars in the model are somewhat shallower than is suggested by the data, 
the overall number of stars in the G- and G/K- dwarf samples is overestimated by \mbox{$\sim20$\%} and \mbox{$\sim15$\%}, respectively. 
The predicted vertical profile of F stars is too steep, and the number of F stars in the volume is underestimated by \mbox{$\sim26$\%}. 
The same \mbox{$\sim25$\%} lack of stars is observed for A stars. 
The inner part of the number density profile of A stars, \mbox{$|z|<300$ pc}, fits the observed profile well, 
however we see a significant discrepancy between the data and model further from the Galactic plane: the predicted number density 
falls off too steeply and is unable to reproduce the change in profile slope that happens at \mbox{$|z| \approx 300-350$ pc}. 

The vertical kinematics of the five \textit{Gaia} samples is illustrated in \mbox{Figure \ref{fig:fw}}. 
Again, the observed and predicted quantities, the normalised vertical velocity distributions $f(|W|)$, 
are plotted with black crosses and coloured lines, respectively. 
As expected, there are non-negligible deviations between the data and the standard model (red curves) 
also in terms of kinematics. 
The main discrepancy reveals itself for the MS stars, where the mismatch between the relative fractions 
of dynamically cold and warm stars is quite significant. This feature was already reported by us in \citet{sysoliatina18}, 
where we also found that this effect is most pronounced close to the Galactic plane. 
Additionally, we observe a small imbalance between the fractions of dynamically cold and warm stars for the F-star sample 
(the relative ratio of dynamically cold stars is slightly overestimated in the model). 
In the case of the RC sample, the situation is reversed: 
the predicted relative fraction of dynamically cold stars is somewhat too low in the standard model. 
Finally, we look at the apparent Hess diagram of the stars observed within the conic volume (\mbox{Section \ref{sect:data_gaia_cone}}). 
Constructed according to the recipe explained in \mbox{Section \ref{sect:method_cone}}, the Hess diagram is displayed in \mbox{Figure \ref{fig:hess}}. 
Different panels in each row show the contributions from the thin disk, thick disk, and halo, as well as the total predicted Hess diagram 
and the logarithmic $\chi^2$ calculated according to \mbox{Eq. (\ref{eq:log_likelihood_h})}. 
Different rows correspond to the data (Figure \ref{fig:cone_hess}, shown here for a visual comparison) and 
the standard, $\mathrm{MCMC1}$ and $\mathrm{MCMC2}$ JJ models.
In the case of the standard model, the most important discrepancies between the modelled and observed apparent Hess diagram are located 
in the triangular area at \mbox{$G_\mathrm{BP}-G_\mathrm{RP} \approx 0.6-0.7$ mag} and \mbox{$G \approx 16-17$ mag}. 
This discrepancy originates from the halo component and is likely to be caused by an overestimated halo surface density. 
Another feature corresponding to the poor model-to-data consistency is observed at the bright end, for \mbox{$G \approx 12-15$ mag} and 
$G_\mathrm{BP}-G_\mathrm{RP} \approx 0.8$ mag. This can be attributed to the overestimated star counts of the thin-disk component. 
The flawed consistency at \mbox{$G_\mathrm{BP}-G_\mathrm{RP} \approx 1$ mag} and \mbox{$G \approx 14-17$ mag}
can be related to both the thin- and thick-disk components. 
Additionally, the standard model performance in the \mbox{$G_\mathrm{BP}-G_\mathrm{RP}>2$ mag} 
regime is also not perfect and we relate this to the mismatch of  
the lower-MS slope between the data and isochrones. We note that the vertical stripes at $G_\mathrm{BP}-G_\mathrm{RP} \approx 1.5, 1.7, 1.9$ 
that are seen for the model predictions constructed with the PARSEC isochrones are related to the interpolation scheme of the stellar evolution code 
and not to the model features. 

The described picture, when viewed as a whole, implies that several changes need to be made in order to improve the model's performance. 
The inconsistencies seen over the Hess diagram in \mbox{Figure \ref{fig:hess}} 
indicate that the surface densities of the model components need to be adjusted.
To improve the vertical profiles of A and F stars, 
a significant fraction of young stars with ages of \mbox{$\tau \lesssim 4-5$ Gyr} has to be added to the standard \mbox{JJ model}. 
However, simply adding more young stars cannot resolve the problem, 
as it is not just the number of A and F stars but also the shapes of their vertical profiles 
that have to be adjusted. In the case of the \mbox{A-star} sample the required profile adjustment is quite significant at \mbox{$|z| \gtrsim 300$ pc}. 
The revealed difference between the observed and predicted A-star vertical profiles means 
that a small fraction of the youngest population represented by A stars is more dynamically heated than 
prescribed by the model AVR. A simple test shows that the observed vertical number density profile of A stars can be 
well fitted by a double-$sech^n$ profile. The main isothermal component of such fit 
(90\% of the sample stars) has a velocity dispersion of \mbox{$\sim6$ km s$^{-1}$}, 
similar to the value given by AVR for the youngest populations. However, the velocity dispersion of the second isothermal component has to be 
\mbox{$\sim12$ km s$^{-1}$}, namely, twice as large. This is what motivated us to add a recent peak  
to the thin-disk SFR with the assumed mean age of \mbox{$\tau_\mathrm{p2} \approx 0.5$ Gyr}\footnote{We number peaks according 
to their position on the Galactic time axis, not the age axis.}
corresponding to the typical age of A stars (\mbox{Figure \ref{fig:loc_age}}). 
Through the analogy to A stars, we added another SF burst centered at older ages, \mbox{$\tau_\mathrm{p1} \approx 3$ Gyr}, which may help 
to improve model-to-data consistency for the F-star sample.
We also decouple the vertical velocity dispersions $\sigma_\mathrm{p1}$ and $\sigma_\mathrm{p2}$ 
of the young populations associated with the two added SF bursts 
from the thin-disk kinematics prescribed by the monotonous AVR. Here, the fraction of the thin-disk populations 
associated with the SF burst at a given Galactic time is defined as a ratio of the SFR function with a peak relative to a monotonously decreasing SFR. 
In terms of the definitions given by \mbox{Eqs. (\ref{eq:sfrd0}) and (\ref{eq:sfrd})}, this is given as 
$SFR^\prime_\mathrm{d}(t)/SFR_\mathrm{d}(t)$. 
Finally, we decided to adapt two AVR parameters in order to improve the velocity distributions $f(|W|)$
and the IMF parameters as the latter is naturally entangled with the SFR. 

\subsubsection{MCMC1 model}\label{sect:results_mcmc1}

In the case of the optimised JJ model $\mathrm{MCMC1}$, most of the issues described above are either resolved or relaxed.

\begin{figure*}[t!]
\centerline{\resizebox{\hsize}{!}{\includegraphics{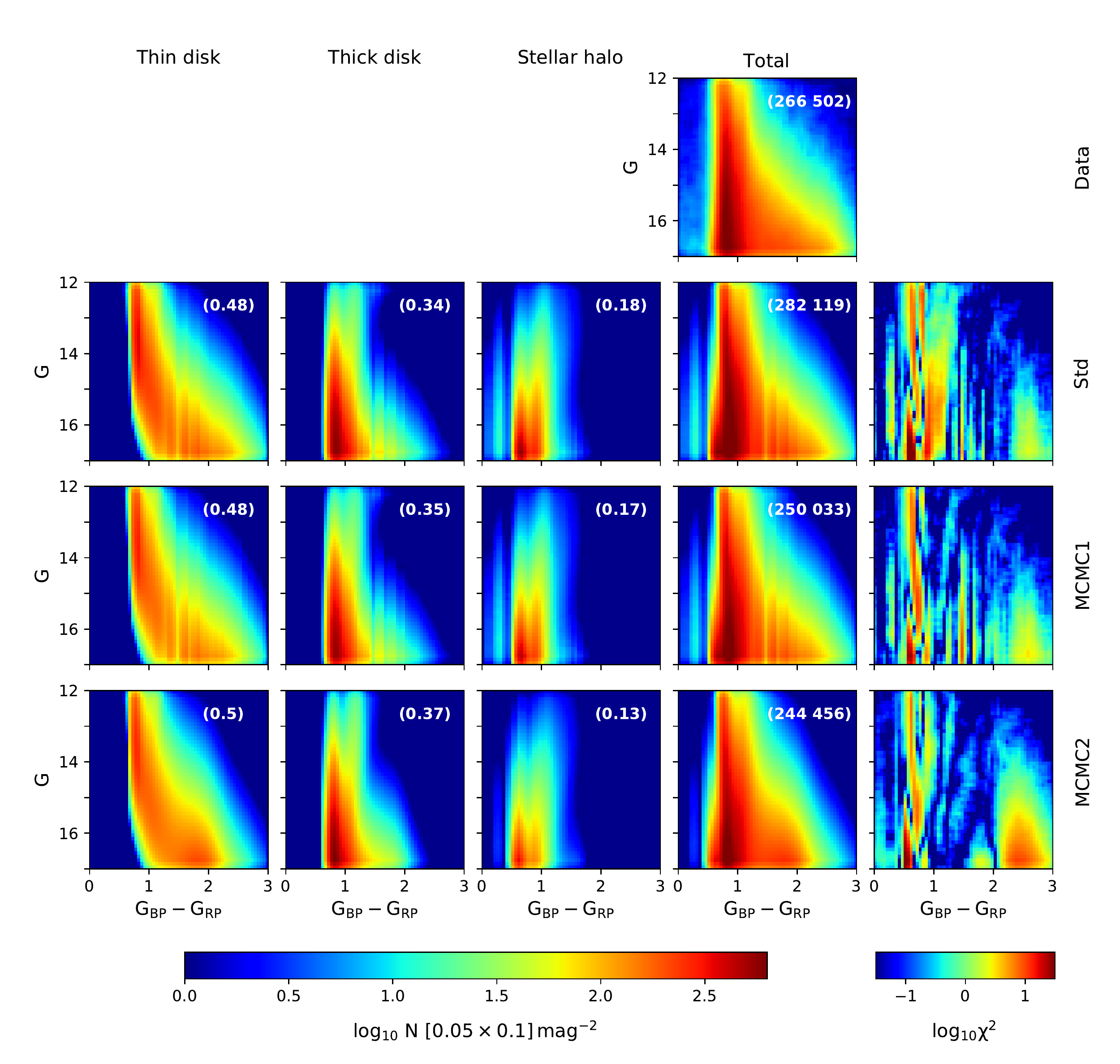}}}
\caption{Observed and predicted smoothed apparent Hess diagram of the \textit{Gaia} conic sample. 
Four rows (top to bottom) show the data and predictions of the standard, the $\mathrm{MCMC1}$, and the $\mathrm{MCMC2}$ models.
The contributions of the thin and thick disk as well as of the halo are shown in three columns on the left. 
Values given there in parenthesis correspond to 
each component's relative contribution to the total Hess diagram in terms of star counts. 
The fourth column shows the total Hess diagram, where the observed or modelled number of stars is given in parentheses.  
The last column displays the logarithmic $\chi^2$ calculated according to \mbox{Eq. (\ref{eq:log_likelihood_h})}. 
}
\label{fig:hess}
\end{figure*}

The predicted number density profiles calculated with the new parameter values (\mbox{Table \ref{tab:best_param}})
are shown in \mbox{Figure \ref{fig:nz}} with blue curves. 
The vertical profiles of A and F stars are now much better reproduced, also the underestimation in their overall star counts is reduced to 
\mbox{$\sim$7\%} and \mbox{$\sim$11\%}, respectively. 
The vertical profiles of the MS stars became somewhat steeper (G- and G/K-dwarf samples), and follow the observed trends more closely. 
The overall excess in the predicted number of stars for these two MS samples is reduced to \mbox{$\sim$10\%} and \mbox{$\sim$6\%}, respectively. 
Also, the $\mathrm{MCMC1}$ model is more consistent with the data in terms of the vertical kinematics of the local stars. 
The corresponding modelled velocity distribution functions are shown in blue in \mbox{Figure \ref{fig:fw}}. 
The velocity distributions $f(|W|)$ of F and RC stars match the observed distributions perfectly, 
and in the case of the G-, G/K-, and K-dwarf samples, the model-to-data fit is noticeably improved. 
Additionally, almost all the aforementioned problematic areas at the apparent Hess diagram appear to be improved after the model calibration against \textit{Gaia}. The overall star count model-to-data mismatch remains on the level of \mbox{$\sim6$\%}
(compare the second and third row in \mbox{Figure \ref{fig:hess}}). 

This improvement of the model performance is achieved as a result of several factors. 
Firstly, the model parameters are optimised in the high-dimensional space. Secondly, 
two additional SF bursts with the special vertical kinematics are added. Thirdly, the AMR of the model is adjusted to be consistent 
with the observed metallicity distribution of the local populations. We find that the Gaussian SF bursts centered at 
ages of \mbox{$\sim3$ Gyr} and \mbox{$\sim0.5$ Gyr} and characterised by the dispersions of \mbox{$\sim0.7$ Gyr} and \mbox{$\sim0.25$ Gyr}
can significantly improve the model performance for the upper MS. 
The derived vertical velocity dispersions of the thin-disk stellar populations related to these episodes of the SF enhancement 
are \mbox{$\sim26$ km s$^{-1}$} and \mbox{$\sim12.6$ km s$^{-1}$}, respectively. 
Using the positions, widths, and amplitude-related parameters of the two SF bursts from \mbox{Table \ref{tab:best_param}}, we can calculate 
local surface densities of these thin-disk populations and find \mbox{$\sim 1.8$ M$_\odot$ pc$^{-2}$} and \mbox{$\sim 0.27$ M$_\odot$ pc$^{-2}$} 
for the older and younger bursts, respectively. 
Thus, these extra heated populations constitute \mbox{$\sim$5.3\%} and \mbox{$\sim$0.8\%} of the total disk surface density at the Solar circle. 

\begin{figure}
\centerline{\resizebox{\hsize}{!}{\includegraphics{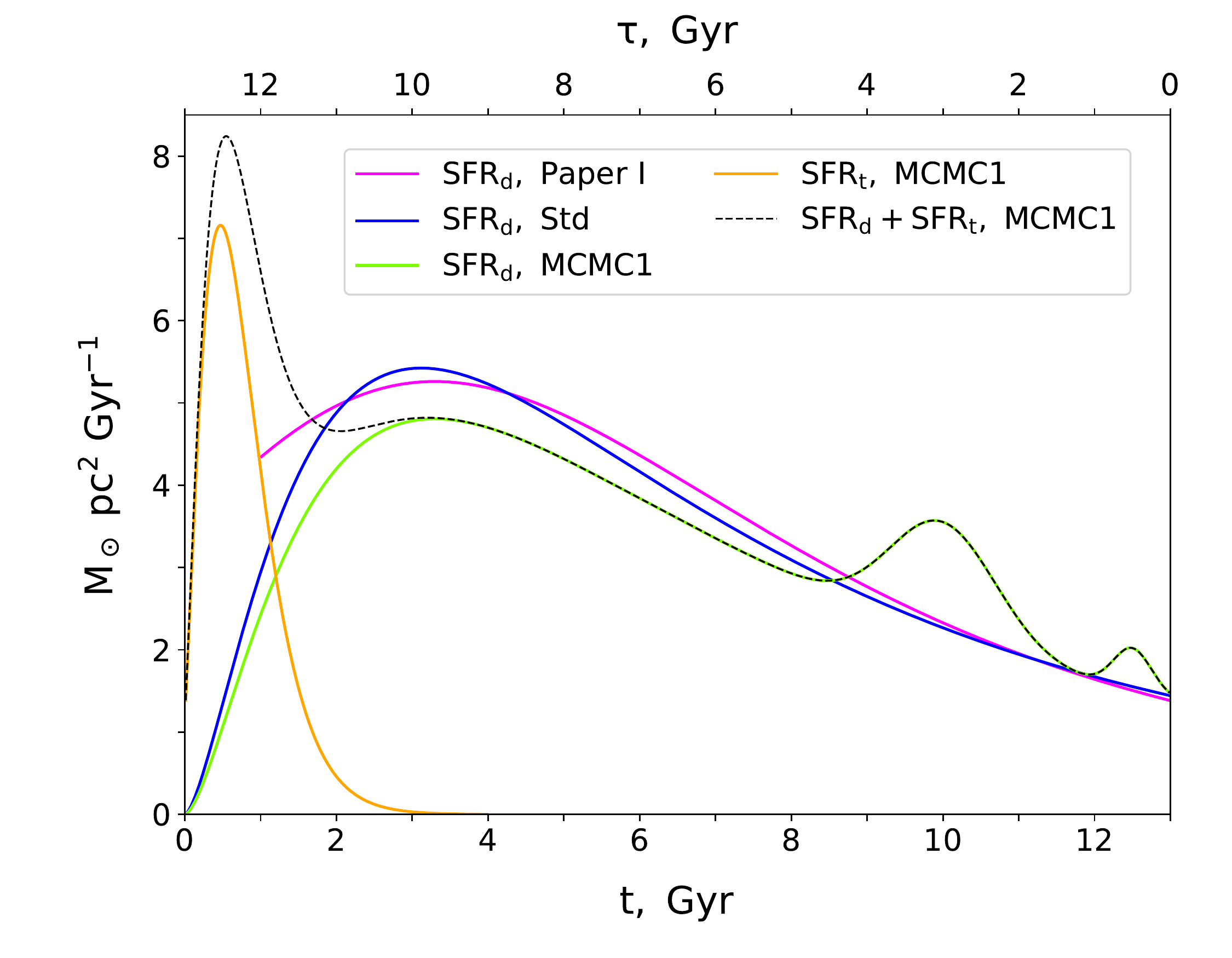}}}
\caption{Derived SFR function of the Galactic disk at the Solar circle. 
Orange and green curves show the thick- and thin-disk SFR calculated according to 
\mbox{Eqs. (\ref{eq:sfrt}) and (\ref{eq:sfrd})} with the parameters of the $\mathrm{MCMC1}$ model. 
The dashed black curve is the corresponding SFR of the total disk. 
The thin-disk SFR of the standard model is shown in blue. For comparison, we also plot the original thin-disk SFR from \citetalias{just10} (magenta curve).}
\label{fig:nsfr}
\end{figure}

The new SFR function with two recent SF bursts is displayed in \mbox{Figure \ref{fig:nsfr}}. 
The thin- and thick-disk SFR, as well as the total SFR, are plotted in 
solid orange, green, and dashed black curves, respectively. The thin-disk SFR of the standard model 
is shown in blue and the magenta curve corresponds to the thin-disk SFR function presented in \citetalias{just10}. 
We see that the new and original thin-disk SFR functions are fully consistent in terms of the overall decline 
and the only important difference between them is the presence of the recent SF bursts confirmed in this work. 

Alongside the general model improvement, we report on the leftover model-to-data deviations. 
Among them, there is a mismatch between the observed and simulated RC star counts: 
the vertical profile and the overall star counts of the RC stars in the volume 
remain essentially the same after the model optimisation. 

In the case of K dwarfs, the shape of the vertical profile is not improved and remains to be shallower than suggested by the \textit{Gaia} data. 
From \mbox{Figure \ref{fig:hess},} we also see that the reddest part of the Hess diagram, \mbox{$G_\mathrm{BP}-G_\mathrm{RP}>2$ mag}, 
populated by the faint low-mass K dwarfs, does not appear to be improved following the optimisation. Additionally, 
there is still a noticeable imbalance between the dynamically cold and hot MS populations of the disk 
(\mbox{Figure \ref{fig:fw}}). The potential sources of these leftover discrepancies, their importance, 
as well as the model limitations and caveats are discussed below in \mbox{Section \ref{sect:discus_lim}}.

\subsubsection{MCMC2 model}\label{sect:results_mcmc2}

The vertical number density profiles and $W$-velocity distribution functions predicted by the $\mathrm{MCMC2}$ model 
are plotted as green dashed curves in \mbox{Figures \ref{fig:nz} and \ref{fig:fw}}, respectively. 
It is easy to see from these plots, as well as from \mbox{Table \ref{tab:best_param}}, that 
the parameters and predictions of the models $\mathrm{MCMC1}$ and $\mathrm{MCMC2}$ are very close to being consistent. 
However, we cannot conclude that the stellar library does not have any impact on our stellar population synthesis 
as the apparent Hess diagram constructed with the MIST isochrones exhibits special features when compared to the 
predictions of the $\mathrm{MCMC1}$ model (compare the third and fourth row in \mbox{Figure \ref{fig:hess}}). 
It is obvious that the MIST isochrones give a less realistic model of the low-mass MS compared to the PARSEC code. 
It is for this reason that we mainly rely on the PARSEC stellar library and only use MIST isochrones for a complementary test.  

\subsubsection{Consistency test}\label{sect:results_loc}

\begin{figure}[t]
\centerline{\resizebox{\hsize}{!}{\includegraphics[scale=0.4]{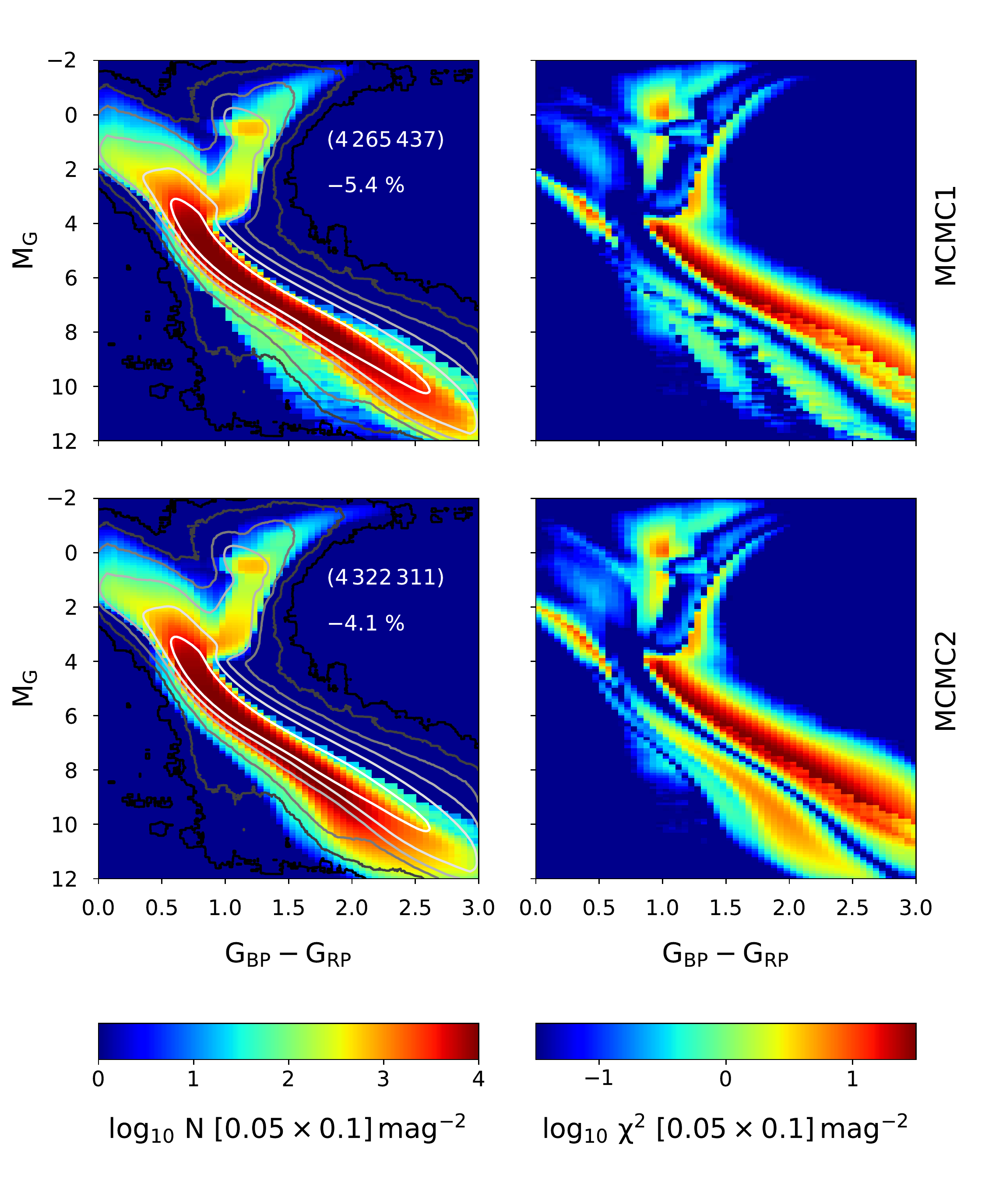}}}
\caption{Absolute Hess diagram of the full local sample (left column) as predicted 
by the $\mathrm{MCMC1}$ and $\mathrm{MCMC2}$ models (top to bottom). 
The black and white contours are added for comparison and show the distribution of stars in the data. 
The overplotted values give the total predicted number of stars in the sample 
and its relative difference to the observed number of stars. 
The right column shows the corresponding logarithmic $\chi^2$.}
\label{fig:loc_all}
\end{figure}

The final step is performing a simple consistency test. We model the full local sample (\mbox{Section \ref{sect:data_gaia_loc}}) 
and compare it to the data over the absolute Hess diagram. 
The Hess diagrams shown in \mbox{Figure \ref{fig:loc_all}} are constructed with the updated JJ models 
$\mathrm{MCMC1}$ and $\mathrm{MCMC2}$ (left column). 
The overplotted black and white contours mark the distribution of stars in the data. 
The right column of \mbox{Figure \ref{fig:loc_all}} shows the logarithmic $\chi^2$ calculated for 
both models according to \mbox{Eq. (\ref{eq:log_likelihood_h})}. 

Again, we see that the $\mathrm{MCMC1}$ model gives a fully realistic representation of the data:
a comparison of the observed and modelled MS positions (\mbox{Figure \ref{fig:loc_all}}, upper left panel) 
shows that they are in good agreement with each other. 
The total number of the observed and predicted stars agrees within {$\sim5$\%}, with a slight deficit of stars in the model relative to the data.  
The corresponding $\chi^2$ plot (\mbox{Figure \ref{fig:loc_all}}, upper right panel) reveals 
a part of the MS that is severely underpopulated in the $\mathrm{MCMC1}$ model (red stripe along the MS). 
However, this feature is expected as it originates simply from the fact that the modelled MS is narrower than the real one. 
This happens because we do not model binaries located right above the MS, and (to a lesser degree) because
no metallicity scatter is added to AMR. The remaining inconsistency between the low-mass MS slopes 
is attributed to the properties of the PARSEC stellar library. 

The predictions of the $\mathrm{MCMC2}$ model (\mbox{Figure \ref{fig:loc_all}}, second row) have the same 
features, but as we already expected based on its predictions of the K-dwarfs vertical number density profiles (\mbox{Figure \ref{fig:nz}})
and the apparent Hess diagram (\mbox{Figure \ref{fig:hess}}, last row), its overall model-to-data consistency 
in terms of the observed and modelled MS slopes is worse 
than in the case where PARSEC isochrones are used. Additionally, with regard to the lower-MS slope problem, 
we see that the modelled MS is shifted to the blue by \mbox{$\sim0.1$ mag}.

\section{Discussion}\label{sect:discus}

\subsection{Limitations and caveats }\label{sect:discus_lim}

Before discussing the most interesting result of this work, the reconstructed SFR, we address potential limitations 
of our approach and estimate their possible impact on our findings. 

\subsubsection{Modelling approach }\label{sect:discus_lim_gen}

The most important limitation of our study is related to the way we define the SF history of the disk. 
Firstly, we assumed a parallel evolution of the thick and thin disk, which might not necessarily be the case.  
There are also two-infall models discussed in the literature \citep[e. g.][]{grisoni18} and more complicated scenarios  
where the thin- and thick-disk components form sequentially, one after another, with a transfer at \mbox{$\sim8$ Gyr} ago \citep{haywood13}. 
Secondly, we did not allow for a variation of the shape of the thick-disk SFR, but we simply scaled it with the $\Sigma_\mathrm{t}$ parameter. 
However, this should not be a significant limitation as the assumed duration of thick-disk formation is relatively short
and in this case, the exact shape of the SFR function is not important. 
Finally, we define the thin-disk SFR in a parametric way and allow only two recent SF bursts inspired by the observed discrepancies 
between the \textit{Gaia} data and the JJ model with a declining SFR. In reality, the SFR shape may be more complicated, with a number 
of peaks and dips which our present model does not account for. However, we tried to adapt many of the SFR parameters 
in order to vary its shape with as much freedom as it is possible within our framework.  

In addition, we highlight once more that no explicit radial migration process is introduced in the current version of the JJ model. 
Although the new model parameters predict a better consistency between the model and data in terms of vertical kinematics, 
we still see a small imbalance between dynamically cold and hot stars in the model. This may point to a more complicated AVR shape 
than a simple power-law function increasing with stellar age as suggested by \mbox{Eq. (\ref{eq:avr})}. 
Since dynamically cold stars are more likely to migrate, it is possible that there is a surplus of them in the data, 
which would correspond to a more pronounced core in the $f(|W|)$ distributions than our model predicts.

\subsubsection{Distances}\label{sect:discus_lim_dist}

Another source of uncertainty in this study is related to the data selection process. As we explained in \mbox{Section \ref{sect:data_gaia_geom}}, 
we adopt inverse parallaxes as stellar distance estimates and use them to calculate the absolute G-magnitudes
as these are needed for the sample selection. Thus, our sample statistics directly depends on the adopted distance set. 
In \citet{sysoliatina18} we used TGAS astrometric data with relative parallax errors as large as \mbox{30\%} at \mbox{$|z| = 1$ kpc}.
As we show, in order to construct robust stellar density profiles and Hess diagrams, a parallax error model has to be included into the analysis. 
However, a proper modelling of the distance errors is challenging and significantly complexifies stellar population synthesis. 
We also want to avoid any distance quality-related cuts as
these are known to introduce distance-dependent biases \citep{luri18}, which can be also challenging to model. 
Alternatively, Bayesian distances similar to those derived in \citet{bailer-jones18} or \citet{mcmillan18} can be used instead of inverse parallaxes. 
However, they should be used with caution, as the Galactic model used as a prior for Bayesian distances can be inconsistent with our Galactic model, 
which would eventually bias our model parameters. Ideally, only the Bayesian distances 
that were derived with this very model as a prior can be safely used to improve a Galactic model. 
In practice, this re-calculation of distances for each model parameter set 
during the model optimisation is hardly feasible and, 
therefore, a compromise between the calculation efficiency and a robustness of the approach 
should be found to perform a Galactic model calibration over the large volumes. 
At a minimum, additional tests have to be performed to check 
the models' consistency and estimate the impact of different Bayesian distance sets on the model predictions.

At the same time, such Bayesian distances as those derived for the bright APOGEE RC stars and used in this work
can be very helpful as they rely on the spectrophotometric modelling of the observed sample and do not depend on
the assumptions about matter distribution across the Galactic disk. For the local volume considered in this work, any Bayesian distances will be very similar to the inverse parallaxes that we used here, 
therefore, distance uncertainties are not expected to influence our results. 
However, these considerations are of great importance for our future work, where the JJ model is to be applied to larger volumes. 

\subsubsection{Stellar evolution library}\label{sect:discus_lim_stel}

As we emphasise throughout this paper, our stellar population synthesis is naturally sensitive to the choice of the stellar evolution library. 
Two sets of the best model parameters in \mbox{Table \ref{tab:best_param},} obtained in the identical MCMC simulations 
based on the PARSEC and MIST isochrones, 
show the impact of the stellar library on our results. In general, the parameter values differ only slightly between the 
$\mathrm{MCMC1}$ and $\mathrm{MCMC2}$ models, typically by $\sim 0.3 \Delta \theta$. 
Both runs indicate that the local stellar populations as observed by \textit{Gaia} are inconsistent with a monotonously declining thin-disk SFR 
and two recent episodes of the enhanced SF are required to reproduce these data in our model framework. 
However, there are several problems related to the stellar libraries that influence the quality of our modelling. 

For both PARSEC and MIST isochrones, we find a significant difference between the predicted and observed 
MS slope in the low-mass regime that corresponds to our K-dwarf sample and the reddest part of the apparent Hess diagram for stars in the cone, 
\mbox{$G_\mathrm{BP}-G_\mathrm{RP} \gtrsim 1.5-2$ mag}. In the case of MIST isochrones, this mismatch is quite severe 
(\mbox{Figures \ref{fig:hess} and \ref{fig:loc_all}}, bottom rows) and this is the reason why we 
chose this stellar library for a complementary test only and we treat the best model parameters obtained on its basis as less trustworthy. 
On the other hand, the MIST code has a better interpolation scheme than PARSEC, which is clearly demonstrated in \mbox{Figure \ref{fig:loc_age}}. 
However, the lack of smoothness over the CMD in the case of PARSEC isochrones does not lead to any additional problems.
Another issue that is possibly related to theoretic stellar evolution is the discrepancy between the modelled and observed RC star counts. 
\mbox{Figure \ref{fig:nz}} shows that our model is able to reproduce the realistic vertical fall-off of the RC density, however, the 
mismatch in the number of stars remains significant, \mbox{$\sim33$\%}, and does not decrease after the model optimisation.  
We suggest that this large difference can be related to the existing uncertainty in the lifetimes of core He-burning stars \citep{jones15}, 
and, therefore, it should not be overinterpreted. However, other possibilities have not been discarded, such as the observed difference in the density of giants, which may indicate that the shapes of the local SF history or 
AVR function are more complicated than those proposed in this work. 

\subsubsection{Dust model}\label{sect:discus_lim_dust}

As our \textit{Gaia} sample selection relies on the de-reddened colours and magnitudes, 
it is intrinsically sensitive to the assumed 3D model of dust distribution. 
In order to investigate how the choice of the dust model influences our samples statistics, 
we tested two different dust maps: a high-resolution dust model from \citet{green19} complemented at $\sim 1/4$ of the sky 
by the lower-resolution dust map from \citet{lallement18}, and the \citet{lallement18} dust map applied for the whole sky. 
We found that the number of stars in the samples changes by a few \% only, so this does not lead to any significant changes 
in the shapes of their vertical number density profiles and velocity distributions and, therefore, it can hardly affect our derived parameter values. 
An example of the impact of the extinction map on the F-star sample statistics is shown in \mbox{Appendix \ref{sect:append_dust}}. 

\subsubsection{AMR}\label{sect:discus_lim_amr}

In Section \ref{sect:method_amr}, we made several simplifying assumptions to derive the AMR from the modelled RC ages and the 
observed metallicity distributions of the APOGEE stars. 

In particular, we assumed that there is a direct and unique correspondence between stellar metallicities and ages, 
which in reality is not the case. 
Numerous studies based on photometric or spectroscopic metallicities and isochrone or chromospheric age estimates 
report a significant, \mbox{$\gtrsim 0.1$ dex}, metallicity scatter at all ages \citep{rocha-pinto00a,haywood06,casagrande11} 
that looses the correlation between age and metallicity. For example, in \citet{bergemann14} the authors used \mbox{\textit{Gaia}-ESO} abundances and 
did not find any significant age-metallicity correlation for stars younger than 8 Gyr in the extended Solar neighbourhood. 
The large scatter in metallicity cannot be entirely assigned to the effect of random errors of spectroscopic observations, but partly has physical origin: 
the effect of radial migration leads to mixing of stars from different radial zones in the Galactic disk and this partly smears out 
the age-metallicity correlation at each radius. In this study, we ignored the physical scatter in the AMR for the sake of saving computational time. 
Indeed, adding a scatter on top of the AMR implies that the number of stellar assemblies increases. 
This can noticeably stretch the calculation time required to achieve a reliable constraint on the model parameters 
(\mbox{Section \ref{sect:method_mcmc}}). On the other hand, adding a scatter in metallicity is expected to have a negligible 
impact on our results in the end: features on the synthetic CMD can only become smoother, 
but the number of stars in our wide colour-magnitude boxes used for the sample selection remains essentially unchanged. 

Additionally, the most metal-rich parts of the high- and low-$\alpha$ populations of the 
RC APOGEE sample were removed from our AMR reconstruction process. Instead, we later add them to the predicted metallicity distributions 
in the form of the Gaussian error model. However, we cannot exclude the possibility that these high values of metallicity are 
realistic -- and not artifacts produced by the pipeline routine applied to the range of metallicities where hardly any standard stars are known.  
In this case, a small fraction of our mock stellar populations, of ages \mbox{$\tau \lesssim 1.0$ Gyr}, 
have metallicites underestimated by up to \mbox{$\sim0.2$ dex}. 
This may have several implications for the population synthesis.
The youngest dwarfs are located at the right (red) brim of the MS on the CMD. As increase in metallicity shifts an isochrone to redder colour, 
this means that in the case of an underestimated metallicity of the youngest stars, the width of our predicted MS is also slightly underestimated. 
For the F-star region on the CMD, the situation is inversed: the youngest F stars lay at the blue edge of the upper MS and, therefore, 
an underestimation of their metallicities means that the edge of the upper MS is shifted to bluer colours. 
Similarly, the RC population will appear too blue if the metallicity of its youngest stars is underestimated. 
However, none of these potential biases on the predicted CMD can introduce any significant difference to our sample statistics. 
Our colour-magnitude windows are wide enough so that small shifts of the MS and other CMD features cannot result in any significant change in 
the number of stellar assemblies falling into them. Therefore, an influence of this assumption on our final results is negligible.

\subsection{Local star-formation history}\label{sect:discus_sfr}

The SFR of the Solar neighbourhood, as a key ingredient of dynamical and chemical evolution models, 
was probed in many studies with different techniques. 
However, there is still no full consistency between the results of these attempts. 

\citet{aumer09} used the local MS and turn-off stars common between \textit{Hipparcos} and GCS, 
and simultaneously constrained the local kinematics and the SFR. To do so, 
they fitted the observed $B-V$ colour distributions while assuming 
a disk heating function that monotonously increases with age along with  
the IMF from \citet{kroupa93}, with its slope for \mbox{$m > 1$ M$_\odot$} varied in a 
small range \mbox{around $-2.7$}. They also adopted a reddening model from \citet{vergely98}, 
corrected for the sample incompleteness, and used Padova isochrones to account for the stellar evolution. 
They tested different SFR shapes, including single- and double-exponential forms, a non-exponential decline, 
and a smooth SFR with an overlaid oscillatory component. 
These authors found that the irregular SFR is not favoured in their analysis and their best fit suggests an exponentially declining SFR. 
Similar results were obtained by \citet{bovy17a}. 
In this study, a local sample from the TGAS catalogue of the \mbox{\textit{Gaia} DR1} was investigated 
and the long-lived \mbox{K dwarfs} were used to reconstruct the local SF history. The observed luminosity function of the sample
was converted to the mass function, assuming the IMFs from \citet{kroupa01} and \citet{chabrier01}, which was then used to derive the local SFR 
with the help of MS lifetimes and correction for the disk thickness. The author found that the local SFR reconstructed from the TGAS data 
can be perfectly fitted with an exponentially declining law, consistently with \citet{aumer09}. 
Another interesting study was presented in \citet{snaith15}, where the time evolution of [Si/Fe]  
was used as a marker of the SF history. For the inner disk, \mbox{$R\lesssim 8$ kpc}, their best-fit SFR 
has a strong peak at ages \mbox{$10-12$ Gyr} 
and about \mbox{1 Gyr} quenching period \mbox{8 Gyr ago,} followed by a quasi-linear behaviour for the last \mbox{7 Gyr}. 
This result was found to be insensitive to the variation of different parameters of the assumed chemical evolution model, including the IMF shape. 
\citet{xiang18} presented a study based on a sample of \mbox{$\sim1$ million} MS \mbox{turn-off} and subgiant stars from the LAMOST Galactic Survey 
with determined Bayesian isochrone ages. The authors performed a detailed reconstruction and analysis 
of the 3D stellar mass distribution of mono-age populations
across the Galactic disk. They fitted vertical stellar distributions with $sech^n$ profiles 
and used the obtained scale heights to convert mid-plane mass densities to the present-day surface density as a function of age. 
With stellar evolution taken into account, this gives the SFR at different Galactocentric distances. 
Their SFR at \mbox{$R=8$ kpc} has a peak \mbox{$\sim 6$ Gyr ago} and declines monotonously up to the present day. 
\citet{frankel19} studied the APOGEE RC stars in a wide range of Galactocentric distances, 
$6 \ \text{kpc} \, < R < 13 \ \text{kpc}$, 
and derived a radially dependent SFR of the $\alpha$-low disk, with radial migration and chemical enrichment included in a parametric form. 
Their SFR was prescribed to cover the last \mbox{8 Gyr} in age and follow an exponential law, 
thus, it was not possible to investigate the SFR shape in this study, as it was fixed. 
But an interesting outcome of this modelling is a prediction of an inside-out growth of the disk $-$ 
a formation scenario that can successfully reproduce elemental abundances across the disk \citep{chiappini01,colavitti09,grisoni18}.

On the other hand, there are a number of works that report a complicated, non-monotonous behaviour of the local SFR for the last \mbox{$5-6$ Gyr}. \citet{hernandez00} used Bayesian technique to reproduce the CMD 
of the \textit{Hipparcos} complete volume-limited sample and determined the local SF history 
over the last \mbox{3 Gyr} with a high \mbox{50-Myr} resolution. 
The authors found that the reconstructed SFR can be represented as a constant continuum with an oscillatory component with a period of \mbox{0.5 Gyr}. 
\citet{rocha-pinto00b} derived the local SFR by applying the scale-height, stellar-evolution, and volume corrections 
to chromospheric ages of \mbox{552 late-type} dwarfs. The obtained SFR has several episodes of SF enhancement, 
at \mbox{$0-1$ Gyr}, \mbox{$2-5$ Gyr}, and \mbox{$7-9$ Gyr} ago, with the oldest burst being uncertain. 
Quite a different result was obtained by \citet{cignoni06} who used bright \textit{Hipparcos} stars within \mbox{80 pc} from the Sun 
and performed a Bayesian fit of the observed CMD with an MCMC technique. The recovered SFR relies on the Pisa evolutionary library \citep{cariulo04}, a
power-law IMF with a slope $-2.35$, and the AMR as observed for GCS stars within \mbox{40 pc} from the Sun. Their best SFR 
is essentially bimodal: epochs of enhanced SF are present for ages \mbox{$2-6$ Gyr} and \mbox{$10-12$ Gyr}, 
with the recent burst \mbox{$\sim2$ times}
more enhanced than the older one. This result is consistent with the SFR determined by \citet{vergely02}, where 
a very similar Bayesian inversion technique was applied to \textit{Hipparcos} stars. 
A bimodal SFR was also obtained by \citet{rowell13}. They constrained the SFR by fitting the SDSS white-dwarf luminosity function
and found broad SF peaks at \mbox{$2-3$ Gyr} and \mbox{$7-9$ Gyr} ago. A comparable result 
was recently presented by \citet{bernard18}, where a recovered dynamically evolved SFR 
demonstrates peaks at ages of \mbox{14 Gyr} and \mbox{$2-3$ Gyr}. 
Again, this analysis was based on fitting the present-day CMD, which was constructed with the TGAS stars within \mbox{250 pc} from the Sun. 
The best fit was based on the BaSTI stellar evolution library \citep{pietrinferni04}, and a correction for the disk thickening was applied to 
extrapolate the result to the whole local cylinder. 

To sum up, most of the authors agree that there was a SF peak in the remote past, \mbox{$10-14$ Gyr} ago. 
A number of authors also agree that there were one or multiple SF enhancement epochs during the last \mbox{$\sim6-7$ Gyr}, 
though the relative weights of the early and recent SF peaks are quite different in different studies. 
We can mention several reasons for this variety among the results. Different authors often use different stellar evolution libraries, 
applying different IMF, AVR, and kinematics as ingredients of their dynamical or chemical evolution models used to fit the observed quantities. 
Also many studies are based on stellar ages, which are known to be challenging to constrain. As a result, the time resolution of SFR may be limited, 
and the SFR shape can be distorted due to systematic errors that affect ages. Commonly, only a small number of model parameters are fitted 
simultaneously, the influence of other ingredients such as the IMF and disk heating is tested post factum, 
which may also lead to non-robust results if the multidimensional PDF in the full parameter space has a 
complicated shape, which cannot be known in advance and revealed with a small number of tests. 

Our best total-disk SFR shown in \mbox{Figure \ref{fig:nsfr}} can be viewed as broadly consistent with studies 
of \citet{snaith15}, \citet{cignoni06}, \citet{rowell13}, and \citet{bernard18} 
as we also see two epochs of active star formation. However, our results do not agree with regard to the details concerning the shape of the overall SFR continuum. 
When compared to \citet{aumer09} and \citet{bovy17a}, we confirm their reported decline of the SFR to the present day, however, in our case, 
the exact shape of this decline is non-monotonous and non-exponential. The positions of the three SFR peaks reported by \citet{rocha-pinto00b} are 
comparable to our two recent SF bursts and a weak peak \mbox{$\sim9.5$ Gyr} ago due to maximum contribution from the thin-disk star formation. 
We also confirm the result presented in \citet{mor19} as our SF peak centered at \mbox{$\tau \approx 3$ Gyr} 
is very similar to the one found in their study; 
a recent SF enhancement at \mbox{$\tau \approx 0.5$ Gyr} found by us could not be recovered by \citet{mor19} due to their limited age resolution. 
Finally, our SFR is not consistent with the local SFR from \citet{xiang18} as both the global shape and small time-scale 
behaviour of two SFR functions are very different. 

Is is necessary to stress here that most of the parameters characterising these recent SF bursts are found quite 
difficult to constrain within our framework. Thus, some uncertainty still remains with respect to the exact shape and duration of the 
SF enhancement periods. According to our results, which show, for example, that the most recent epoch of the enhanced SF may be actively happening now
as \mbox{$\tau_\mathrm{p2} \approx 0$ Gyr} is not excluded (see the likelihood PDF for this parameter in \mbox{Figure \ref{fig:full_corner}}). 

Finally, we find it necessary to put our results in the framework of the radial migration problem. 
As we mentioned in \mbox{Section \ref{sect:intro}}, our local SFR can be roughly viewed as a disk formation history 
averaged over a radial annulus \mbox{R$_\odot\pm\Delta$R}, where $\Delta$R represents the effect of the radial migration and 
may be significantly larger than the half-width of the local annulus of \mbox{0.15 kpc}, 
where the data were selected (\mbox{Section \ref{sect:data_gaia_geom}}). 
However, this is a simplified picture as we know that the radial migration effect accumulates with time, such that 
the oldest populations observed locally are expected to be significantly more contaminated by migrators than the youngest ones. 
If we adopt the radial migration strength of 2.6 $ \mathrm{kpc}\times \sqrt{\tau/6 \ \mathrm{Gyr}}$ from \citet{frankel20}, 
then for the mean ages of our recent SF bursts of \mbox{0.5 Gyr} and \mbox{3 Gyr}, we get \mbox{$\sim$0.75 kpc} and \mbox{$\sim$1.8 kpc}. 
It is easy to convert this to the distributions over R$_{\mathrm{birth}}$ within the local annulus \mbox{R$_\odot \pm0.15$ kpc}, 
for simplicity assuming an exponential radial density fall-off with the scale length of \mbox{2.0 kpc} 
(consistent with the value for the metal-rich thin-disk population found by us in \citealp{sysoliatina18a}). 
Due to the negative density gradient with R, 
local R$_{\mathrm{birth}}$ distributions will be shifted towards the inner disk, with older stars representing, on average, smaller R: 
for \mbox{$\tau=0.5$ Gyr} distribution peaks at \mbox{$\sim8$ kpc}, and for \mbox{$\tau=3$ Gyr} the peak is at \mbox{$\sim7.4$ kpc}. 
From the cumulative distributions, we estimate that 68\% of the local \mbox{0.5-Gyr} and \mbox{3-Gyr} old stars were born 
within \mbox{$\sim$0.55 kpc} and \mbox{$\sim$1.5 kpc} from R$_\odot$, respectively. 
Thus, the additional assumption on the strength of radial migration does not disprove our findings, 
but simply gives us an idea of the range of Galactocentric distances represented by our `local' SFR at the given ages.

The recent epochs of an increased SF activity in the Solar neighbourhood recovered in our study 
may indicate the recent gas infall episodes. Also, the most recent increase of the SF may result from an increase of the SF 
efficiency related to an exhaustion of the local gas reservoir. In order to put more constraints on the recent SFR 
and clarify the origin of the SF bursts, a further and in-depth study of the local young populations is required.

\section{Conclusions}\label{sect:final}

In this work, we present an updated version of our semi-analytic dynamic model of the Milky Way disk. 
The new features of the the JJ model are summarised in the following:
\begin{enumerate}[label={(\arabic*)}]
 \item We introduced a new simple gas model that consists of two isothermal populations representing molecular and atomic gas components.
 They are characterised by the local surface densities and scale heights that are adopted from \citet{mckee15} and \citet{nakanishi16}, respectively. 
 \item We linked the $W$-velocity dispersion of a zero-age thin-disk stellar population to the $W$-velocity dispersion of the molecular gas, 
 with the latter determined during an iterative solving of the Poisson-Boltzmann equation. 
 This allows us to set a constraint on one of the AVR parameters and, therefore, to reduce the total number of free parameters in the model. 
 \item Assuming a parallel evolution of the thin and thick disk, we defined an extended in time thick-disk SFR. 
 It has a rapid power-law increase, 
 peaks at age of \mbox{$\sim12.5$ Gyr} and declines exponentially, such that the thick-disk SF drops to zero \mbox{$\sim10$ Gyr} ago 
 (\mbox{Figure \ref{fig:nsfr}}, orange curve). 
 \item A new power-law analytic form of the thin-disk SFR is presented, 
 such that the new \mbox{JJ model} with an updated disk treatment is consistent with the results of the previous works in this series, 
 \citetalias{just10}-\citetalias{rybizki15}.  
 We also allowed two recent SF bursts at ages \mbox{$0-1$ Gyr} and \mbox{$2-4$ Gyr} added to the thin-disk SFR in the form of Gaussian peaks. 
 Additionally, in order to reproduce the observed vertical number density trends of A and F stars, 
 we decoupled the vertical kinematics of the thin-disk populations associated with the assumed SF bursts 
 from the kinematics of the rest of thin-disk populations prescribed by the AVR. 
 To do so, we introduced parameters, $\sigma_\mathrm{p1}$ and $\sigma_\mathrm{p2},$ 
 which give the $W$-velocity dispersion of the stellar populations associated with the SF excess during the 
 recent epochs of the enhanced SF. 
\end{enumerate}

We used stars from \textit{Gaia} DR2 selected with simple colour-magnitude cuts on the CMD. 
The six samples included young stars of A and F spectral classes, mixed-age sample dominated by RC giants, 
and three sets of the long-lived G and K dwarfs probing the lower MS region. An additional sample of \textit{Gaia} stars 
was selected in two cones directed towards the northern and southern Galactic poles 
(see \mbox{Section \ref{sect:data_gaia}} and \mbox{Table \ref{tab:data}}). 

Using the local APOGEE RC stars, we constrained the AMR in the Solar annulus, 
assuming a separate chemical evolution of the thin and thick disk (\mbox{Figure \ref{fig:get_amr}}). 
The derived AMR is used in the stellar population synthesis with the PARSEC stellar library. 
Then, working within a Bayesian framework, we simultaneously optimised the \mbox{JJ model} performance in terms of 
the vertical kinematics and number density profiles of the selected populations, as well as 
in terms of the apparent Hess diagram of the conic sample. 
We searched for a maximum probability region in a multidimensional model parameter space using the MCMC technique 
and self-consistently constrained 22 model parameters. These parameters include the local surface densities of the four Galactic components, 
thin- and thick-disk vertical kinematics, the shape of the IMF, and the thin-disk SFR function. 
These derived parameter values were also brought into consistency with the model-dependent AMR constrained with the APOGEE data. 

We find that a monotonously declining SFR is inconsistent with the \textit{Gaia} data and the observed local star counts imply 
two SF bursts in the recent past centered at ages \mbox{$\sim0.5$ Gyr} and \mbox{$\sim3$ Gyr}. 
These bursts are characterised by \mbox{$\sim30$\%} and \mbox{$\sim55$\%} increase of the SF relative 
to the underlying declining continuum, respectively. 
The thin-disk stellar populations associated with this SF excess were found to have $W$-velocity 
dispersions of \mbox{$\sim12.5$ km s$^{-1}$} and \mbox{$\sim26$ km s$^{-1}$}, which is $\sim1.8$ times higher  
than the velocity dispersion prescribed by the monotonously declining AVR for the thin-disk populations of the same age. 
The relative contribution of these heated populations to the overall local surface density of the disk 
is estimated as \mbox{$\sim0.8$\%} and \mbox{$\sim5.3$\%}. 
The new local JJ model is able to reproduce the overall star counts with \mbox{$\sim5$-\%} accuracy for \mbox{$|z| \lesssim 600$ pc} 
(\mbox{Figure \ref{fig:loc_all}}) and with \mbox{$\sim 6-8$-\%} accuracy in the conic volume towards the Galactic poles 
(\mbox{Figure \ref{fig:hess}}).   

In summary, the spatial distribution and motions of the local stars as observed by \textit{Gaia} 
can be well reproduced within a framework of our plane-symmetric semi-analytic model 
and a calibration of the model against the high-quality astrometric and spectroscopic data reveals 
previously unknown details of the Milky Way's disk evolution.

\section*{Acknowledgements}

{\tiny{This work was supported by the Deutsche Forschungsgemeinschaft 
(DFG, German Research Foundation)  -- Project-ID 13871353 --
SFB 881 (`The Milky Way System', subproject A06). 

This work has made use of data from the European Space Agency (ESA)
mission {\it Gaia} (\url{https://www.cosmos.esa.int/gaia}), processed by
the {\it Gaia} Data Processing and Analysis Consortium (DPAC,
\url{https://www.cosmos.esa.int/web/gaia/dpac/consortium}). Funding
for the DPAC has been provided by national institutions, in particular
the institutions participating in the {\it Gaia} Multilateral Agreement.

Besides the software mentioned in the text,
this research made use of Python packages SciPy \citep{scipy-nmeth20} and NumPy \citep{numpy11}, a Python library for publication quality graphics 
matplotlib \citep{hunter07}, a community-developed core Python package for Astronomy Astropy \citep{astropy13,astropy18}, 
a Python mini-package \texttt{fast-histogram} (\url{https://github.com/astrofrog/fast-histogram}), as well as
an interactive graphical viewer and editor for tabular data TOPCAT \citep{taylor05}. 

We also thank Jan Rybizki for the helpful discussion and 
Alex Razim for many useful comments on the paper structure and tips on the text style. 

Finally, we thank organisers of the inspiring seminar `The Science Cloud' (12-15.01.2020, Bad Honnef, Germany) funded  
by the Wilhelm and Else Heraeus-Foundation.}}

\bibliographystyle{./aa.bst}
\bibliography{gdr2_calibr.bib}

\begin{appendix}

\section{TAP queries}\label{sect:append_tap}

\begin{minipage}{17.0cm}
\centering
{\tiny
\begin{tabular}{m{17.0cm}} 
\hhline{-} \smallbreak
\texttt{\textbf{SELECT} source\_id, ra, dec, pmra, pmdec, parallax, radial\_velocity,phot\_g\_mean\_mag, bp\_rp,}\\
\quad \texttt{ra\_error, dec\_error, pmra\_error, pmdec\_error, parallax\_error, radial\_velocity\_error,}\\ 
\quad \texttt{dec\_parallax\_corr, dec\_pmdec\_corr, dec\_pmra\_corr, parallax\_pmra\_corr, parallax\_pmdec\_corr,}\\ 
\quad \texttt{pmra\_pmdec\_corr, ra\_dec\_corr, ra\_parallax\_corr, ra\_pmdec\_corr, ra\_pmra\_corr,}\\
\quad \texttt{phot\_bp\_rp\_excess\_factor, astrometric\_excess\_noise, astrometric\_excess\_noise\_sig,}\\
\quad \texttt{RADIANS(b) as b\_rad, RADIANS(l) as l\_rad,}\\
\quad \texttt{\mbox{QUAD($R_\odot$*$R_\odot$ + COS(b\_rag)*COS(b\_rad)/parallax/parallax - 2*$R_\odot$*COS(b\_rad)*COS(l\_rad)/parallax) as rg}}\\
\texttt{\textbf{FROM} gaiadr2.gaia\_source}\\
\texttt{\textbf{WHERE} \textbf{AND} phot\_g\_mean\_mag > 7 \textbf{AND} phot\_g\_mean\_mag < 17 \textbf{AND} bp\_rp > 0 \textbf{AND} bp\_rp < 3.3} \\
\quad \texttt{1000/parallax < 600 \textbf{AND} rg > 8.05 \textbf{AND} rg < 8.35}\\[0.5em]  
\hhline{-}
\end{tabular} 
\label{tab:query1}}
\end{minipage} 

\vspace*{0.5cm}

\begin{minipage}{17.0cm}
\centering
{\tiny 
\begin{tabular}{m{17.0cm}} 
\hhline{-} \smallbreak 
\texttt{\textbf{SELECT} source\_id, l, b, ra, dec, parallax, phot\_g\_mean\_mag, bp\_rp, phot\_bp\_rp\_excess\_factor}\\
\texttt{\textbf{FROM} gaiadr2.gaia\_source} \\
\texttt{\textbf{WHERE} ABS(b) > 80 \textbf{AND} bp\_rp > 0 \textbf{AND} bp\_rp < 3.5 
                                   \textbf{AND} phot\_g\_mean\_mag > 12 \textbf{AND} phot\_g\_mean\_mag < 17 }\\[0.5em] 
\hhline{-}
\end{tabular} 
\label{tab:query2}}
\end{minipage} 


\section{Impact of dust maps}\label{sect:append_dust}

\begin{figure*}[hb!]
\centering
\includegraphics[scale=0.5]{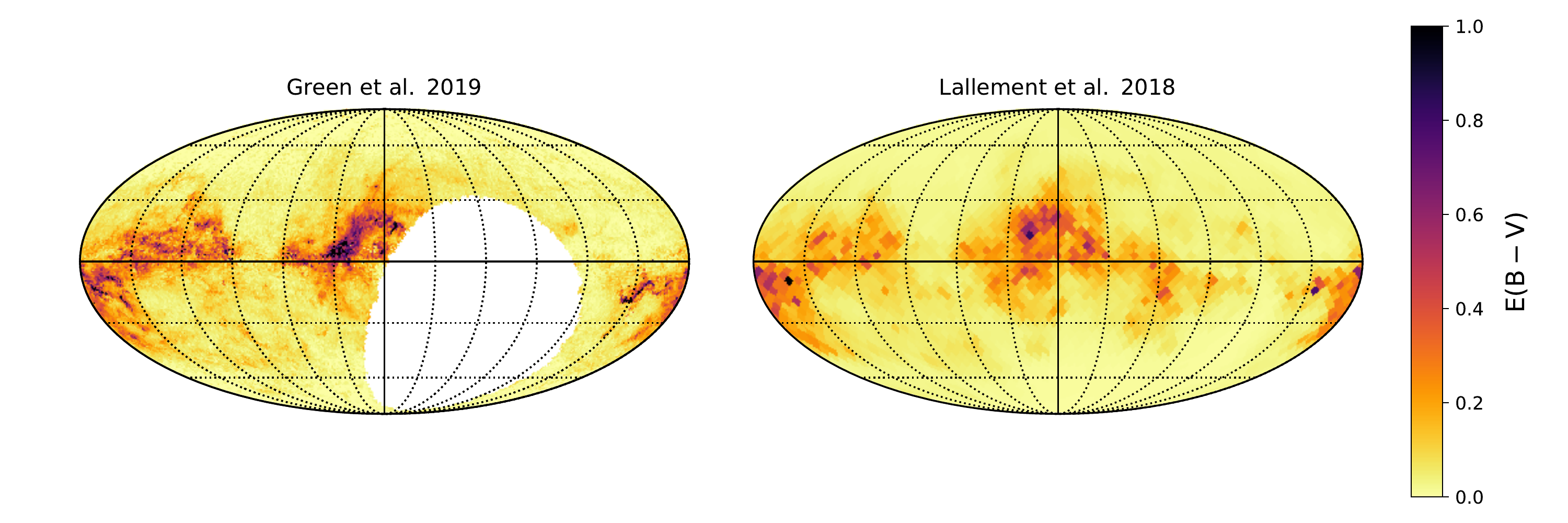}
\caption{Extinction maps used in this study. Colour-coding displays the cumulative 
reddening up to \mbox{500 pc} from the Sun.}
\label{fig:ext_maps}
\end{figure*}

We use a probabilistic extinction map constructed using \textit{Gaia} parallaxes 
and photometry from 2MASS and PanSTARRS \citet{green19} (\mbox{Figure \ref{fig:ext_maps}}, left panel). 
This map has a high resolution, on the order of $\sim10^\prime$, 
and covers a large range of heliocentric distances, up to \mbox{$\sim10$ kpc}. 
However, it does not cover the whole sky, so we complement it with another 3D dust map from \citet{lallement18}. 
This map is based on the APOGEE red giant stars with 
reddening estimated from \textit{Gaia} and 2MASS photometry and photometric distances. 
It has significantly lower resolution and smaller spatial coverage:
it extends \mbox{2 kpc} away from the Sun in the Galactic plane, but goes only as high as \mbox{$|z|=600$ pc} in the vertical direction. 
We downloaded the distance reddening curves using the online tool \textit{Stilism}\footnote{\url{https://stilism.obspm.fr/}. \newline 
An updated version of the map has been recently presented in \citet{lallement19} where \textit{Gaia} parallaxes 
were used as a proxy of distance to the APOGEE target stars, but this new dust map is not yet available in \textit{Stilism}.}
for \mbox{$3 \, 072$ line-of-sights} that corresponds to the angular resolution of $\sim3.66^\circ$  
(\mbox{Figure \ref{fig:ext_maps}}, right panel). Both maps have similar large-\newpage 
\vspace*{102mm}

scale features, but the map from \citet{green19} provides many small-scale details 
and, therefore, it is more preferable. 

\begin{figure*}
\begin{minipage}[c]{0.5\textwidth}
 \includegraphics[scale=0.55]{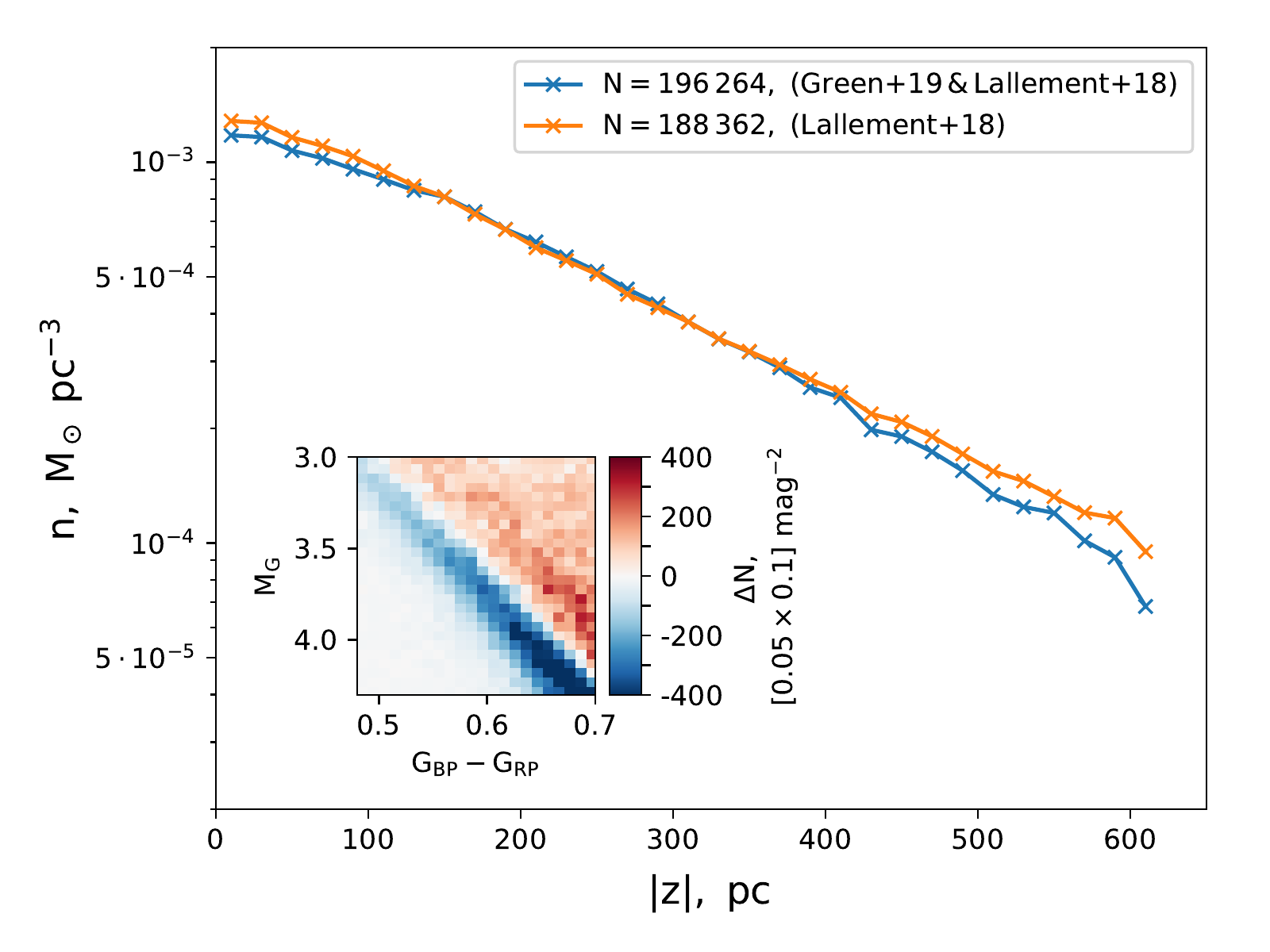}
\end{minipage}\hfill
\begin{minipage}[c]{0.4\textwidth}
    \caption{Vertical number density profiles of two F-star samples. The samples are selected in the colour-magnitude window 
    defined in \mbox{Table \ref{tab:data}}
    using \textit{Gaia} DR2 colours and magnitudes de-reddened with two different dust maps from \mbox{Figure \ref{fig:ext_maps}}. 
    The inset plot shows the absolute difference between these two F-star samples over the Hess diagram.}\label{fig:ext_maps_f_impact}
\end{minipage}
\end{figure*}

We selected our \textit{Gaia} samples (\mbox{Table \ref{tab:data}}) using both maps and then compare 
the sample statistics. We find that the overall  
number of stars in the samples may change by \mbox{$\sim4$\%} maximum, 
with the largest impact observed for the F-star sample. 
Indeed, in the case of F stars we select a CMD region with a strong gradient in the stellar density 
(\mbox{Figure \ref{fig:ext_maps_f_impact}}, inset), 
such that a small shift along colour and magnitude axes due to a change in reddening and extinction leads 
to a significant change in the number of stars within this region. 
\mbox{Figure \ref{fig:ext_maps_f_impact}} shows that the corresponding difference in the vertical profile shape is small.
We do not find any impact from this effect on the velocity distributions of our samples.

In summary, changing the dust model may produce only a small impact on some of the analysed quantities and, therefore, this step 
would not  influence the optimised values of the \mbox{JJ model} parameters determined in this work. 

\vspace*{0.1cm}

\section{Full MCMC output}\label{sect:append_mcmc}

\begin{figure*}[hb!]
  \centering
  \vspace*{-20mm}
  \includegraphics[scale=0.133]{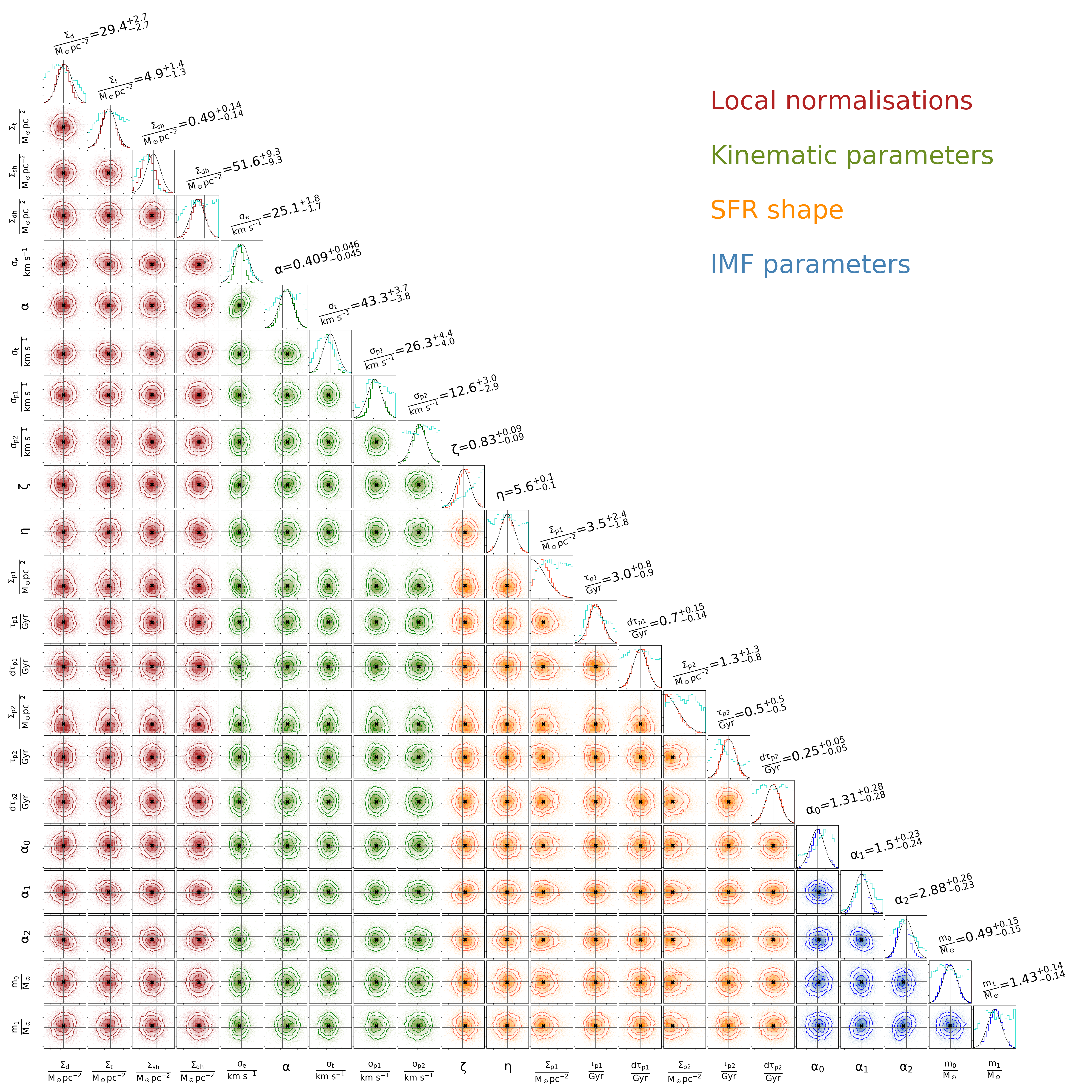} 
\caption{Same as in Figure \ref{fig:sub_corner}, but the full posterior PDF.}
\label{fig:full_corner}
\end{figure*}

\end{appendix}

\end{document}